\documentclass[%
aps,
prd,
amssymb,
amsmath,
amsfonts,
eqsecnum,
showpacs,
nofootinbib,
floatfix,
twocolumn,
twoside,
a4paper,
superscriptaddress,
longbibliography
]{revtex4-2}

\usepackage{graphicx}
\usepackage{dcolumn}
\usepackage{bm}
\usepackage{tabularx}
\usepackage{multirow}
\usepackage{capt-of}
\usepackage{color}
\usepackage{url}
\usepackage[utf8]{inputenc}
\usepackage{xcolor} 
\usepackage[table]{xcolor}
\usepackage[dvipsnames]{xcolor}
\usepackage[svgnames]{xcolor}
\usepackage[x11names]{xcolor}
\usepackage{booktabs}

\usepackage[nolist,nohyperlinks]{acronym}
\usepackage{xspace}
\usepackage[colorlinks]{hyperref}
\usepackage[normalem]{ulem}
\usepackage{longtable}
\usepackage{makecell}

\usepackage[caption=false]{subfig}

\DeclareMathAlphabet{\mathbfsf}{\encodingdefault}{\sfdefault}{bx}{sl}

\newcommand{\be}{\begin{equation}}
\newcommand{\ee}{\end{equation}}
\newcommand{\bea}{\begin{eqnarray}}
\newcommand{\eea}{\end{eqnarray}}

\newcommand{\phX}{\texttt{IMRPhenomXAS}\xspace}

\newcommand{\phXPHM}{\texttt{IMRPhenomXPHM}\xspace}
\newcommand{\phT}{\texttt{IMRPhenomT}\xspace}

\newcommand{\phXE}{\texttt{IMRPhenomXE}\xspace}

\newcommand{\phTE}{\texttt{IMRPhenomTE}\xspace}
\newcommand{\phTEHM}{\texttt{IMRPhenomTEHM}\xspace}
\newcommand{\teobEHM}{\texttt{TEOBResumS-Dalí}\xspace}
\newcommand{\seobEHM}{\texttt{SEOBNRv5EHM}\xspace}
\newcommand{\seobE}{\texttt{SEOBNRv5E}\xspace}

\definecolor{dodgerblue}{HTML}{1E90FF}
\definecolor{viennared}{HTML}{DA0A14}
\definecolor{ctorange}{HTML}{FF6C0C}
\definecolor{granadagreen}{HTML}{078931}
\definecolor{wales}{HTML}{ff0038}
\definecolor{valenciacfred}{HTML}{ee3524}
\definecolor{barcelonafcgold}{HTML}{edbb00}
\definecolor{jam}{HTML}{A50B5E}
\definecolor{austriawien}{HTML}{441678}

\AtBeginDocument{%
  \hypersetup{
    citecolor=dodgerblue,
    linkcolor=dodgerblue,   
    urlcolor=dodgerblue}}

\newcommand{\UIB}{Departament de F\'isica, Universitat de les Illes Balears, IAC3 -- IEEC, Crta. Valldemossa km 7.5, E-07122 Palma, Spain}

\newcommand{\Nikhef}
{National Institute for Subatomic Physics (Nikhef), Science Park 105, 1098 XG, Amsterdam, The Netherlands}

\allowdisplaybreaks[1]

\begin{document}

\title[IMRPhenomXE]{Fast frequency-domain phenomenological modeling of eccentric aligned-spin binary black holes}
 
\author{Antoni Ramos-Buades}
\affiliation{\UIB}


\author{Quentin Henry}
\affiliation{\UIB}


\author{Maria Haney}
\affiliation{\Nikhef}

\date{\today}

\begin{abstract}
We present the  \phXE frequency-domain phenomenological waveform model for the dominant mode of 
inspiral-merger-ringdown non-precessing binary black holes in elliptical orbits. 
\phXE extends the quasi-circular \phX waveform model for the dominant $(\ell, |m|) =$ (2,2) modes to eccentric binaries.
For the inspiral part, orbit-averaged equations of motion within the quasi-Keplerian parametrization up to third post-Newtonian order,
including spin effects, are evolved, and the waveform modes are computed using the stationary phase approximation 
on eccentricity expanded expressions up to $\mathcal{O}(e^{12})$.
The  model assumes circularization at merger-ringdown, where it adopts the underlying quasicircular \phX baseline. 
We show that \phXE reduces to the accurate \phX model in the quasi-circular limit. 
Compared against 186 public numerical relativity waveforms from the Simulating eXtreme Spacetimes catalog with initial eccentricities up to $~0.8$, \phXE provides values of unfaithfulness below $3\%$ for $72\%$ of simulations with initial eccentricities below 0.4. For larger eccentricities, the unfaithfulness degrades up to $\gtrsim 10\%$ due to the underlying small eccentricity expansions and additional modelling approximations. In terms of speed, \phXE outperforms any of the existing inspiral-merger-ringdown eccentric waveform models.
We demonstrate the efficiency, robustness, and modularity of \phXE through injections into zero noise and parameter-estimation analyses 
of gravitational-wave events, showing that \phXE is a ready-to-use waveform model for gravitational-wave astronomy 
in the era of rapidly growing event catalogs.
\end{abstract}


\maketitle


\section{Introduction}
\label{sec:Introduction}
Roughly a decade from the first direct observation of gravitational waves (GWs) in September 2015 \cite{LIGOScientific:2016aoc}, the LIGO-Virgo-KAGRA collaboration \cite{LIGOScientific:2018mvr,LIGOScientific:2019lzm,LIGOScientific:2020ibl,LIGOScientific:2021usb,LIGOScientific:2021djp,LIGOScientific:2025slb}
and independent groups \cite{Venumadhav:2019lyq,Nitz:2021uxj,Nitz:2021zwj,Olsen:2022pin,Wadekar:2023gea,Mehta:2023zlk} have detected more than two hundred signals. All of these events are consistent with compact binary coalescences of binary black holes (BBHs), 
binary neutron stars (BNSs)
 or neutron-star black-hole (NSBH) systems.

The origin of the compact objects observed by the LVK detectors~\cite{KAGRA:2013rdx,LIGOInstrumentWhitePaper,VirgoInstrumentWhitePaper} 
is generally associated with two main astrophysical scenarios: isolated binary evolution and dynamical
 formation ~\cite{Bethe:1998bn,Belczynski:2001uc,Belczynski:2014iua,mennekens2014massive,Belczynski:2016obo,Eldridge:2016ymr,Marchant:2016wow,Stevenson:2017tfq,Giacobbo:2018etu,Kruckow:2018slo,Kruckow:2018slo,Mandel:2018hfr,Zevin:2020gbd,Mapelli:2020vfa,Karathanasis:2022rtr,Bouffanais:2021wcr,LIGOScientific:2021psn}. While isolated binary evolution predicts binaries with
 negligible orbital eccentricity by the time they enter the frequency band of ground-based detectors \cite{Peters:1964zz}, 
 dynamical formation is the main mechanism to form binaries in active astrophysical 
 environments such as globular clusters, and can lead to eccentric 
 binaries emitting GWs in the detectors' frequency band. Thus, orbital eccentricity
  is one of the smoking guns to decipher the origin of the observed population of compact binaries. 

Several studies have been performed to infer signatures of orbital eccentricity in detected 
GW events. Early works ~\cite{Romero-Shaw:2020aaj,Romero-Shaw:2019itr,Romero-Shaw:2021ual, Romero-Shaw:2022xko} 
found evidence of at least three GW signals, the events GW190620, GW191109, and GW200208\_22. More recent studies, 
Ref.~\cite{Gupte:2024jfe} analyzed 57 GW events finding support of eccentricity for the events GW190701, GW200129, and GW200208\_22,  
Ref.~\cite{Planas:2025jny} analyzed 17 GW events and found support GW200129, GW200208\_22, GW190701 and GW190929, 
while Ref.~\cite{Iglesias:2022xfc} analyzed several GW events such as  GW190929 and GW190521 without finding 
evidence for eccentricity. Furthermore, some studies have focused on measuring the eccentricity of particularly
 interesting individual GW events, such as  GW190521 where Refs.~\cite{Romero-Shaw:2020thy,Gayathri:2020coq,Gamba:2021gap} 
 found evidence for eccentricity or a hyperbolic capture of a BBH. In contrast 
 Refs.~\cite{Iglesias:2022xfc,Bonino:2022hkj,Ramos-Buades:2023yhy,Gupte:2024jfe,Planas:2025feq} have shown no evidence of eccentricity.  Recently, there 
 has been also some further investigations on the eccentric nature
  of GW200208\_22 \cite{Romero-Shaw:2025vbc}. Moving to systems with matter content,
  Ref. \cite{Morras:2025xfu} found evidence of orbital eccentricity 
  in GW200105 using an inspiral-only eccentric precessing-spin waveform model, 
  and Ref.~\cite{Planas:2025plq} obtained support for eccenticity using a 
  inspiral-merger-ringdown (IMR) eccentric non-precessing spin model. Recently, Refs. \cite{Kacanja:2025kpr,Jan:2025fps,Tiwari:2025fua} also analyzed GW200105 
  and found support for eccentric signatures in GW200105.
  All these studies show the need to develop accurate and computationally efficient waveform models that include the effects of eccentricity and which can be used to gauge 
  the systematics in the measurements of eccentricity. 

The construction of accurate eccentric waveform models relies on calibration 
and comparisons to numerical-relativity (NR) simulations with eccentricity effects \cite{Huerta:2019oxn,Ramos-Buades:2019uvh,Ramos-Buades:2022lgf,Healy:2022wdn,Bonino:2024xrv,Nee:2025zdy,Knapp:2024yww,Scheel:2025jct}.
 Nevertheless, the increase of dimensionality of the binary parameter 
 space due to eccentricity (2 additional parameters) and the cost of
  producing these simulations has mostly stimulated the development of 
  models which combine numerical results with perturbative solutions of the Einstein equations.

Over the last years, many eccentric waveform models have been constructed. 
Based on post-Newtonian (PN) theory~\cite{Junker:1992kle,Gopakumar:1997bs,Gopakumar:2001dy,Memmesheimer:2004cv,Damour:2004bz,Konigsdorffer:2006zt,Arun:2007rg,Arun:2007sg,
Arun:2009mc,Mishra:2015bqa,Boetzel:2019nfw,Ebersold:2019kdc,Henry:2023tka,
Boetzel:2017zza, Paul:2022xfy}, inspiral-only eccentric models, some of them
 including spin-precession effects have been developed~\cite{Yunes:2009yz,
 Cornish:2010cd,ShapiroKey:2010cnz,Huerta:2014eca,Moore:2016qxz,
 Loutrel:2017fgu,Tanay:2016zog,Tanay:2019knc,Tiwari:2020hsu,Tiwari:2019jtz,
 Moore:2018kvz,Moore:2019xkm,Klein:2018ybm, Klein:2021jtd, Sridhar:2024zms,
 Morras:2025nlp}. Furthermore complete IMR eccentric models were developed 
 in Refs.~\cite{Huerta:2016rwp,Huerta:2017kez,Hinder:2017sxy,Islam:2024tcs,
 Ramos-Buades:2019uvh,Chattaraj:2022tay,Manna:2024ycx,Paul:2024ujx}.

State-of-the-art IMR eccentric models can be divided in three main families: the Effective-One-Body (EOB) 
formalism~\cite{Buonanno:1998gg,Buonanno:2000ef,Damour:2000we,Buonanno:2005xu}, including the 
SEOBNR models~\cite{Bohe:2016gbl, Cotesta:2018fcv,Ossokine:2020kjp,Ramos-Buades:2021adz,Pompiliv5,RamosBuadesv5, Gamboa:2024hli, Liu:2019jpg, Cao:2017ndf,Estelles:2025zah} 
and TEOBResumS models~\cite{Akcay:2020qrj, Nagar:2018plt, Nagar:2018zoe, Nagar:2020pcj,Nagar:2024oyk,Albanesi:2024xus, Gamba:2024cvy}; 
the NRSurrogate approach~\cite{Blackman:2017dfb,Varma:2018mmi,Varma:2019csw,Islam:2022laz,Rink:2024swg,Islam:2024zqo,Maurya:2025shc,Nee:2025nmh} 
and the IMRPhenom framework~\cite{Husa:2015iqa,Khan:2015jqa,
London:2017bcn,Pratten:2020ceb,Garcia-Quiros:2020qpx,Estelles:2021gvs,
Husa:2015iqa,Khan:2015jqa,London:2017bcn,Pratten:2020fqn,Pratten:2020ceb,
Garcia-Quiros:2020qpx,Estelles:2020twz, Estelles:2020osj, Estelles:2021gvs,Thompson:2023ase,Colleoni:2024knd,Planas:2025feq}. 
The EOB eccentric waveform models have reached a state of maturity both 
in the \texttt{SEOBNR} and \texttt{TEOBResumS} families, with the 
construction of the accurate multipolar aligned-spin \seobEHM \cite{Gamboa:2024imd} and \teobEHM \cite{Nagar:2024oyk} models, 
as well as the development of the first precessing-spin eccentric models \cite{Liu:2023ldr,Gamba:2024cvy}. 
The eccentric NRSurrogate models \cite{Islam:2022laz,Islam:2024zqo,Maurya:2025shc,Nee:2025nmh} interpolate directly NR data 
making them the most accurate models, however, regarding eccentric binaries 
the lack of NR simulations covering the eccentric parameter space is limiting their accuracy and applicability. Finally, the phenomenological 
family has recently produced the time-domain eccentric aligned-spin multipolar 
\phTEHM model \cite{Planas:2025feq}. The \phTEHM model is based on the quasi-Keplerian 
parametrization (QKP) to describe eccentricity effects during the inspiral phase, and 
assumes circularization at merger-ringdown, similar to the IMR EOB models. The \phTEHM 
model achieves an unfaithfulness lower than $2\%$ against eccentric NR 
waveforms \cite{Planas:2025feq}, it accurately describes eccentric binaries up to an 
eccentricity $e=0.4$ at 10Hz, and it is the state-of-the-art in computational efficiency
 of eccentric models in time-domain.  We henceforth use \phTE to refer to the \phTEHM model including only the $(2,|2|)$-modes.

All the previously described eccentric IMR models are in time-domain, 
in this paper we develop the first frequency-domain eccentric IMR waveform model, 
\phXE, which describes the dominant $\{(\ell,|m|)=(2,2)\}$ modes for non-precessing 
spin binaries with two eccentric parameters. The model extends the accurate quasicircular
 \phX model \cite{Pratten:2020fqn} to eccentric binaries. For the inspiral part eccentric 
 effects are included  by performing a numerical evolution of the 3PN quasi-Keplerian 
 evolution equations \cite{Henry:2023tka}, consistent with the one present in the \phTEHM model \cite{Planas:2025feq}, and then computing numerically the Stationary Phase Approximation (SPA) 
 on the third post-Newtonian (PN) time-domain waveform expanded up to $\mathcal{O}(e^{12})$ 
 order. At merger-ringdown \phXE assumes circularization of the binary, and it employs 
 the quasicircular ringdown model from \phX. We find that \phXE has a quasicircular 
 limit consistent with the underlying quasicircular model, and when compared against 
 186 public eccentric simulations from the Simulating eXtreme Spacetimes (SXS) catalog
\cite{Scheel:2025jct} the model provides an unfaithfulness $<3\%$ for $72\%$ of
simulations, with initial eccentricities $\lesssim 0.4$. 

Computational efficiency is a key feature of \phXE, which is implemented 
in the highly efficient \texttt{phenomxpy} Python package \cite{Garcia-Quiros:2025usi}. 
We show that \phXE is the most efficient IMR eccentric waveform model compared to the 
state-of-the-art eccentric IMR models. We demonstrate its applicability by performing
 Bayesian inference studies with NR injection on zero noise, and analyzing three 
 GW events GW150914, GW151226 and GW190521 with the \texttt{Bilby} python package \cite{Ashton:2018jfp,Romero-Shaw:2020owr}. We find consistent inference of the source parameters
  for these signals with previous results in the literature 
  \cite{Ramos-Buades:2023yhy,Gamboa:2024hli,Planas:2025feq}, 
  and no evidence of eccentricity for any of them.

The paper is organized as follows: 
We first discuss our notation and conventions
in Sec.~\ref{sec:notation}. We present the model building blocks in Sec. \ref{sec:model}, with an overview of the model in Sec.  \ref{sec:model_overview}, then the description of the time-domain dynamics in Sec. \ref{sec:model_td}, followed by the application of the stationary phase approximation to the eccentric inspiral in Sec. \ref{sec:model_fd}, and the construction of the IMR model in Sec. \ref{sec:model_imr}.  We then focus on the validation of the model
in Sec.~\ref{sec:validation}, where we introduce the notation for the unfaithfulness calculation in Sec. \ref{sec:NRunfaithfulness}, 
the study of the quasicircular limit in Sec. \ref{sec:QC_limit}, the accuracy assessment against eccentric NR simulations
in Sec. \ref{sec:ecc_NRcomparison}, and timing benchmarks in Sec. \ref{sec:benchmarks}.
Bayesian inference results are presented in Sec. \ref{sec:PE}, with studies on 
zero noise injections in Sec. \ref{sec:PE_nr} and real GW data in Sec. \ref{sec:PE_gw}.
We conclude the paper in Sec.~\ref{sec:conclusions}, and provide a list of 
the NR simulations used in this work in Appendix \ref{sec:nr_list}.

\section{Notation}\label{sec:notation}
We use geometric units $G=c=1$ unless explicitly stated.
We define the mass ratio $q = m_1/m_2 \geq 1$, total mass $M=m_1+m_2$, and symmetric mass ratio $\eta = m_1 m_2 /M^2$, where $m_{1,2}$ are the individual component masses.
We introduce the chirp mass $\mathcal{M} = M \eta^{3/5}$, a relevant combination of masses employed in GW data analysis. 

In this work we restrict to individual components of the spin vectors $\bm{S}_{1,2}$ aligned or anti-aligned with the direction of the orbital angular momentum $\hat{\bm{L}}$ of the binary. Henceforth, denoted as nonprecessing-spin binaries, and characterized by the dimensionless spin components, 
\begin{equation}
\chi_i = \frac{\bm{S}_i \cdot \hat{\bm{L}}}{m_i^2}, \quad i \in {1,2},
\label{eq:eqChiz}
\end{equation}
which range the interval $[-1,1]$.

In the source frame the GW signal of a nonprecessing-spin, 
eccentric binary can be described by the following set of 
intrinsic parameters $\Theta = \{m_{1,2}, \chi_{1,2}, e, l\}$, 
where $e$ denotes the initial orbital eccentricity parameter and $l$ the initial mean anomaly parameters\footnote{We note that the choice of orbital eccentricity and
 mean anomaly to describe the ellipticity of the orbit is not unique, and 
 there are other parametrizations of the orbit possible in terms of different anomalies, 
 radial parameters, etc.}. In vacuum General Relativity the total mass of
  the system is a scale parameter, and it is common in the waveform modelling community to work equivalently with the the mass ratio or symmetric mass ratio as the set of intrinsic parametes, i.e.,  $\Theta = \{q, \chi_{1,2}, e, l\}$.


\section{Model construction}\label{sec:model}

In this Section we describe the new \phXE waveform model. We provide an overview of the model in Sec. \ref{sec:model_overview}. A detailed description of the eccentric dynamics and the time-domain eccentricity effects in the waveform is provided in Sec. \ref{sec:model_td}. In Sec. \ref{sec:model_fd} we apply the stationary phase approximation to the inspiral time-domain waveform, and in Sec. \ref{sec:model_imr} we explain the procedure to construct the full eccentric IMR waveform.

\subsection{Overview}\label{sec:model_overview}
The frequency-domain GW polarizations $\tilde{h}_{+,\times}$ can be represented by the complex strain $h$, which at the same time can be decomposed in terms of the spin-weighted $-2$ spherical harmonic, $Y^{-2}_{lm}$, basis
\begin{equation}
\tilde{h}_+-i\tilde{h}_\times  \equiv h(f, \Theta) =  \sum_{\ell=2}^{\infty} \sum_{m=-\ell}^{\ell} Y^{-2}_{\ell m}(\iota, \varphi) \tilde{h}_{\ell m}(f, \Theta),
\label{eq:eqStrain}
\end{equation}
where $\tilde{h}_{\ell m}(f, \Theta)$ denote the GW modes, $f$ indicates the Fourier frequency, $(\iota, \varphi)$ are the inclination and azimuthal angle which describe the angular position of the 
line of sight measured in the source frame, and  $\Theta = \{m_{1,2}, \chi_{1,2}, e, l\}$ indicates 
the intrinsic parameters of the source\footnote{We indicate the frequency-domain waveform quantities 
with a tilde symbol, i.e. $\tilde{h}(f)$, to differentiate from the 
corresponding time-domain quantity $h(t)$.}. 

For binaries on a planar motion, i.e. nonprecessing-spin, the negative-$m$ modes can be recovered from the positive ones through the relation (in frequency domain)~\cite{Garcia-Quiros:2020qpx}
\begin{equation}
\tilde{h}_{\ell m} (f) = (-1)^\ell \tilde{h}^*_{\ell -m}(-f),
\label{eq:eqhlmNeg}
\end{equation}
where $^*$ denotes complex conjugation. As a consequence, 
we henceforth refer only to the positive $(\ell,m)=(2,2)$-mode 
taking into account that the negative $m$-mode can be directly obtained 
from the positive one through Eq. \eqref{eq:eqhlmNeg}.

The \phXE waveform model is built upon the accurate quasicircular 
nonprecessing-spin waveform model \phX \cite{Pratten:2020ceb} for 
the dominant $(\ell,|m|) = (2,2)$ mode in frequency domain. 
Specifically, \phX describes the $(2,2)$-mode as
\begin{equation}
h_{22}(f) = A_{22} \text{e}^{-i \phi_{22}},
\label{eq:eqh22}
\end{equation}
where $A_{22}$ and $\phi_{22}$ indicate the amplitude and phase of the (2,2)-mode,
 respectively. The modelling approach of the \phX model consists in splitting both
  quantities in three regions: inspiral, merger and ringdown. 
In each region phenomenological models  are produced by calibrating them 
  to NR and extreme mass ratio waveforms \cite{Pratten:2020fqn}. 

In the \phXE model, eccentricity effects are included to describe 
the inspiral up until merger using PN results. At merger, 
the system is assumed to have circularized and the \phX 
ringdown models for the phase and amplitude are employed.

The dynamics within the PN framework is described through the QKP
 which parametrizes the conservative motion of the binary. 
 The problem is parametrized in terms of the four time-dependent 
 variables ${x,e,l,\lambda}$, with $x=\omega^{2/3}$, 
 where $\omega$ is the orbital frequency, $l=n(t-t_0)$ is the mean anomaly,
$n$ is the mean motion \cite{Henry:2023tka} 
and $t_0$ is some reference time, 
  $e=e_t$ is the time-eccentricity, which corresponds to the eccentric 
  parameter appearing in the equation of the mean anomaly.
   Finally, the orbital phase~$\phi$ can be split into a secularly 
   growing part, $\lambda= K l$ ($K$ being the periastron advance),
   and an oscillatory part, $W$, which can be expressed as a function of $x,e,l$.

The \phXE model includes the PN information of the QKP for nonspinning 
and (aligned) spin effects up to 3PN in modified 
harmonic (MH) coordinates~\cite{Memmesheimer:2004cv,Boetzel:2017zza,Henry:2023tka}.
It solves the orbit-averaged equations, i.e. their secular contributions, denoted by $\bar{x}, \bar{e}, \bar{l}$ and $\bar{\lambda}$.
For the time evolution, \phXE employs the NR-calibrated quasicircular \phT phase \cite{Estelles:2020twz} with the same approach as the eccentric \phTEHM model \cite{Planas:2025feq} to 
increase the accuracy of the evolution. Similarly as with \phTEHM \cite{Planas:2025feq},
 we include the EOB secular evolution equations for $\bar{x}, \bar{e}$ which are 
 the default options for the evolution of the dynamics. 
 In order to simplify notation we henceforth use $\{x, e, l, \lambda \} \equiv  \{\bar{x},\bar{e}, \bar{l}, \bar{\lambda} \}$. 

The secular dynamics,   $\{x, e, l, \lambda \}$ is then used to compute the 3PN time-domain waveform using an eccentricity expanded form in terms of the mean anomaly, which takes the symbolic form
\begin{equation}
h^{\rm PN}_{22} (t) = \text{e}^{- 2i \lambda} \sum_{j=-N_e}^{N_e} A_{j} \text{e}^{i j l },
\label{eq:eqhPN}
\end{equation}
where $A_j$ are eccentricity expanded, 3PN accurate expressions 
of $(x,e,\eta,\chi_{1,2})$. In this paper, we have extended PN results 
to the order $\mathcal{O}(e^{12})$, fixing $N_e=12$.\footnote{See Sec.~III~C.~3. of 
Ref.~\cite{Henry:2025uta} for a discussion on the convergence 
of the eccentricity expanded series.} 

Once the time-domain waveform is available, we apply the SPA to 
the individual mean anomaly harmonics in Eq.~\eqref{eq:eqhPN} 
to obtain expressions for the frequency domain phase and amplitude (see Sec. \ref{sec:model_fd} for details),
\begin{equation}
\Psi^{(j)} =  2 \pi f t - 2\lambda + j l \pm\frac{\pi}{4}\,,\quad \mathcal{A}^{\rm SPA}_j  = \sqrt{\frac{2\pi}{|\ddot{\Psi}^{(j)}|}} A_j,
\label{eq:eqSPA}
\end{equation}
with the corresponding waveform written as
\begin{equation}
\tilde{h}_{22} = \sum_{j=-N_e}^{N_e} \mathcal{A}^{\rm SPA}_j  \text{e}^{i \Psi^{(j)}} \equiv  \sum_{j=-N_e}^{N_e} \tilde{h}_j.
\label{eq:eqSPAh22}
\end{equation}
The only time-domain amplitude $A_j$ containing terms non-proportional 
to eccentricity is the $j=0$ harmonic. As a consequence, in the limit 
$e\rightarrow 0$ the only remaining harmonic is the $j=0$ one. Thus, 
we use the $j=0$ inspiral harmonic together with the merger-ringdown model 
of \phX to construct an IMR harmonic. The final \phXE model 
is a sum of the IMR $j=0$ harmonic with the rest of eccentric inspiral modes
\begin{equation}
\tilde{h}^{\rm XE}_{22} =   \tilde{h}^{\rm IMR}_{j=0} + \sum_{\substack{|j|\leq N_e \\ j\neq 0}}  \tilde{h}_j\,.
\end{equation}
For further details on the model construction we refer the reader to Secs. \ref{sec:model_td}, \ref{sec:model_fd} and \ref{sec:model_imr}.

\subsection{Eccentric inspiral in time-domain}\label{sec:model_td}

Using the QKP from the PN derivations mentioned above, the \phXE model follows a similar strategy as the time-domain \phTE model \cite{Planas:2025feq}: 
\begin{itemize}
\item[1)] Evolve the orbit-averaged dynamics for ${x,e,l,\lambda}$.
\item[2)] Evaluate the waveform in time-domain using the evolved dynamics. 
\end{itemize} 
Benefitting from sharing the same python infrastructure in the 
\texttt{phenomxpy} package we evolve the same dynamics as in the 
case of the \phTE model. The explicit expressions can be found in
 Sec. \ref{subsec:ecc_dynamics}. While in the case of 2) we use the
  time-domain PN expressions for $h_{22}$ in the literature and we 
  express them in an eccentricity expansion up to $\mathcal{O}(e^{12})$ 
  and in terms of a slowly varying radial phase parameter, the mean anomaly,
   so that one can afterward apply the SPA onto the time-domain modes. 
   The details of the latter calculation are provided in Sec. \ref{sec:td_waveforms}.

\subsubsection{Eccentric dynamics}\label{subsec:ecc_dynamics}
Similarly to the time-domain \phTE model \cite{Planas:2025feq}, 
the frequency-domain \phXE evolves 3PN spinning evolution equations 
for the orbit-average dynamics of an elliptical binary with non-precessing spins.
 The dynamics is described using the QKP, with evolving variables being 
 the PN $x$ parameter, related to the orbit-average orbital frequency,
  $x = \bar{\omega}^{2/3}$, the time eccentricity $e_t \equiv e$, the
   mean anomaly $l$ and the orbit-average phase parameter $\lambda$.

For this evolution we employ two sets of evolution equations
 for $\{\dot{x},\dot{e}\}$ in different coordinates, the modified harmonic
  (MH) coordinates evolution in the PN framework  \cite{Blanchet:2013haa,
  Henry:2023tka} and the EOB coordinates derived in Ref. \cite{Gamboa:2024imd}. 
  We refer the reader therein for details on the derivation of such expressions.

As in the case of \phTE model \cite{Planas:2025feq} we employ 
the (2,2)-mode frequency, $\omega^{\rm T}_{22}$, of the
 quasicircular \phT model \cite{Estelles:2020osj} to improve the 
 accuracy of the evolution, as  $\omega^{\rm T}_{22}$ has
  been calibrated to nonprecessing-spin quasicircular NR waveforms. 
  On an orbit-average the $(2,2)$-mode frequency and the orbital 
  frequency are related as\footnote{In the PN framework, this relation holds up to 3.5PN. 
  At 4PN, the relation between the GW and orbital frequencies 
  contains a logarithm, see Eq.~(6.8) of Ref.~\cite{Blanchet:2023sbv}.  
   This is due to the fact that the GW propagates in a Schwarzschild background,
    shifting the apparent frequency.}
\begin{equation}
\bar{\omega}^{\rm T}_{\rm orb}(t) = \frac{1}{2}\bar{\omega}^{\rm T}_{22}(t).
\label{eq:eqOm22}
\end{equation}
The direct use of the analytical expression $\dot{\bar{\omega}}_{\rm orb}^T(t)$ requires 
a map between the times used in the quasicircular evolution and the times of the a priori 
unknown eccentric evolution. In order to avoid this problem related to the time 
coordinate, we work with orbit-average orbital frequencies and create an 
interpolant  $\dot{x}^T_{\rm QC}(x_{\rm QC})$, add this term to the right-hand side 
of the evolution equation for $\dot{x}$ and complement the right-hand side with a 
term, $\delta \dot{x}$, adding the eccentricity effects as described in EOB/PN coordinates, 
\begin{equation}
\delta \dot{x}(x,e) =  \dot{x}(x,e) -  \dot{x}(x,e=0).
\label{eq:eqDeltaX}
\end{equation}
Next, we solve for the system of coupled ODEs in a given gauge (PN or EOB)
\begin{align}
\begin{split}
\dot{x}^{\rm EOB/PN} &= \dot{x}^{T}_{\rm QC} + \delta \dot{x}^{\rm EOB/PN}_{\rm 3PN}\,,\\
\dot{e}^{\rm EOB/PN} &= \dot{e}^{\rm EOB/PN}_{\rm 3PN}\,. 
\label{eq:eqEv}
\end{split}
\end{align}
In practice we evolve Eqs. \eqref{eq:eqEv}, together with the PN evolution equations, 
\begin{subequations}\label{eq:ldlambdad}
\begin{align}
\dot{l}(x,e) 
&= \bar{n}(x,e)\,,\label{eq:eqMeanAno}
\\
\dot{\lambda} (x,e) &= \bar{\omega} = x^{3/2}\,,
\end{align}
\end{subequations}
using an explicit Runge-Kutta method of 8th order 
 implemented in \texttt{scipy} \cite{10.5555/153158,2020SciPy-NMeth}  
 until the peak frequency of the quasicircular \phT model, i.e., 
 the frequency of the peak of the amplitude of the (2,2)-mode.
 Since the right-hand side of Eqs.~\eqref{eq:eqMeanAno} 
is expressed in MH coordinates, we first transform the solution $e^\text{EOB}$ computed in Eq.~\eqref{eq:eqEv} back to MH using Eq.~(A2a) of Ref.~\cite{Gamboa:2024imd} after integrating Eq.~\eqref{eq:ldlambdad}.

The explicit expressions used for the evolution equations 
described in EOB coordinates can be found in Appendix D of 
Ref. \cite{Gamboa:2024imd} and in the Supplementary material therein. 
While in PN harmonic coordinates the results are splitted 
in different references \cite{Blanchet:2013haa,Henry:2023tka}.

\subsubsection{Time-domain waveform}\label{sec:td_waveforms}

The amplitude modes in time-domain, $h_{\ell m}$, from BBHs on elliptical orbits have 
been computed within the PN formalism in the literature \cite{Konigsdorffer:2006zt,Mishra:2015bqa,Boetzel:2017zza,Boetzel:2019nfw,Ebersold:2019kdc,Paul:2022xfy,Henry:2023tka}. 
Here we focus on the $(\ell,m)=(2,2)$ mode and nonprecessing-spin eccentric BBHs. 
The $(2,2)-$mode can be expressed as \cite{Henry:2023tka}
\begin{equation}
h_{22}(t) = h^{\text{inst}}_{22} + h^{\text{tail}}_{22} + h^{\text{mem}}_{22}, 
\label{eq:eq16}
\end{equation}
where $h^{\text{inst}}_{22}$, $h^{\text{tail}}_{22}$ and $h^{\text{mem}}_{22}$ indicate the instantaneous, tail and memory contributions, respectively. 

As the final goal is to implement these PN information in 
an existing IMR model, which does not include 
contributions due to memory yet, we focus first 
on the instantaneous and tail contributions 
and leave the implementation of the oscillatory memory contributions for future work.

The (2,2)-mode can be written as \cite{Boetzel:2019nfw,Henry:2023tka} 
\begin{equation}
h_{22} = 8 \eta x\sqrt{\frac{\pi}{5}} H^\psi_{22} \,\text{e}^{-2 i \psi}\,,
\label{eq:eq17}
\end{equation}
where the phase $\psi$ corresponds to the phase associated with the observable GW frequency. It is defined as
\begin{equation}
\psi = \phi(l=\xi)= \lambda_\xi + W(\xi)
\,,
\label{eq:eq18}
\end{equation}
where the redefined mean anomaly $\xi$ is introduced in order to reabsorb the tail gauge constant $b_0$ appearing in the modes, and $\lambda_\xi = K \xi$. At the 3PN order, it reads 
\begin{equation}
\xi = \bar{l} - 3\left(1-\eta \frac{x}{2}\right)x^{3/2} \ln \left( \frac{x}{x'_0}\right)\,.
\label{eq:eq19}
\end{equation}
We refer to Ref.~\cite{Boetzel:2019nfw} for more details, notably regarding the inclusion of
 the post-adiabatic corrections, and where the value of $x_0'$ is given in Eq.~(64) therein. 
 The (2,2)-mode symbolically takes the form $H_{22}^\psi =  F(x, e, \xi, \eta, \chi_S, \chi_A)$. 
 The 3PN nonspinning terms can be found in Refs.~\cite{Mishra:2015bqa,Boetzel:2019nfw, 
 Ebersold:2019kdc}, and the spinning terms in Ref.~\cite{Henry:2023tka} 
 where both results are expanded to $\mathcal{O}(e^6)$.

Here we extend the expression up to $\mathcal{O}(e^{12})$ in MH coordinates. The procedure to obtain the expressions is the same as in Refs. \cite{Boetzel:2017zza,Boetzel:2019nfw, Ebersold:2019kdc,Henry:2023tka}  and we refer therein for details. In the following we outline the main steps of the calculation at Newtonian order and provide the full expressions up to 3PN and expanded up to $\mathcal{O}(e^{12})$ in the supplementary material~\cite{SuppMaterial}.

Let us now explain the procedure by expliciting the Newtonian order. 
The (2,2)-mode in terms of the eccentric anomaly, $u$, 
and without eccentricity expansions, can be expressed, using the Keplerian parametrization~\cite{Mishra:2015bqa}, as
\begin{align}\label{eq:eq21}
h^\text{Newt}_{22} =& 8 \eta x \sqrt{\frac{\pi}{5}} H^\text{Newt}_{22}\, \text{e}^{-2 i \phi}\,, \\
H^\text{Newt}_{22} =&\, \frac{1}{2} + \frac{1}{2(1-e\cos u)}  + \frac{1-e^2}{(1-e\cos u)^2} \nonumber\\
& + \frac{i e\sqrt{1-e^2}\sin u}{(1-e\cos u)^2}\,. 
\end{align}
As the ultimate goal is the application of the SPA to obtain frequency domain waveforms,
 the time-domain waveform modes need to be expressed in terms of a non-oscillatory phase, 
 i.e., the dependence of Eq. \eqref{eq:eq21} on the eccentric anomaly, $u$, 
 needs to be replaced by the mean anomaly, $l$, at the cost of doing eccentricity
  expansions. This reduces to expressing the two terms $e^{-2 i \phi}$ and
   $H^{\rm Newt}_{22}$ as infinite series expansions in terms of $l$.  
   Using the results of Ref.~\cite{Boetzel:2017zza}, the term with the 
   orbital phase can be expressed as
\begin{equation}
     \text{e}^{-2 i \phi}  = \text{e}^{-2 i (\lambda + W)} = \text{e}^{- 2 i \lambda} \text{e}^{- 2 i W},
\label{eq:eq22}
\end{equation}
where the first term is already written in terms of smooth functions, while the second term can be decomposed with
\begin{align}
	\text{e}^{imW} &= \sum_{n=-\infty}^{\infty} \mathcal{P}_n^{mW} \text{e}^{inl} ,
  \label{eq:eq23}
\end{align}
where $m=-2$ for the (2,2)-mode. The Fourier coefficients $ \mathcal{P}_n^{mW}$
 are given in Appendix E of Ref. \cite{Boetzel:2017zza} for non-spinning binaries 
 and we have augmented them with the spinning terms of Ref. \cite{Henry:2023tka}.

The series expansion of $H^{\rm Newt}_{22}$ requires to express the following terms 
\begin{equation}
\frac{1}{(1-e \cos u )}\,,\,\frac{1}{(1-e \cos u )^2}\,,\, \frac{\sin u}{(1-e \cos u )^2}\,,
\end{equation}
in terms of harmonics of $l$. These expansions can be found in Ref. \cite{Boetzel:2017zza},
\begin{subequations}
	\begin{align}
		\frac{1}{(1-e\cos u)^n} =&\; \sum_{j=0}^{\infty} \mathcal{A}_j^n \cos(j l)\,,\\
		\mathcal{A}_j^n =&\; \sum_{k=0}^{\infty} b_k^n \zeta_j^{ku} \,,
	\end{align}
\end{subequations}
where the coefficients $ b_k^n$ come from the expansion 
of $(1-e\cos u)^n$ in terms of $\cos(j u)$ and are functions of $e$  \cite{Boetzel:2017zza}.

The coefficients $\zeta_j^{ku}$ correspond to the coefficients of the Fourier series of $\cos(ju)$ as a function of the mean anomaly
~\cite{Boetzel:2017zza}
\begin{subequations}
	\begin{align}
	\cos(ju) =&\; \sum_{s=0}^{\infty}\zeta_s^{ju} \cos(sl) \,, \\
	\zeta_0^{ju} =&\; \frac{1}{2}\left(-e\,\delta_{j1} + \alpha_j j\right) \,,\\
	\zeta_s^{ju} =&\; \frac{j}{s}\left\{ J_{s-j}(s e) - J_{s+j}(s e)  \right\} \nonumber\\
	&+\frac{j}{2} \sum_{i=1}^{\infty}\alpha_i \big\{ J_{s-j+i}(s e) - J_{s-j-i}(s e) \nonumber\\
	&- J_{s+j+i}(s e) + J_{s+j-i}(s e)\big\}\,. \label{eq:zetasju}
	\end{align}
 \label{eq:eq24}
\end{subequations}

Similarly, one can express $\sin(ju)$ through 
\begin{subequations}
	\begin{align}\label{eq:sinju}
		\sin(ju) =&\; \sum_{s=1}^{\infty}\sigma_s^{ju} \sin(sl) \,,\\
		\sigma_s^{ju} =&\; \frac{j}{s}\left\{ J_{s+j}(s e) + J_{s-j}(s e)  \right\} \nonumber\\
		&+\frac{j}{2} \sum_{i=1}^{\infty}\alpha_i \big\{ J_{s+j+i}(s e) - J_{s+j-i}(s e) \nonumber\\
		&+ J_{s-j+i}(s e) -J_{s-j-i}(s e)\big\}\,.\label{eq:sigmasju}
	\end{align}
 \label{eq:eq25}
\end{subequations}
The coefficients $\alpha_i$ present in Eqs. ~\eqref{eq:zetasju} and~\eqref{eq:sigmasju} 
are of 3PN order and come from the inversion of the 3PN accurate generalized Kepler equation  
\begin{align}\label{eq:GKE}
	l = u - e \sin u + \sum_{j=1}^{\infty} \alpha_j \sin(ju)\,.
\end{align}
Their explicit expressions at 3PN (in the notations of, e.g. Ref.~\cite{Boetzel:2017zza}) read
\begin{align}\label{eq:alpha3PN}
	\alpha_j =& 2\beta_{\phi}^j \frac{\sqrt{1-e_{\phi}^2}}{e_{\phi}^3} \Bigg\{(f_{4t} 
	+ f_{6t}) e_{\phi}^2 +\frac{(g_{4t} + g_{6t}) e_{\phi}^3}{j\sqrt{1-e_{\phi}^2}} \nonumber\\
	& + 2i_{6t}e_{\phi} \left(j\sqrt{1-e_{\phi}^2}-1\right) \\
	& + h_{6t} \left(4-e_{\phi}^2-6j\sqrt{1-e_{\phi}^2} + 2j^2(1-e_{\phi}^2)\right) \Bigg\}\,,\nonumber
\end{align}
with $\beta_{\phi} = (1-\sqrt{1-e_{\phi}^2})/e_{\phi}$, while $e_\phi$ 
is the phase eccentricity, related to the time eccentricity $e$, 
see e.g. Ref.~\cite{Memmesheimer:2004cv} for more details. 

In our calculation we include consistently the aligned-spin results from Ref.~\cite{Henry:2023tka} in $e_\phi$, in the functions $g_{4t}$ and $f_{4t}$, 
which correspond to the expressions for $f_{v-u}$ and $f_v$ in
 Eqs. (2.26b), (A3a) and (A3b) in \cite{Henry:2023tka}. 
 One should be aware that Ref.~\cite{Boetzel:2019nfw} includes 
 the periastron advance in the definition of $g_{4t}$ and $f_{4t}$
  while Ref.~\cite{Henry:2023tka} does not.

Using the above Fourier series of the dynamical variables, we can rewrite 
Eq.~\eqref{eq:eq21} as
\begin{align} 
h^\text{Newt}_{22} &= 8 \eta x \sqrt{\frac{\pi}{5}} \text{e}^{-2 i \lambda} \left( \sum_{n=-N_e}^{N_e}\mathcal{P}_n^{-2W} \text{e}^{inl} \right) \nonumber \\
& \times \Bigg\{ \frac{1}{2} + \sum_{j=0}^{N_e} \left(  \frac{1}{2} \mathcal{A}_j^1 +(1-e) \mathcal{A}_j^2 \right)\cos (jl) \label{eq:h22Newt}  \\
&  + i e \sqrt{1-e^2}  \left(\sum_{s=1}^{N_e}\sigma_s^{1u} \sin(sl) \right)\left(  \sum_{j=0}^{N_e} \mathcal{A}_j^2 \cos(j l) \right)  \Bigg \},\nonumber
\end{align}
%
where the index $N_e$ corresponds to the number of terms 
($\equiv$ to the order in eccentricity expansion) included in the series expansions.
 Eq. \eqref{eq:h22Newt} in its current form implies the 
 multiplication of three sums, which can be linearized 
via trigonometric relations\footnote{Note that analytically one can
 compute the product of the series using the expressions 
 provided in Appendix D of Ref.~\cite{Boetzel:2017zza}.}.

In practice, we expressed all relevant quantities in terms of $(x,e,l)$ and expand $h_{22}$ up to certain eccentricity order via a Taylor expansion. By doing so, the (2,2)-mode is simply written as a finite sum of $2N_e+1$ harmonics
\begin{equation}
h_{22} (t)= 8 x \eta \sqrt{\frac{\pi}{5}} \text{e}^{-2 i \lambda_\xi} \sum_{j=- N_e}^{N_e}  a_j \text{e}^{ i j \xi}\,.
\label{eq:eqh22NewtSimp}
\end{equation}
As shown in Ref. \cite{Planas:2025feq}, the use of $\xi$ does not 
ameliorate the accuracy of the model and implies additional calculations 
which can affect the computational cost,
 thus in practice, we use the expressions of $a_j$ computed with the phase redefinition,
  but use the evolution of $l$ and $\lambda$ instead of $\xi$ and $\lambda_\xi$. 

In this paper, we expand consistently the quantities to the $\mathcal{O}(e^{12})$ order to increase the coverage in the eccentricity parameter space. At leading PN order, the first 3 coefficients $a_j$ read
\begin{widetext}
\begin{subequations}\label{eq:aj}
\begin{align}
a_{-1} &=  \frac{9 e}{4} -\frac{171 e^3}{32}+\frac{963 e^5}{256}-\frac{4311 e^7}{4096}+\frac{58689 e^9}{327680} -\frac{93753 e^{11}}{13107200} + \mathcal{O}\left( \frac{1}{c^2}\right) +\mathcal{O}(e^{13})\,,\\
a_{0} &= 1-\frac{5 e^2}{2}+\frac{23 e^4}{16}-\frac{65 e^6}{288}+\frac{85 e^8}{2304}+\frac{1007 e^{10}}{115200}+\frac{293 e^{12}}{41472} + \mathcal{O}\left( \frac{1}{c^2}\right) +\mathcal{O}(e^{14}) \,,\\
a_{+1} & = -\frac{3 e}{4}+\frac{13 e^3}{32}+\frac{5 e^5}{768}+\frac{227 e^7}{12288}+\frac{34349 e^9}{2949120}+\frac{2957173 e^{11}}{353894400} + \mathcal{O}\left( \frac{1}{c^2}\right) +\mathcal{O}(e^{13})\, .
\end{align}
\end{subequations}
\end{widetext}
As illustrated in Eqs.~\eqref{eq:aj}, the eccentricity-expanded coefficients $a_j$ are of order $\mathcal{O}(e^{|j|})$. Thus, only $a_0$ has a non-vanishing value in the quasicircular limit ($e \rightarrow 0 $), whose value coincides with the well-known value of $h_{22}$~\cite{Henry:2022dzx,Blanchet:2023sbv,Blanchet:2023bwj} at 3PN.

It is important to remark that, as discussed in Sec. III C 3 
of Ref.~\cite{Henry:2025uta}, the power series used to invert 
Eq. \eqref{eq:GKE} diverges for $e>e_{\rm max} \sim 0.6627434$.
This means that the eccentricity expanded PN results are not valid beyond 
$e_{\rm max}$ and are expected to have a low accuracy for eccentricities 
close or beyond this value. This is one of the reasons why in the comparison
against eccentric NR waveforms in Sec. \ref{sec:ecc_NRcomparison}, the mismatches
of the \phXE model increase when the initial eccentricity of the 
NR waveforms increases. This behaviour is displayed 
in Fig. \ref{fig:nr_mismatches}, where for $e>e_{\rm max}$
the values of the mismatches increase significantly.

The extension of the coefficients $a_j$ to higher PN orders follows the same logic as in the Newtonian case explained above, but it involves larger calculations with more complicated expressions which we skip and refer to Refs.~\cite{Ebersold:2019kdc,Henry:2023tka} for more details. 
The expression of $H_{22}^\psi(x,e,\xi)$ at 3PN including non-spinning and spinning contributions, expanded to $\mathcal{O}(e^{12})$, in MH coordinates is provided in the supplementary material~\cite{SuppMaterial}. 

\subsection{Eccentric waveforms in frequency-domain}\label{sec:model_fd}

Once obtained the time domain expression for the (2,2)-mode as detailed in
 Sec.~\ref{sec:model_td}, we apply the stationary phase approximation (SPA) to obtain the frequency domain waveforms. 
 In the following we outline the SPA and its application.

Given a time-domain signal which can be expressed as 
\begin{equation}
h(t) = B \text{e}^{-i \theta},
\label{eq:eq_spa1}
\end{equation}
where $B(t)$ and $\theta(t)$ are functions of times, while $i$ denotes the imaginary unit, the Fourier transform of the signal can be written as 
\begin{equation}
\Tilde{h}(f) = \int_{- \infty}^\infty h(t) \text{e}^{2 \pi i f t} dt = \int_{- \infty}^\infty B(t) \text{e}^{i (2 \pi f t - \theta(t))} dt\,.
\label{eq:eq_spa2}
\end{equation}
If the amplitude in the integrand of Eq. \eqref{eq:eq_spa2} varies much slower than the phase $\theta$, i.e., $\dot{B}/B \ll \dot{\theta}$, then for most of the values of $t$ the integrand is rapidly oscillating. However, there exist a time in which the phase of the integrand is approximately constant and it thus contributes significantly to the integral. This point in time is called the stationary time $t_S$. Under these conditions it can be shown that the stationary phase condition is satisfied when 
\begin{equation}
2 \pi f -  \dot{\theta}_S  = 0,
\label{eq:eq_spa3}
\end{equation}
which provides a mapping between the Fourier frequency $f$ and the time derivative of the phase $\theta$. As a consequence a Taylor expansion of the phase to first order around the stationary time provides~\cite{Hughes:2021exa}, 
\begin{equation}
\Tilde{h}(f) = B_S \sqrt{\frac{2 \pi}{|\Ddot{\theta}_S|}} \text{e}^{i[2 \pi f t_S - \theta_S- \text{sign}(\ddot{\theta}_S)\pi/4]}\,,
\label{eq:eq_spa4}
\end{equation}
which is an approximation to compute the Fourier transform of the original time domain signal. See Refs. \cite{Hughes:2021exa,Moore:2018kvz,Buonanno:2009zt} for details in the calculations above.

In the case of a time domain signal of the form of Eq.~\eqref{eq:eqh22NewtSimp}, 
one can apply the SPA  individually to each harmonic in mean anomaly. 
This supposes to introduce one stationary time $t_S^{(j)}$ for each harmonic. We obtain 
\begin{align}
\label{eq:eq_spa_h22}
\tilde{h}_{22}(f) &=  \sum_{j=-N}^N  \mathcal{A}_j \text{e}^{i \Psi^{(j)}}, \\
\label{eq:eq_spa_amp22}
 \mathcal{A}_j &= 8 \pi x \eta \sqrt{\frac{2}{5|\ddot{\Psi}^{(j)}|}} a_j \,, \\
\label{eq:eq_spa_phas22}
\Psi^{(j)} &=  2 \pi f t - 2\lambda + j l + \text{sign}\bigl(\ddot{\Psi}^{(j)}\bigr)\frac{\pi}{4}\,, \\
\ddot{\Psi}^{(j)} &= - 2\dot{\bar{\omega}} + j \dot{\bar{n}}\,,\label{eq:eq_spa_d2ph}
\end{align}
where all quantities have to be evaluated at the time $t=t_S^{(j)}$, $\bar{\omega}$ 
and $\bar{n}$ are the orbit-averaged orbital frequency and mean motion and 
the coefficients $a_j = a_j(x,e,\eta, \chi_1, \chi_2)$ represent the 3PN expressions 
for the time-domain modes. Notice that for QC orbits, the sign of $\ddot{\Psi}^{(j)}$ is
 always -1 because it corresponds to the mode $j=0$ only and $\dot{\omega}>0$.
 But in the present case, its sign could eventually change depending on the considered harmonic.

In principle, one could provide PN-expanded expressions for the SPA
 amplitudes $\mathcal{A}_j$ as a function of the dynamical variables 
 $x,e$ (as well as the intrinsic parameters of the binary), however, 
 the same is not possible for the SPA phase in Eq. \eqref{eq:eq_spa_phas22} 
 that would require expansions in the initial eccentricity $e_0$ (see Refs.
  \cite{Moore:2016qxz,Tanay:2016zog,Tiwari:2019jtz,Moore:2019xkm}). Such 
  expansions in $e_0$ set restrictions in the parameter space coverage
   with a rapid degradation in accuracy for 
   eccentricity beyond 0.1 \cite{Moore:2016qxz,Tiwari:2019jtz}.

In order to avoid the parameter space restriction 
in eccentricity due to the use of analytical expressions 
we compute the SPA amplitude and phase numerically using the 
solution of the evolved dynamics. In the case of the SPA amplitude, 
we first interpolate the dynamical quantities using cubic splines and
 then compute the corresponding time derivatives entering Eq. \eqref{eq:eq_spa_d2ph}. 

\subsection{IMR model}\label{sec:model_imr}
The description of the inspiral via a decomposition in mean anomaly harmonics 
and the calculation of the SPA numerically as 
described in Secs. \ref{sec:td_waveforms} and \ref{sec:model_fd} 
impose strong constraints on the construction of the IMR model. 
Specifically, we construct 
an IMR phase and amplitude based on the $j=0$ mean anomaly harmonic, 
which is the only
 non-vanishing harmonic in the quasicircular limit, 
 and add on top of that the rest of inspiral-only mean anomaly harmonics. 

\subsubsection{IMR phase}\label{sec:phase_j0}
The eccentric inspiral evolution of the \phXE model finishes at a certain dimensionless 
frequency that we call $\rm{Mf}_{\rm last}$, while the underlying quasicircular \phX model
splits both amplitude and phase into different regions. Specifically, the \phX phase 
model has two transition frequencies between the inspiral and intermediate, 
$\rm{Mf}^{\phi}_{\rm IN}$, and intermediate and ringdown $\rm{Mf}^{\phi}_{\rm IM}$ regions. 
Thus, the inspiral description based on PN cannot overcome $\rm{Mf}^{\phi}_{\rm IM}$, 
and additionally it has to reach it with a small eccentricity value. As a consequence 
we impose some constraints such that no waveform is generated if
the eccentricity at the end of the inspiral is $e\geq 0.2$, and that 
the maximum frequency up to which to apply the SPA is $0.9 \rm{Mf}^{\phi}_{\rm IM}$. 
Thus, $\rm{Mf}_{\rm last}$ is defined as,
\begin{equation}
f_{\rm last} = \left\{
\begin{aligned}
& f^{\phi}_{\rm IN}, && f_{0}< f^{\phi}_{\rm IN}  \\
& 0.9 f^{\phi}_{\rm IM},   && f_{0}> f^{\phi}_{\rm IN} \quad \&  \quad f_{0}< 0.9 f^{\phi}_{\rm IM} \\
\end{aligned}
\right.,
\label{eq:eq_flast}
\end{equation}
where $f_0$ is the starting frequency of waveform generation, 
and we have omitted the total mass in Eq. \eqref{eq:eq_flast} to save space.

Furthermore, in order to match the intermediate and ringdown regions 
of the \phX phase, which have been calibrated to the analytical
 quasicircular SPA \cite{Pratten:2020fqn}, we rescale the
  numerically computed phase $\Psi_{j=0}$,
\begin{equation}
\Phi_{j=0} = -\eta (\Psi_{j=0} + \Delta \phi),
\label{eq:eq_phj0_rescale}
\end{equation}
where $\Delta \phi$ is the time-shift and phase offset in
 the \phX model calibrated to quasicircular NR simulations \cite{Pratten:2020fqn}. 

This rescaled phase computed in the coarse grid,
 outcome of the ODE evolution, is then interpolated to be
  evaluated at a finer grid, typically specified by the user, and 
  at $\rm{Mf}_{\rm last}$, where the phase and the frequency derivative
   are computed. These latter values are needed to obtain the new intermediate 
   connection coefficients of the phase $C^{'}_{\rm int,1}$ and 
   $C^{'}_{\rm int,2}$  which ensure that the phase is smooth 
   and continuous between regions. 

With the intermediate coefficients  $C^{'}_{\rm int,1}$ and $C^{'}_{\rm int,2}$ 
 we construct an intermediate phase region up to $\rm{Mf}^{\phi}_{\rm IM}$,
  where we compute new connection coefficients for the 
  ringdown region, $C^{'}_{\rm RD,1}$ and $C^{'}_{\rm RD,2}$, which 
  are used to construct the final ringdown phase. 

The connection coefficients in the \phXE model at a given dimensionless frequency $\rm{Mf}_1$ can be computed as 
\begin{equation}
\begin{split}
C^{'}_{\rm X,2} &= \frac{dY}{df}(\rm{Mf}_1)  - \frac{d\phi^{\rm XAS}_{\rm X}}{df}(\rm{Mf}_1), \\
C^{'}_{\rm X,1} &= Y(\rm{Mf}_1)  - \phi^{\rm XAS}_{X}(\rm{Mf}_1) - C^{'}_{\rm X,2} \rm{Mf}_1, \\
\mathcal{C}^{'}_{\rm X}(\rm{Mf}) &=  C^{\rm '}_{\rm X,1} + C^{'}_{\rm X,2} \rm{Mf}_1,
\end{split}
\label{eq:eq_c12_int}
\end{equation}
where $X=\{\rm{int},{RD}\}$ denotes the region intermediate or ringdown,
 $Y$ is the \phXE phase at the frequency $\rm{Mf}_1$, which takes 
 the values $\rm{Mf}_1 = \{\rm{Mf}_{\rm last}, \rm{Mf}^\phi_{\rm IM}\}$.

In summary the IMR j=0 phase can be expressed as 
\begin{equation}
\phi^{\rm IMR}_{j=0} =  \left \{
\begin{aligned}
 &\Phi, \hspace{2.65cm}  \rm{if}\hspace{0.2cm}  \rm{Mf} \in (0, 0.9 \rm{Mf}^\phi_{\rm IM}],\\
 & \phi^{\rm XAS}_{\rm int} + \mathcal{C}^{'}_{\rm int}(\rm{Mf}), \hspace{0.26cm} \rm{if } \hspace{0.2cm} Mf \in [0.9  Mf^\phi_{\rm IM}, Mf^\phi_{\rm IM}], \\
 & \phi^{\rm XAS}_{\rm RD} + \mathcal{C}^{'}_{\rm RD}(\rm{Mf}), \hspace{0.25cm}  \rm{if} \hspace{0.2cm} \rm{Mf} \in [\rm{Mf}^\phi_{\rm IM},\rm{Mf}_{\rm max}], \\
\end{aligned}
 \right.
\label{eq:eq_phj0_imr}
\end{equation}
where $\rm{Mf}_{\rm max}$ is the maximum frequency specified by the user and the $\mathcal{C}^{'}$ functions are defined in Eq. \eqref{eq:eq_c12_int}.

\subsubsection{IMR amplitude}\label{sec:amp_j0}
The construction of the IMR amplitude is simpler than the phase 
due to the lack of the phase offset and time-shift ambiguity 
in frequency-domain. Our procedure here consists in first computing 
the IMR \phX amplitude, $\mathcal{A}^{\rm XAS}_{22}=  |\tilde{h}^{\rm XAS}_{22}|$,
 and adding on top of that the eccentric corrections based on the numerically computed SPA.

To be consistent with the phase, the amplitude based on the numerical
 SPA is computed up to $\rm{Mf}_{\rm last}$. 
Specifically, we compute the residual between the SPA amplitude of the 
time-domain (2,2)-mode computed 
from the eccentric dynamics and the quasicircular counterpart.
For the latter we employ the (2,2)-mode amplitude of 
the \phT model \cite{Estelles:2020osj,Estelles:2020twz},  $A^{\rm PhenomT}_{22}$, and 
the quasicircular frequency evolution which we have already used during the eccentric evolution,
\begin{equation}
\mathcal{A}^{\rm QC}_{j=0} =   \sqrt{2 \pi} (|2 \dot{\bar{\omega}}_{\rm QC}|)^{-1/2}  A^{\rm PhenomT}_{22},
\label{eq:eq_amp_spa_qc}
\end{equation}
where the time derivative of the frequency evolution of \phT, $\dot{\bar{\omega}}_{\rm QC}$, is computed numerically.

For the eccentric SPA amplitude we compute the difference between the PN time-domain eccentric and quasicircular amplitudes,
\begin{equation}
\delta A^{\rm ecc}_{j=0} = A^{\rm PN}_{j=0} - A^{\rm PN}_{j=0}(e=0) ,
\label{eq:eq_diff_amp_spa}
\end{equation}
then we add this difference onto the time-domain quasicircular \phT amplitude
\begin{equation}
A^{\rm ecc}_{j=0}= A^{\rm PhenomT}_{22} +\delta A^{\rm ecc}_{j=0}.
\label{eq:eq_amp_td_imr}
\end{equation}
As in the quasicircular case we compute the SPA amplitude as, 
\begin{equation}
\mathcal{A}^{\rm ecc}_{j=0} =   \sqrt{2 \pi} (|2 \dot{\bar{\omega}}|)^{-1/2}  A^{\rm ecc}_{j=0},
\label{eq:eq_amp_spa_ecc}
\end{equation}
where $\dot{\bar{\omega}}$ is the time derivative of the 
orbit-averaged frequency obtained from the ODE evolution. 
The derivative is computed numerically. 
Once we have the SPA amplitude, we compute the difference 
between Eqs. \eqref{eq:eq_amp_spa_qc} and \eqref{eq:eq_amp_spa_ecc},
\begin{eqnarray}
\delta \mathcal{A}_{j=0} &= \mathcal{A}^{\rm ecc}_{j=0} - \mathcal{A}^{\rm QC}_{j=0}.
\label{eq:eq_diff_amp_spa_j0}
\end{eqnarray}
Finally, the IMR $j=0$ amplitude is computed adding Eq. \eqref{eq:eq_diff_amp_spa_j0} to the IMR \phX amplitude, 
\begin{equation}
\mathcal{A}^{\rm IMR}_{j=0} =  \mathcal{A}^{\rm XAS}_{22} + w \delta \mathcal{A}_{j=0}, 
\label{eq:eq_ampj0_imr}
\end{equation}
where $ w$ is a window function of the form,
\begin{equation}
w(\rm{Mf}, \beta, \rm{Mf}_{\rm last} ) = 1/\left(1+ \text{e}^{\beta (\rm{Mf}-\rm{Mf}_{\rm last})}\right),
\label{eq:eq_amp_window}
\end{equation}
with the parameter $\beta = 10^3$ chosen empirically. 

\subsubsection{Full waveform}\label{sec:full_imr}
Once the $j=0$ phase and amplitude are computed we can combine them to obtain the full harmonic as, 
\begin{equation}
\tilde{h}^{\rm IMR}_{j=0} = \mathcal{A}^{\rm IMR}_{j=0} \text{e}^{i \phi^{\rm IMR}_{j=0} }.
\label{eq:eq_hj0}
\end{equation}
The rest of the mean anomaly harmonics are computed during the inspiral up to $\rm{Mf}_{\rm last}$ using the 3PN eccentric aligned-spin expressions for the time-domain amplitudes and the SPA, 
\begin{equation}
\label{eq:eq_spa_hj}
\tilde{h}_j = \mathcal{A}_j \text{e}^{-i \Psi^{(j)}},
\end{equation}
where $ \mathcal{A}_j$ and $\Psi_j$ are given by Eqs. \eqref{eq:eq_spa_amp22} and \eqref{eq:eq_spa_phas22}, respectively. 

Then, the contribution of the rest of mean anomaly harmonics can be expressed as, 
\begin{equation}
\Delta h = \sum_{\substack{|j|\leq N_e \\ j\neq 0}}  \tilde{h}_j\,, \\
\label{eq:eq_spa_sum}
\end{equation}
where $N_e$ denotes the number of $|j|>0$ mean anomaly harmonics included
$N_e\in[0,12]$. This number is useful as it is proportional to the leading order power in eccentricity included in the harmonic, and it is related to the total number of harmonics in the model as 
\begin{equation}
N_{\rm harm} = 2 N_e + 1. \\
\label{eq:eq_harmonics}
\end{equation}
For instance, when setting $N_e=1$ implies the use of the $j=\{0, \pm 1\}$
harmonics, i.e., $N_{\rm harm}= 3$. 
$N_e$ is a freely specifiable parameter in \phXE with a maximum value of 12,
determined according to the underlying eccentricity 
expansions of time-domain amplitudes up to $\mathcal{O}(e^{12})$. 
The choice of the default value of mean anomaly harmonics in the \phXE model
depends on the target parameter space. We find in Sec. \ref{sec:validation} 
that including mean anomaly harmonics above 9 ($N_e=4$) does not improve 
substantially the accuracy of the model for eccentricities up to 0.4, 
while going up to 13 ($N_e=6$) increases the accuracy of the model 
for eccentricities around 0.8. As a consequence we set 
$N_e=6$ as the default value in the model.

Finally, the (2,2)-mode in the \phXE model can be represented as 
\begin{equation}
\tilde{h}_{22} = \tilde{h}^{\rm IMR}_{j=0}  + \Delta h\,.
\end{equation}


\section{Model Performance and Validation}
\label{sec:validation}

In this section, we assess the accuracy of the eccentric \phXE waveform model 
by comparing its quasi-circular limit to the NR-calibrated 
quasicircular \phX and \phT models. 
In the eccentric case, we compare \phXE to eccentric NR waveforms. 

\subsection{Faithfulness function}
\label{sec:NRunfaithfulness}

An eccentric, aligned-spin BBH system is described by thirteen parameters. 
Six of these are intrinsic source parameters: the mass ratio $q$, the total mass $M$, 
the spin components in the direction of the orbital angular momentum $\chi_1$ and $\chi_2$, 
and two parameters describing the ellipse \textemdash we have chosen the orbital eccentricity
 $e$ and the mean anomaly $l$ at a reference time. The other seven are extrinsic parameters 
 that relate the source frame to a detector frame: the inclination and reference orbital 
 phase $(\iota,\varphi_0)$, the sky location $(\theta,\phi)$, the polarization angle $\psi$, 
 the luminosity distance $D_L$, and the time of arrival $t_c$.

The strain measured by a GW detector can be written as
\begin{equation}
\begin{split}
h(t) &= F_{+}(\theta,\phi,\psi)\,h_{+}(\iota,\varphi_0,D_L,\boldsymbol{\Theta},t_c;t)\\
&\quad + F_{\times}(\theta,\phi,\psi)\,h_{\times}(\iota,\varphi_0,D_L,\boldsymbol{\Theta},t_c;t),
\end{split}
\label{eq:detector_strain}
\end{equation}
where $\boldsymbol{\Theta}=\{m_{1,2},\chi_{1,2},e,l\}$ is the intrinsic parameter vector, and $F_{+,\times}(\theta,\phi,\psi)$ are the detector antenna pattern functions. 

The complex polarizations can be decomposed into spin-weighted $-2$ spherical harmonics,
\begin{equation}
h_{+}(t)-i h_{\times}(t)
= \sum_{l=2}^{\infty}\sum_{m=-l}^{l} {}_{-2}Y_{lm}(\varphi,\iota)\,h_{lm}(\boldsymbol{\Theta};t),
\label{eq:mode_decomposition}
\end{equation}
where $h_{lm}(\boldsymbol{\Theta};t)$ denotes the individual waveform modes and ${}_{-2}Y_{lm}$ are the spin-weighted spherical harmonics.

To compare two waveforms $h_1$ and $h_2$ in the presence of detector noise, 
the usual noise-weighted inner product is employed:
\begin{equation}
\langle h_1 | h_2 \rangle = 4\,\mathrm{Re}\int_{f_{\min}}^{f_{\max}}
\frac{\tilde{h}_1(f)\,\tilde{h}_2^{*}(f)}{S_n(f)}\,df,
\label{eq:inner_product}
\end{equation}
where tildes denote Fourier transforms, an asterisk denotes complex conjugation, 
and $S_n(f)$ is the one-sided noise power spectral density (PSD) of the detector. 
In this work, we are using the zero-detuned high-power PSD of Advanced LIGO at design sensitivity \cite{Barsotti:2018}. 
When both waveforms span the detector band, the integral limits are set to $f_{\min}=10\,$Hz and $f_{\max}=2048\,$Hz; 
for NR waveforms that start at higher frequencies, 
the lower limit is set to $f_{\min}=1.35\,\bar{f}_{\rm start}$, 
where $\bar{f}_{\rm start}$ is the initial orbit-average 
GW frequency of the NR waveform.

The faithfulness between a signal waveform $h_{\rm s}$ and 
a template waveform $h_{\rm t}$ is defined as the normalized 
inner product maximized over nuisance parameters. For quasi-circular 
multimodal waveforms, one typically optimizes over the relative phases
 or the effective polarization angle for different values of the inclination 
 angle. When comparing models with only (2,2)-mode content, it is sufficient
to optimize over a phase offset and a time shift at a fixed inclination (we set $\iota_s=0$). 

For eccentric binaries, however, additional optimizations are needed,
 due to the
gauge-dependent nature of eccentricity in GR. The $e$ and $l$ parameters
chosen for \phXE are gauge-dependent quantities, and signal and template
waveforms may be relying on different choices for the eccentric orbit 
parameterization. When comparing eccentric waveforms that employ different
eccentricity definitions, we therefore need to construct an appropriate mapping 
between them. For example, Ref. \cite{Ramos-Buades:2022lgf} introduces a waveform-based
definition of eccentricity that reduces to the eccentricity of the Newtonian limit,
while Refs. \cite{Shaikh:2023ypz,Shaikh:2025tae} use the same definition to construct
 an algorithm that maps the eccentricity evolution across waveform models.

In our comparisons with NR simulations, we instead adopt an optimization procedure to determine the best-matching waveform for a given eccentric NR dataset. Specifically, we jointly optimize over the mean anomaly and eccentricity at the initial orbit-average frequency following a similar procedure as in Ref. \cite{Planas:2025feq}. Thus, the faithfulness between two non-precessing spin dominant-mode eccentric waveforms can be expressed as 
\begin{equation}
\begin{aligned}
\mathcal{F}_\Xi = \max_\Xi 
\frac{\langle h_{\rm s}|h_{\rm t}\rangle}
{\sqrt{\langle h_{\rm s}|h_{\rm s}\rangle\,\langle h_{\rm t}|h_{\rm t}\rangle}}.
\end{aligned}
\label{eq:faithfulness}
\end{equation}
where $\Xi = \{t_c, \varphi,e, l\}$. 
The unfaithfulness or mismatch is defined as,
\begin{equation}
\mathcal{M} \equiv \mathcal{M}_\Xi  = 1 - \mathcal{F}_\Xi.
\label{eq:unfaithfulness}
\end{equation}
The mismatch  $0<\mathcal{M}<1$ represents the degree of disagreement between two waveforms, with values close to 1 indicating large discrepancies and values close to 0 manifesting good agreement between both signals.

\subsection{Quasi-circular limit}\label{sec:QC_limit}

\begin{figure*}[!]
    \centering
    \includegraphics[width=1\linewidth]{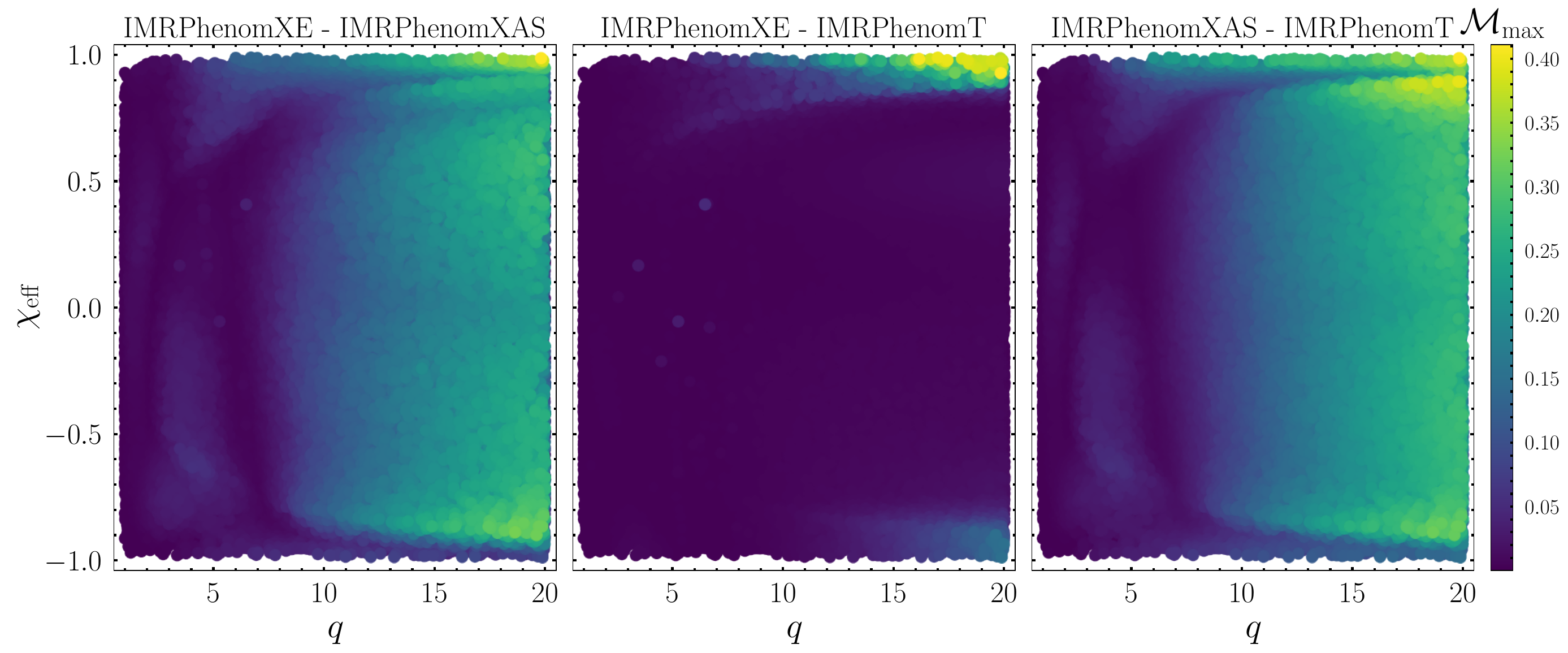}
    \caption{
   Mismatch distribution in the effective-spin parameter and mass ratio plane between the \phX, \phXE and \phT models for $10^5$ quasi-circular configurations randomly sampled within the \phX and \phT validity regions: mass ratio $q\in[1,20]$, total mass $M\in[10,200]\ M_{\odot}$, dimensionless spin components $\chi_i\in[-0.99,0.99]$, using an azimuthal phase $\varphi=0^\circ $ and inclination angle $\iota=0^\circ$. Each configuration is color-coded by its maximum value of the mismatch within the total mass range. From left to right, the panels correspond to the \phXE{-}\phX, \phXE{-}\phT and \phX{-}\phT comparisons.}
    \label{fig:qc_limit}
\end{figure*}

We validate the eccentric aligned-spin model \phXE in the quasi-circular (QC) limit 
by comparing it to its underlying frequency-domain non-eccentric baseline, \phX, 
and to the time-domain quasi-circular model \phT, which underlies the frequency 
evolution of the numerical evolution implemented in \phXE. The \phXE model is 
constructed on top of the \phX model within the new \texttt{phenomxpy} Python 
package~\cite{Garcia-Quiros:2025usi}, which reproduces the version
 implemented in LALSuite~\cite{Pratten:2020fqn} up to numerical error. 

A key difference between the frequency-domain quasi-circular \phX 
and the eccentric \phXE models is the treatment of the inspiral. While 
\phX uses analytical Ans\"atze calibrated to NR waveforms, 
\phXE applies the SPA approximation to a time-domain signal constructed from the 
frequency evolution of the \phT model, see Eq. \eqref{eq:eqEv}. As a consequence,
 in the quasi-circular limit \phXE is closer to the \phT model 
 and can exhibit the same underlying waveform systematics between 
 boths quasicircular models.

In Fig.~\ref{fig:qc_limit}, we display the mismatch between models (\phXE{-}\phX, \phXE{-}\phT and \phX{-}\phT) as
 function of the effective-spin parameter, $\chi_{\rm eff}$, and the mass ratio, 
 $q$, for $10^5$ randomly chosen quasi-circular configurations in the following 
 parameter space:  $q\in[1,20]$, total mass $M\in[10,200]\ M_{\odot}$, dimensionless
  spin components $\chi_i\in[-0.99,0.99]$, azimuthal phase $\varphi=0^\circ $, and inclination 
  angle $\iota=0^\circ$. We color-code each configuration by its maximum 
  value of mismatch in the chosen total mass range. 

The results show that mismatches between \phXE and \phX significantly 
degrade above $q \sim 10$, especially at high positive spins where the
 mismatch can reach $\approx 40\%$. The comparison of \phXE and \phT reveals 
 much lower values of mismatch ($\sim 10^{-3}$) even beyond $q>10$, with 
 the exception of high positive and negative spins. 
 This better agreement is due to the different inspiral prescriptions of \phX and \phXE.
  In order to confirm this behavior, we also compare mismatches between the quasi-circular 
  \phX and \phT models and observe results that are qualitatively similar to the comparison 
  of \phXE and \phT. This demonstrates that differences in the quasi-circular limit come 
  from unresolved systematics between the time-domain and frequency-domain phenomenological
   models in regions of parameter space where the NR data is scarce. 

The results in this section, combined with the demonstrated quasi-circular accuracy of the \phX
 and \phT models against NR simulations \cite{Pratten:2020fqn,Estelles:2020twz}, confirm that 
 the eccentric \phXE model robustly and faithfully reproduces the QC limit. The differences with
  respect to \phX are substantial for mass ratios $q>10$, but remain small and controlled when
   compared to the time-domain \phT model, which highlights the need to populate the high mass 
   ratio region with NR simulations in order to reduce the waveform systematics between models.

\subsection{Comparison with eccentric NR waveforms}\label{sec:ecc_NRcomparison}

\begin{figure}[!]
    \centering
    \includegraphics[width=\linewidth]{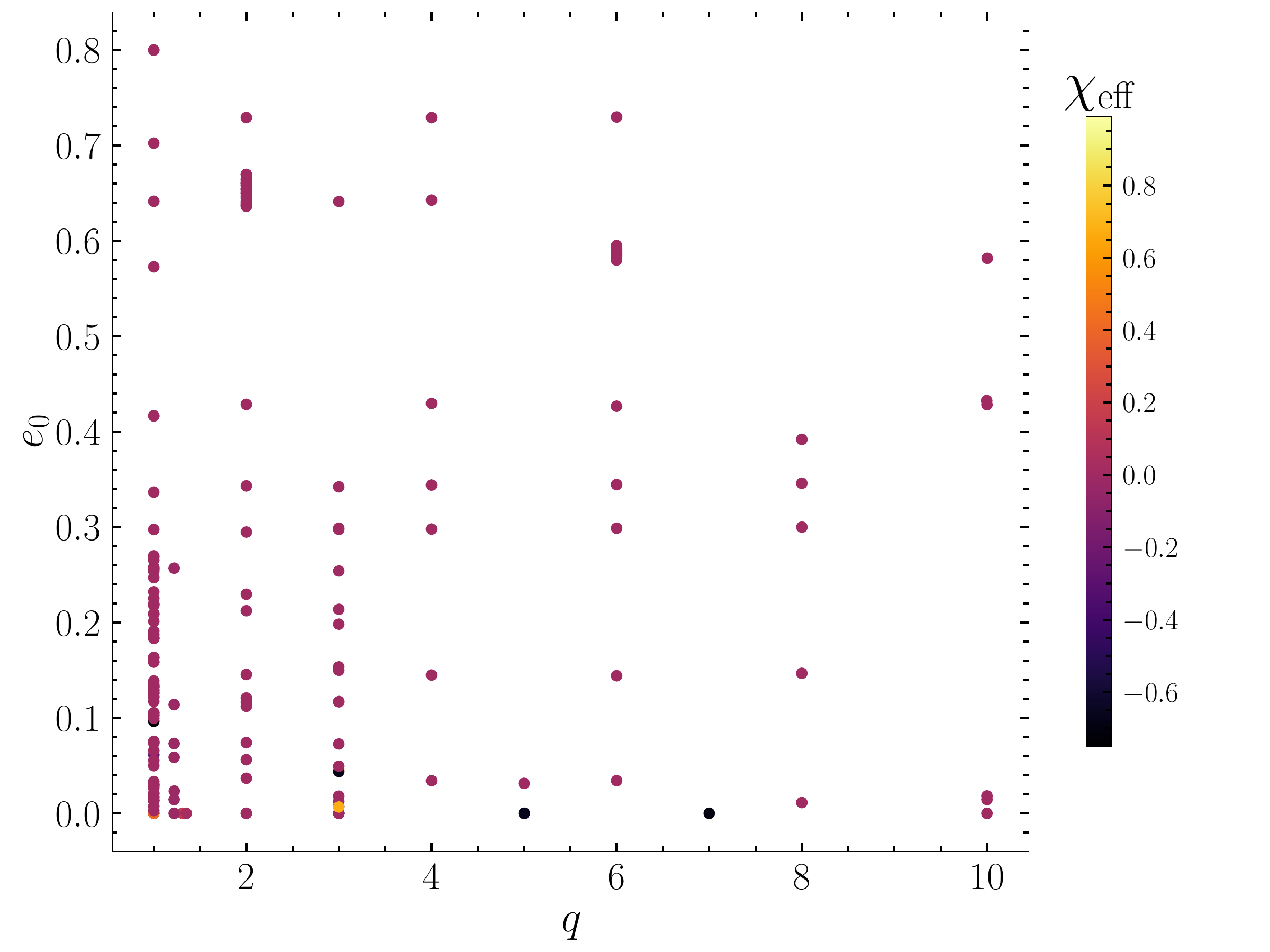} 
    \caption{Parameter space distribution (in initial eccentricity $e_0$, mass ratio $q$, and 
    effective-spin parameter $\chi_{\rm eff}$) for the 186 NR simulations from the public SXS 
    catalog used in Sec. \ref{sec:ecc_NRcomparison}. }
        \label{fig:nr_param_space}
\end{figure}

We assess the accuracy of the \phXE model by computing mismatches against a dataset
 of 186 eccentric BBH NR simulations\footnote{
 In the third release of the SXS catalog \cite{Scheel:2025jct} there are 184 eccentric
 simulations, but we also include \texttt{SXS:BBH:1363} and \texttt{SXS:BBH:1370}, which have
 been deprecated, in order to compare with previous results in the literature
  \cite{Nagar:2021gss,Ramos-Buades:2023yhy,Gamboa:2024hli,Planas:2025feq}.
 }
  generated with the Spectral Einstein Code (SpEC) 
 code~\cite{Hinder:2017sxy,Islam:2021mha, Ramos-Buades:2022lgf} and publicly available in the 
 SXS catalog \cite{Boyle:2019kee,Scheel:2025jct}. Figure \ref{fig:nr_param_space} displays the 
 distribution of the initial eccentricity, mass ratio and effective-spin parameter of the simulations 
 considered. These parameters are extracted from the metadata of the simulations, and they are computed
  from the simulations as described in Ref. \cite{Scheel:2025jct}. The available NR waveforms are mostly 
  non-spinning and concentrated at initial eccentricities smaller than 0.5 (measured at the start of the 
  NR waveform). There is a sparse distribution of simulations with initial eccentricities larger than 0.5,
   reaching a maximum of 0.8 for an equal-mass configuration.

\begin{figure*}
\centering
\includegraphics[width=\textwidth]{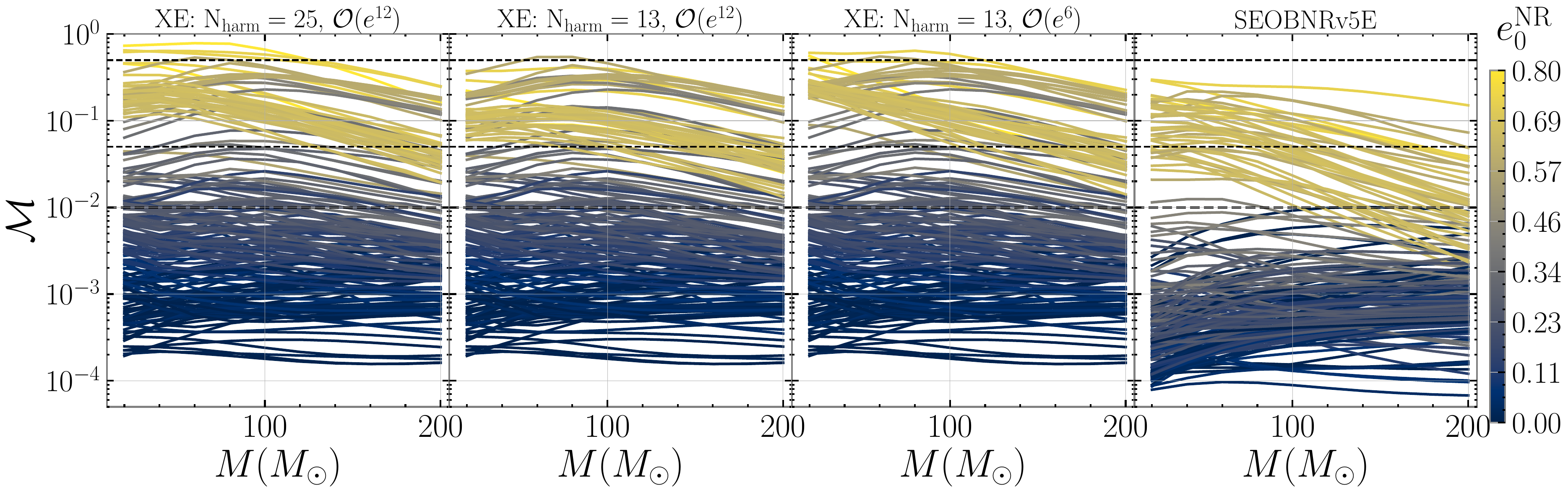}
\caption{Mismatches of the \phXE and \seobE models against the 186 SXS eccentric simulations in Fig.~\ref{fig:nr_param_space}. From left to right, 
\phXE mismatches computed with (N$_{\rm harm},e^X)=\{(25,e^{12}),(13,e^{12}),(13,e^{6})\}$, respectively, 
where N$_{\rm harm}$ corresponds to the number of mean anomaly harmonics and $e^X$ to the highest
order in the eccentricity expansions considered in the waveform. 
The last panel corresponds to the \seobE model.
Each curve corresponds to a NR simulation 
containing the $(2,|2|)$-modes, color-coded by the initial eccentricity $e^{\mathrm{NR}}_0$ 
extracted from the metadata. The mismatches are calculated over a total 
mass range of $M\in[20,200]\ M_{\odot}$.
}  
\label{fig:nr_mismatches}
\end{figure*}

Fig. \ref{fig:nr_mismatches} shows the unfaithfulness, see Eq. \eqref{eq:unfaithfulness}, of
 the \phXE model against all the NR simulations from Fig. \ref{fig:nr_param_space}. Additionally, 
 we compute the mismatch between NR waveforms and the effective-one-body model \seobE, i.e., the state-of-the-art 
 time-domain model \seobEHM \cite{Gamboa:2024hli,Gamboa:2024imd}, but restricted to its $(2,|2|)$ modes. For 
 each NR waveform, we compute the $(2,2)$-mode mismatches by optimizing over the initial eccentricity and 
 mean anomaly\footnote{In the case of \seobE we optimize over the relativistic anomaly parameter, which is 
 the default radial phase parameter of the model.} specified at the initial orbit-averaged frequency 
 measured from the NR waveform, as well as the coalescence phase $\varphi_c$ and coalescence time $t_c$. 
 We optimize over initial eccentricity and mean anomaly only for the lowest total mass considered ($20 M_\odot$) 
 and use those optimal values over the rest of the mass range, for which we optimize over time shifts and phase offsets\footnote{Instead of optimizing over eccentricity and mean anomaly for each total mass, we chose this procedure in order to save computational resources, especially due to the use of the expensive \seobE model. A similar procedure is followed in Ref. \cite{Gamboa:2024hli}.}. 

To help interpret the results presented in Fig. \ref{fig:nr_mismatches}, 
we provide in Table \ref{tab:nr} the SXS IDs of the simulations and the
 maximum mismatch across the mass range for each individual simulation 
with the default version of the \phXE model including $N_{\rm harm}=13$
and expansions in the waveform up to $\mathcal{O}(e^{12})$ (second panel
from the left in Fig. \ref{fig:nr_mismatches}). 

 The unfaithfulness 
 of \phXE against NR is below $3\%$ for $72 \%$ of the simulations. These low mismatch 
 values correspond to simulations with initial eccentricities below 0.4, while high mismatches 
 above $3 \%$ ($28\%$ of cases) result from simulations with larger eccentricities and asymmetric masses.
  The equal-mass simulation  \texttt{SXS:BBH:2527} is an exception, but also has the highest initial eccentricity
   in the catalog, at $e \sim 0.8$. 

   In contrast to the \phXE results, we observe that the mismatches of 
   the \seobE model are typically lower than $1\%$ for eccentricities below 0.4,
   confirming the high accuracy of the model against NR that was already reported in Ref. \cite{Gamboa:2024imd}. 
   We also note a larger set of simulations
   with higher mismatches than these reported in Ref. \cite{Gamboa:2024imd}. This discrepancy is explained because in 
   Ref. \cite{Gamboa:2024imd} mismatches are computed enforcing the length of the \seobE model to be the same as
   the one of NR to avoid the leakage of frequency content due to the difference in length between 
   the model and NR, while here we do not impose such constraint when computing the unfaithfulness as 
   the \phXE is a frequency domain model and the determination of the time-domain length is more involved. Thus,
   we refer the reader to Fig. 4 in Ref. \cite{Gamboa:2024imd} for a more precise estimate of the unfaithfulness
   of \seobE against a similar NR dataset.

\begin{figure*}[!]
\centering
\includegraphics[width=\linewidth]{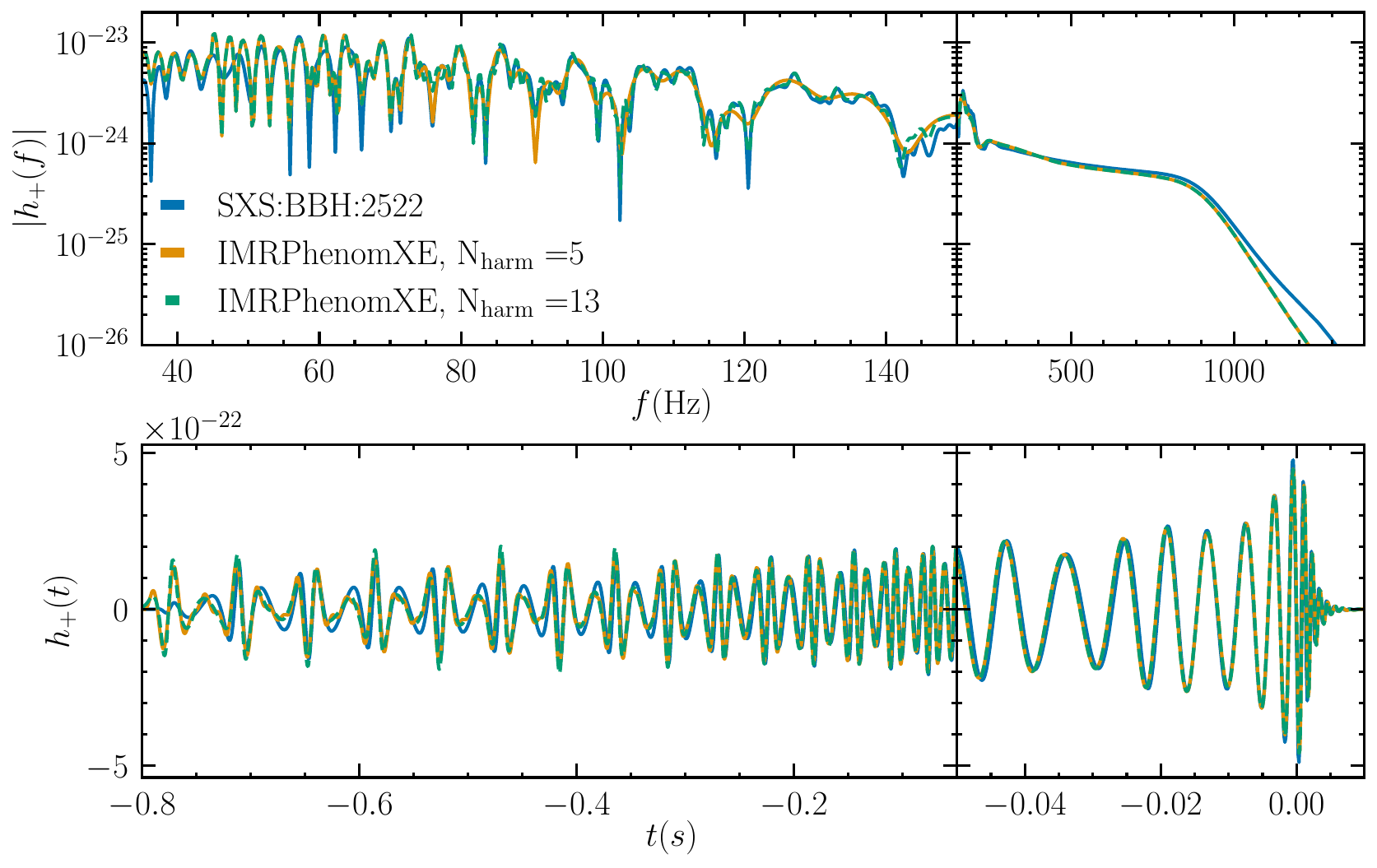}
\caption{Waveform comparison between the NR waveform \texttt{SXS:BBH:2522} (blue) and the best-fitting \phXE waveforms 
generated with 5 (orange) and 13 (green dashed) mean anomaly harmonics (N$_{\rm harm}$). Top panels
 show the frequency-domain amplitude of the $h_+$ polarization, while the lower row shows the 
 time-domain $h_+$ polarization. In each row, the right panel zooms into the merger-ringdown of the waveform. 
 The NR waveform corresponds to an equal-mass, non-spinnning configuration 
 with initial eccentricity $e_0=0.4$ (see Table \ref{tab:nr} for details).}
\label{fig:wf_plot_xe_nr}
\end{figure*}

The difference in accuracy between both models can be explained by the different modeling strategies 
they employ. \seobE combines orbit-averaged evolution equations for the eccentricity and relativistic anomaly
 with an evolution of the instantaneous dynamics through the EOB equations of motion. \phXE, on the other hand,
  relies solely on orbit-averaged evolution equations for the eccentricity and mean anomaly. Additionally,
   \seobE uses noneccentricity-expanded and resummed waveform modes, while \phXE relies on time-domain amplitudes that are 
   eccentricity-expanded up to $e^{12}$\footnote{In a private communication, we have been shown that the 
   calculation of mismatches against the same NR dataset but using the \phTE model \cite{Planas:2025feq} 
   produces results that are similar to the ones obtained with \phXE.}.
    On top of its time-domain amplitude description already impacting accuracy, \phXE performs the SPA approximation
     on the time-domain signal in order to obtain frequency-domain waveforms, which further limits the accuracy 
     of the model. We leave for future work the investigation and implementation of possible 
     strategies to mitigate and overcome such limitations.

\begin{figure}[!]
\centering
\includegraphics[width=\linewidth]{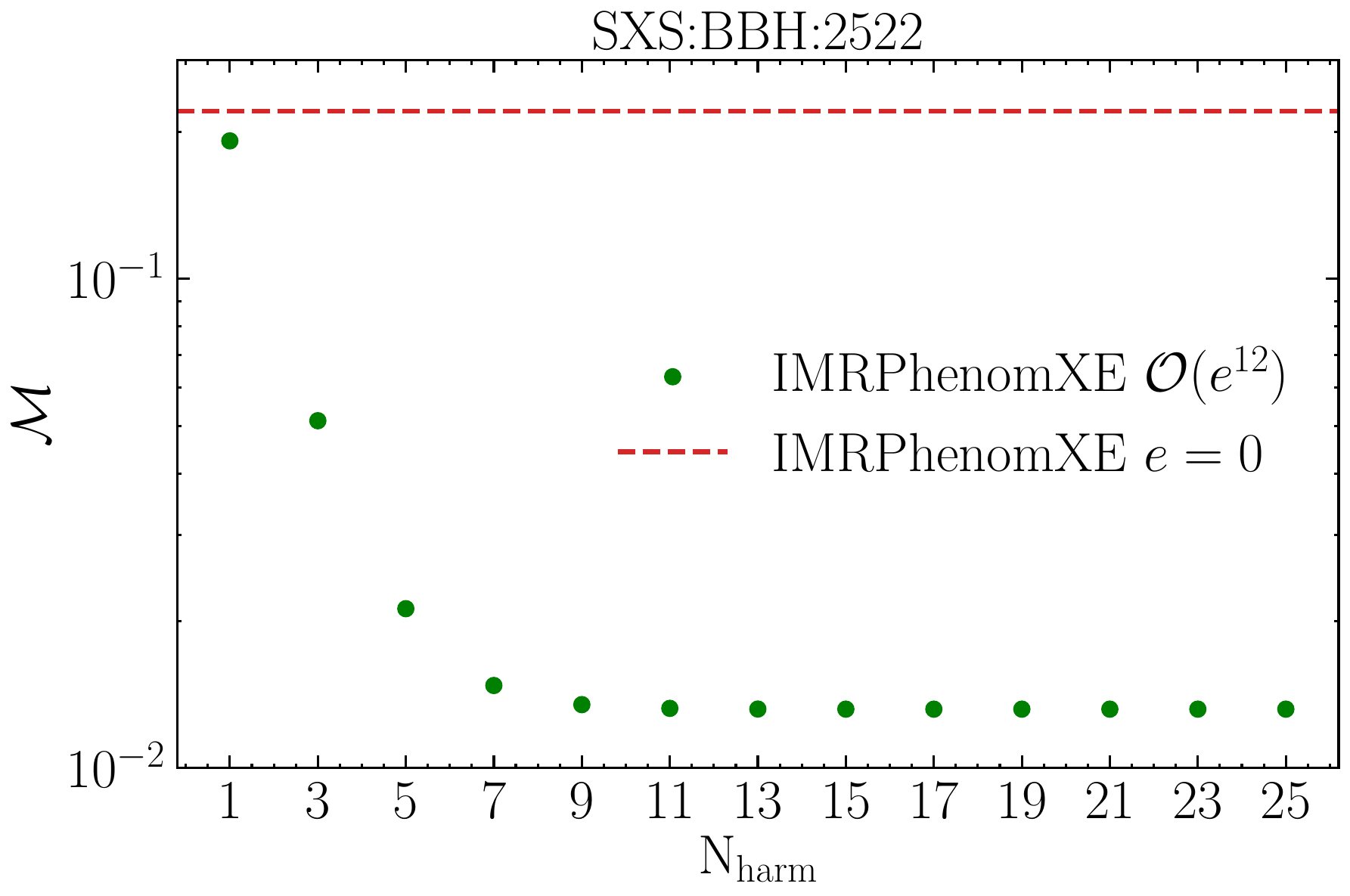}
\includegraphics[width=\linewidth]{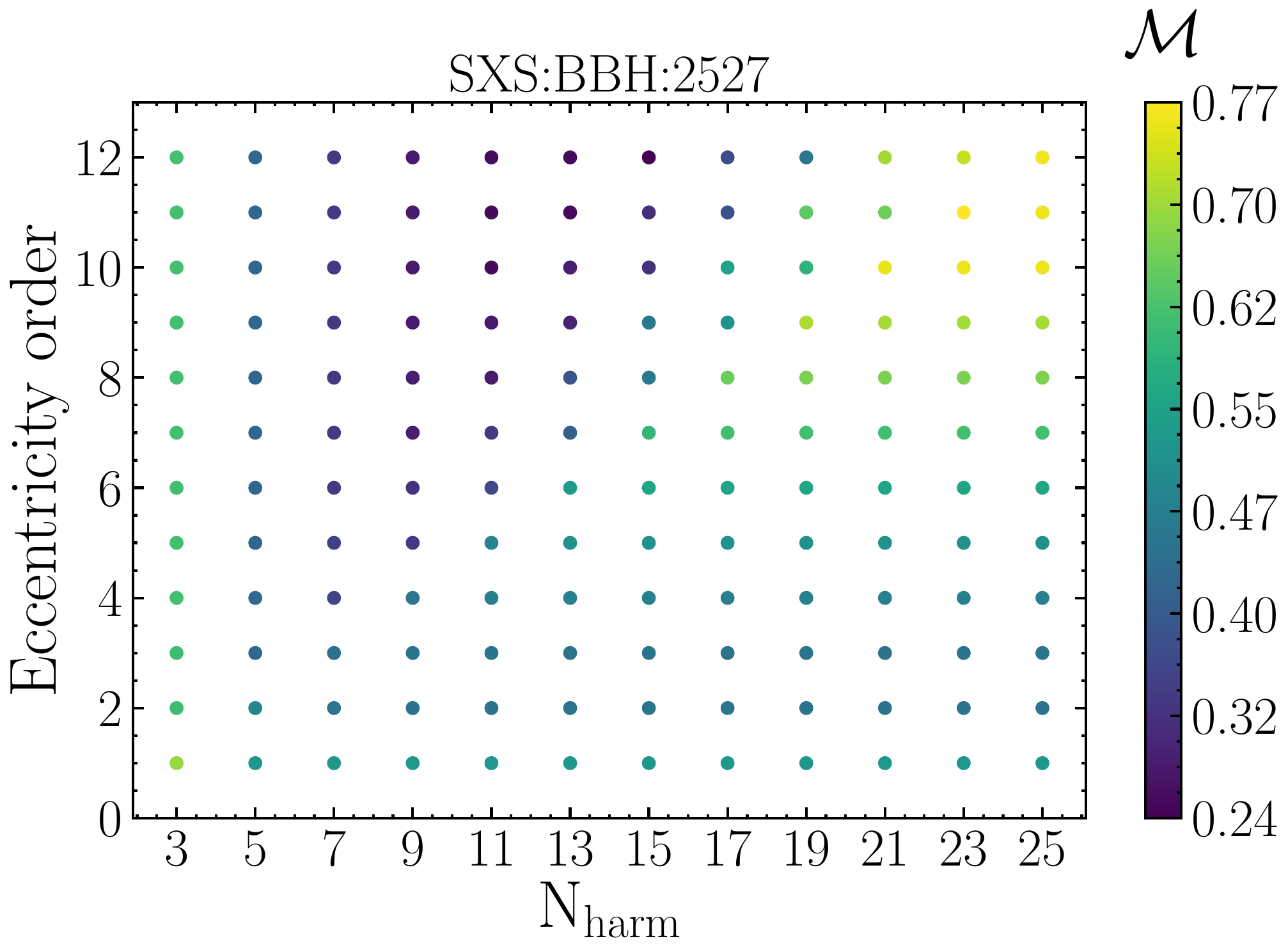}
\caption{\textit{Upper plot:} Mismatch of \phXE against the NR waveform \texttt{SXS:BBH:2522} as a function of the mean anomaly harmonics, N$_{\rm harm}$, for time-domain amplitudes expanded up to $\mathcal{O}(e^{12})$. The horizontal dashed line corresponds to the mismatch of \phXE evaluated with $e=0$. \textit{Lower panel:} Mismatch of \phXE against the NR waveform \texttt{SXS:BBH:2527}, as a function of the mean anomaly harmonics, N$_{\rm harm}$, and eccentricity order, i.e. $\mathcal{O}(e^j)$ with $j=3,5,...,25$, in the time-domain amplitudes. Each point is color-coded by its value of mismatch. In both panels the total mass is fixed to be $20 M_\odot$.
}
\label{fig:harmonics_plot_xe_nr}
\end{figure}

In Figs. \ref{fig:wf_plot_xe_nr} and \ref{fig:harmonics_plot_xe_nr}, we explore the 
impact of the number of mean anomaly harmonics on the accuracy of \phXE. 
Fig. \ref{fig:wf_plot_xe_nr} displays frequency- and time-domain representations of the \phXE model and 
the NR simulation \texttt{SXS:BBH:2522} (see Table \ref{tab:nr} for details), a binary with a moderately 
high initial eccentricity of $e_0 \sim 0.4$ and a total mass of $20 M_\odot$. 
The upper panel shows the frequency-domain amplitude of the plus polarization, $|h_+(f)|$, while the 
lower panel displays the time-domain polarization $h_+(t)$. 

The upper panel of Fig. \ref{fig:harmonics_plot_xe_nr} presents the variation of the 
mismatch between \phXE and the simulation \texttt{SXS:BBH:2522} as a function of 
the number of mean anomaly harmonics included in the \phXE model for a fixed eccentricity 
expansion order of $\mathcal{O}(e^{12})$. 
The results show that the inclusion of 9 instead of 5 mean anomaly harmonics 
decreases the value of the mismatch (from $\sim 2.1\%$ to $\sim 1.3 \%$).
Increasing the number of included mean anomaly harmonics thus improves the accuracy of the model, 
up to the 9th-harmonic\footnote{Note that we are including both the positive and negative harmonics 
from the $j=0$-harmonic. For instance, the first point in Fig. \ref{fig:harmonics_plot_xe_nr}, N$_{\rm harm}=1$, 
corresponds to the inclusion of the $j=0$ harmonic, while the inclusion of the $j=\{0,\pm 1\}$ harmonics corresponds
to N$_{\rm harm}=3$, and similarly for higher-order integers in Fig. \ref{fig:harmonics_plot_xe_nr}.}, 
beyond which we observe that the value of the mismatch gets dominated by the underlying inaccuracies of the model and converges to $1.36 \%$.

Visually the impact of the inclusion of higher harmonics in \phXE can also be observed 
in the upper panel of Fig. \ref{fig:wf_plot_xe_nr}, where the \phXE waveform with 
N$_{\rm harm}=13$ mean anomaly harmonics resembles the early-inspiral features 
in the NR waveform more accurately than the \phXE waveform with N$_{\rm harm}=5$.
For this particular case going to N$_{\rm harm}=13$ (default value of the model) is not 
necessary and one could restrict to N$_{\rm harm}=9$. 
However, if we study cases with higher eccentricity such as \texttt{SXS:BBH:2527}, which 
has an initial eccentricity of 0.8 (see Table \ref{tab:nr}), the use of only  N$_{\rm harm}=9$
is not sufficient. In the lower panel of Fig. \ref{fig:harmonics_plot_xe_nr} we show the distribution
of mismatch as a function of the number of mean anomaly harmonics, N$_{\rm harm}$, and the 
order in the eccentricity expansion of the time-domain amplitudes for a total mass of $20 M_\odot$.
The results show that the use of expansions up to $e^{12}$ can reduce the mismatch from $>50\%$ to 
$\sim 20\%$. The lower plot of Fig. \ref{fig:harmonics_plot_xe_nr} also demonstrates that the use 
of expansions up to $\mathcal{O}(e^6)$ with N$_{\rm harm}=13$ typically developed in the 
literature \cite{Boetzel:2019nfw,Ebersold:2019kdc,Henry:2023tka} can lead to large values of unfaithfulness
for such high eccentricity NR simulations. Furthermore, we observe that the inclusion of higher harmonics
and high eccentricity harmonics can degrade the accuracy of the model, which is potentially related to the 
radius of convergence of the power series used to invert Eq. \eqref{eq:GKE},
which is set to $e_{\rm max} \sim 0.6627434$, and which is propagated throughout
 the calculation of the time-domain eccentric amplitudes in the \phXE model.

A similar study can be performed with the full dataset of NR waveforms available in Table \ref{tab:nr}, 
and this is shown in the three leftmost panels in Fig. \ref{fig:nr_mismatches}. 
The conclusions are similar as in Fig. \ref{fig:harmonics_plot_xe_nr}. 
In Fig. \ref{fig:nr_mismatches} from left to right in the first three panels we
display the mismatches of the \phXE model including all the harmonics and the highest eccentricity 
expansion available, i.e. N$_{\rm harm}=25$ and $\mathcal{O}(e^{12})$, 
reducing the number of harmonics to N$_{\rm harm}=13$ and keeping $\mathcal{O}(e^{12})$,
and with  N$_{\rm harm}=13$ and $\mathcal{O}(e^{6})$, respectively. 
The results show that the inclusion of harmonics up to 25 degrades the accuracy for the high eccentricity 
simulations, potentially due to the convergence issues of the inversion of Kepler equation, Eq. \eqref{eq:GKE},
but also due to the fact that for these high harmonics only the leading order eccentricity corrections are 
included. We leave for future work investigating the inclusion of higher order eccentricity corrections 
in these harmonics as well as the use of resummation techniques to avoid explicit eccentricity expansions.

When reducing the number of harmonics while keeping the expansions in eccentricity up to $e^{12}$ we observe
a reduction of the values of the mismatches for the high eccentricity cases below $<50\%$, while the reduction 
in the eccentricity expansion order to $e^{6}$  with N$_{\rm harm}=13$ causes an increase of the mismatches 
against the high eccentricity NR simulations which is especially
noticeable at low total masses.

As a consequence, we set as N$_{\rm harm}=13$ a the default number of harmonics 
for the \phXE model with eccentricity expansions in the waveform up to $e^{12}$.
We also leave N$_{\rm harm}$ as parameter that the user can freely modify accordingly
to the eccentric parameter which is being targeted.
In Sec. \ref{sec:benchmarks}, we study the impact of this choice on the computational 
efficiency of the model, and investigte further implications for the accuracy of the model 
through parameter estimation studies on mock signals and real GW events in Sec. \ref{sec:PE}.

\section{Benchmarks} \label{sec:benchmarks}
\begin{figure}[!]
\centering
\includegraphics[width=\linewidth]{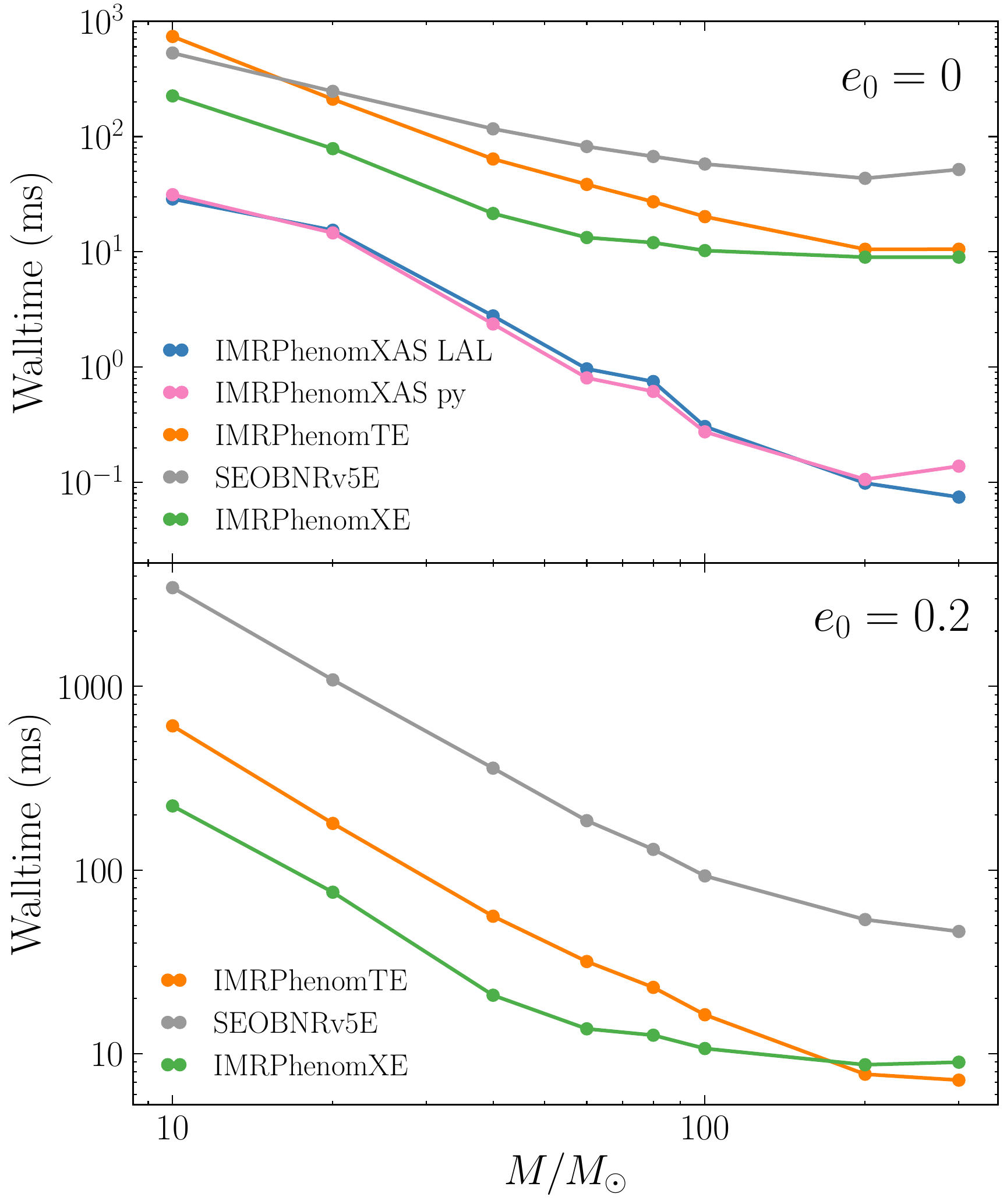}
\caption{Walltime (in milliseconds) as a function of the total mass for several waveform models. The upper panel corresponds to a quasi-circular binary ($e_0=0$), while the lower panel to an eccentric binary with initial eccentricity $e_0=0.2$, and mean anomaly $l_0=1.2$~rad. In both panels the rest of the binary parameters are identical and correspond to mass ratio $q=3$, spins $\chi_1 = 0.4$ and $\chi_2=0.3$, with a starting frequency of $10$ Hz. We compare the eccentric time-domain phenomenological \phTE model (orange), \seobE  (gray) and \phXE (green). Additionally, we compare the quasi-circular \phX implemented in LALSuite (blue) and its implementation in \texttt{phenomxpy} (pink) for the case $e=0$. The benchmark is performed over a range of total masses, $M=\{10,20,40,60,80,100,200,300\}M_{\odot}$. }
\label{fig:benchmark}
\end{figure}

One of the main applications of waveform models is their use for Bayesian inference analyses which 
require the generation of millions of waveforms over different regions of parameter space. 
Consequently, computational efficiency is a critical requirement for any waveform model intended for large-scale data analysis.

A distinctive feature of the \phXE model, compared to other inspiral–merger–ringdown (IMR) time-domain eccentric waveform models~\cite{Islam:2021mha, Setyawati:2021gom, Gamboa:2024hli, Gamba:2024cvy,Planas:2025feq}, is its computational efficiency through its implementation in the highly modular and efficient \texttt{phenomxpy} python package \cite{Garcia-Quiros:2025usi}.  

Eccentric waveform models typically incur a high computational cost due to the complexity of the orbital dynamics and the evolution of eccentricity-related quantities. For instance, the state-of-the-art \seobE model requires, in addition to solving the EOB Hamiltonian and equations of motion, the numerical integration of secular evolution equations involving extended PN expressions evaluated at each time step. While such prescriptions yield a high-fidelity description of eccentric dynamics, they are computationally intensive.
In contrast, the \phXE model adopts a phenomenological framework based on the efficient eccentric dynamics previously implemented in the phenomenological time-domain \phTE waveform model \cite{Planas:2025feq}, combined with an efficient waveform construction in frequency-domain, enabling significantly faster waveform generation. Although this approach does not capture eccentric effects with the same level of accuracy as state-of-the-art EOB models, it offers a substantially more efficient alternative that is well suited for large-scale parameter estimation and population studies.

We present benchmark results\footnote{Benchmarks were performed using a 16-core AMD Ryzen 7 PRO 7840U CPU.} in 
Fig.~\ref{fig:benchmark} for a representative configuration with mass ratio $q=3$, component 
spins $\chi_1=0.4$ and $\chi_2=0.3$, and mean anomaly $l=1.2$~rad, for two initial 
eccentricities ($e=0$ and $e=0.2$) defined at a starting frequency of $10$~Hz. The figure 
reports the walltime in milliseconds required to generate each waveform as a function 
of the total mass for a mass range of $M=[10-300]M_\odot$. We compare the \phTE, \seobE and
 \phXE eccentric waveform models\footnote{We do not include the \texttt{TEOBResumS-Dal\'i} model
  \cite{Nagar:2021gss} in the benchmark 
study to save computing resources. We note that in Ref. \cite{Planas:2025feq} it was
 found to be computationally more costly for low masses and comparable in speed for 
 high masses to the \phTE model.}. In the quasi-circular case ($e_0=0$), we add to the comparison 
 the \phX model implemented in LALSuite \cite{Pratten:2020fqn} and its implementation in \texttt{phenomxpy}, 
 which underlies the \phXE model. 
 
The upper panel in Fig. \ref{fig:benchmark} corresponds to the quasi-circular limit ($e_0=0$). 
At low total masses, \phXE has a walltime of $\sim 200$ms, making it $\sim \times 4$ faster than the 
time-domain \phTE and \seobE models. For these masses, both \phTE and \seobE perform similarly, 
with walltimes of $\sim 800$ms, indicating that the time-domain models are being dominated by 
the interpolation of the waveform into a constant time step in order to perform the Fourier transform. 
At high total masses, the efficient waveform evaluation of \phTE with analytical closed-form expressions 
for the waveform makes it more efficient than the more complex \seobE model, and comparable to the 
\phXE model with walltimes of $\sim 200$ ms. We note that \phXE is, however, still substantially 
slower than the underlying quasi-circular \phX model, which has walltimes ranging from $\sim 30$ms 
to $\sim 0.1$ms for total masses in the $10 M_\odot$ to $300 M\odot$ range. 
\phX is expected to outperform \phXE in terms of speed, due to the use of analytical expressions
 to compute the \phX phase and amplitude. In contrast, \phXE relies on the numerical evolution of the 
 eccentric dynamics and a more complex waveform evaluation. Additionally, we observe very similar
  timings when comparing the \phX implementation in \texttt{phenomxpy} with its counterpart in 
  LALSuite \cite{Pratten:2020fqn}. This indicates that there are no additional penalties in 
  computational cost for the Python implementation in the new \texttt{phenomxpy} package, 
  compared to the previous C99 implementation in LALSuite.

Turning to the case with $e_0=0.2$ in the lower panel of Fig. \ref{fig:benchmark}, 
we observe that \phXE outperforms both \seobE and \phTE at low total masses. 
The walltimes 
of \phXE are $\sim 200$ms at $10 M_\odot$, compared to $\sim 800$ms for \phTE and $\sim 1.5$s 
for \seobE. At total masses greater than $100 M_\odot$, the performance of the \phXE and \phTE models
 becomes comparable with walltimes $\sim 10$ms, while \seobE is $\sim \times 10$ slower. These similar 
 walltimes of \phXE and \phTE are expected, as their computational cost is dominated by the evolution 
 of the eccentric dynamics which is common in both models implemented in \texttt{phenomxpy}. 
 
 We note that eccentricity is a gauge dependent parameter, which
can imply distinct merger times for different models. However, \phXE and \phTE evolve the 
same orbit-average equations describing the eccentric dynamics, thus, their merger times are
expected to be almost identical, while with respect to \seobE differences may arise due to the 
fact that the orbit-average equations are combined with the instantaneous EOB equations \cite{Gamboa:2024hli}. Hence,
when interpreting the results in Fig. \ref{fig:benchmark} additional caution needs to be taken accounting 
for possible small differences in waveform length for the different models considered. 

In conclusion, the benchmarks demonstrate the unique computational efficiency of \phXE. This, in turn, enables parameter-estimation studies over a wide range of source configurations, including low total masses for which inspirals are longer and eccentricity effects may show prominently, as shown in Sec.~\ref{sec:PE}.


\section{Bayesian inference studies}\label{sec:PE}
A primary application of waveform models is the Bayesian inference of source parameters from GW data. 
In this section, we evaluate the performance of the eccentric, aligned-spin \phXE model through parameter-estimation (PE) analyses. We perform synthetic zero-noise injections of three NR waveforms introduced in Sec.~\ref{sec:PE_nr}, and we analyze three observed GW events reported by the LVK Collaboration — GW150914~\cite{LIGOScientific:2016aoc}, GW151226~\cite{LIGOScientific:2016sjg} and GW190521~\cite{LIGOScientific:2020iuh}. Our results are compared with existing studies in the literature, in particular those obtained using the \seobE ~\cite{Gamboa:2024hli} and \phTE \cite{Planas:2025feq} models. 

We perform all analyses using the Python package \texttt{Bilby}~\cite{Ashton_2019,Romero-Shaw:2020owr} with the nested sampling algorithm \texttt{dynesty}~\cite{Speagle:2019ivv}. 
The reference eccentricity $e_{\rm ref}$ and mean anomaly $l_{\rm ref }$ are assigned uniform priors over $l_{\rm{ref}}\in [0,2\pi ]$ and $e_{\rm ref} \in [0,0.4]$, where the upper bound of 0.4 in the eccentricity prior is chosen according to the accuracy results in Sec. \ref{sec:NRunfaithfulness}, where we observed substantial accuracy loss against NR above that value.

The priors on the inverse mass ratio ($1/q$) and chirp mass ($\mathcal{M}$) are chosen to yield uniform sampling in the component masses. 
For the spin parameters $\chi_i$, we adopt priors corresponding to the projection of an isotropic spin distribution onto the direction perpendicular to the orbital plane~\cite{Veitch:2014wba}. The luminosity distance prior is taken proportional to $d_L^2$~\cite{LIGOScientific:2018mvr,LIGOScientific:2020ibl,LIGOScientific:2021djp,LIGOScientific:2025slb}, 
except for GW190521, for which we assume a prior uniform in comoving volume following
Ref.~\cite{LIGOScientific:2020iuh}. All remaining priors, including those on extrinsic parameters and
 orbital phase $\varphi$, are consistent with Ref.~\cite{LIGOScientific:2025slb}. The full prior ranges are specified in the corresponding sections.

Because eccentricity is gauge dependent in general relativity, meaningful comparisons across waveform
 models—or across coordinate choices within a given model—require a common, gauge-invariant prescription.
  While the GW eccentricity $e^{\mathrm{GW}}$ and mean anomaly $l^{\mathrm{GW}}$ defined in Refs.~\cite{Ramos-Buades:2022lgf,Shaikh:2023ypz,Shaikh:2025tae} provide a suitable framework for such
 comparisons, their measurement is typically performed on time-domain waveforms, whereas \phXE is
  constructed in the frequency domain.  We find that a direct inverse Fourier transform of \phXE may
  introduce some numerical noise and minor artifacts that would require additional filtering and
  conditioning before applying the \texttt{gw\_eccentricity} package~\cite{Shaikh:2023ypz,Shaikh:2025tae}. 
 Developing a robust procedure for time-domain reconstruction and post-processing of the \phXE waveforms is left for future work.

A summary of all PE runs is given in Table~\ref{tab:pesummary}, including the waveform models employed,
and the associated computational cost. All analyses use 
 \texttt{naccept}=60 and \texttt{nlive}=1000 in \texttt{dynesty}, with distance and phase marginalization
  enabled. We employ different sampling frequencies for the different analyses, 1024Hz for GW150914,
   16834Hz for GW151226 and 4096Hz for GW190521 as well as the NR injections. 
The computational efficiency of \phXE enables systematic exploration of modeling choices and their impact
 on waveform systematics. Specifically, we explore the choice of number of harmonics for some runs.

The \phXE model allows the starting frequency to be specified independently of the reference frequency. Consequently, changes in $f_{\mathrm{start}}$ driven by the inclusion of highermean anomaly
harmonics do not alter the physical parameters of the source. The inclusion of high positive 
eccentric harmonics in band requires that the \phXE waveform is started at lower frequencies \cite{Planas:2025plq},
\begin{equation}
f^{\rm wf}_{\mathrm{min},j} \approx  \frac{2 }{2 + j }f_{\rm start},
\label{eq:fstart}
\end{equation}
where $f^{\rm wf}_{\mathrm{min},j}$ indicates the starting frequency of waveform generation of \phXE in order to have the j-harmonic starting at  $f_{\rm start}$, which  is the starting frequency of the analysis. Eq. \eqref{eq:fstart} is an approximation based on the SPA condition, and after neglecting a frequency dependent term, this is why the approximation is not valid for $j \leq -2$.  Equation \eqref{eq:fstart} is only valid for the $(2,2)$-mode and provides an estimate of the starting frequency of the waveform such that the $j$-harmonic is fully contained at the starting frequency.

\begin{table}[ht]
\renewcommand{\arraystretch}{1.3}
\centering
\begin{tabular}{@{}cccc@{}}
\toprule
\textbf{Event} 
& \textbf{Model} 
& \begin{tabular}{c} \textbf{Computing} \\ \textbf{resources} \end{tabular} 
& \textbf{Runtime} \\
\midrule

\texttt{SXS:BBH:1355} 
    & \begin{tabular}{c}\phXE \\ (N$_{\rm harm}$=5)\end{tabular}
  & 128 $\times$ 1 
  & 43 min \\
  & \begin{tabular}{c}\phXE \\ (N$_{\rm harm}$=13)\end{tabular}
  & 128 $\times$ 1 
  & 48 min \\
\midrule
\texttt{SXS:BBH:1359}  
    & \begin{tabular}{c}\phXE \\ (N$_{\rm harm}$=5)\end{tabular}
  & 128 $\times$ 1 
  & 57 min \\
  & \begin{tabular}{c}\phXE \\ (N$_{\rm harm}$=13)\end{tabular}
  & 128 $\times$ 1 
  & 63 min \\
\midrule
\multirow{3}{*}{\texttt{SXS:BBH:1363}  }
    & \begin{tabular}{c}\phXE \\ (N$_{\rm harm}$=5)\end{tabular}
  & 128 $\times$ 1 
  & 69 min \\
  & \begin{tabular}{c}\phXE \\ (N$_{\rm harm}$=9)\end{tabular}
  & 128 $\times$ 1 
  & 82 min \\
\midrule \midrule
\multirow{2}{*}{\textbf{GW150914}}
  & \phX
  & 128 $\times$ 1 
  & 8 min \\
  & \phXE
 & 128 $\times$ 1 
  & 68 min \\

\midrule

\multirow{2}{*}{\textbf{GW151226}}
  & \phX
  & 128 $\times$ 1 
  & 84 min \\
  & \phXE
  & 128 $\times$ 1 
  & 235 min \\

\midrule
\multirow{2}{*}{\textbf{GW190521}}
  & \phX
  & 128 $\times$ 1 
  & 8 min \\
  & \phXE
  & 128 $\times$ 1 
  & 23 min \\

\bottomrule
\end{tabular}
\caption{Summary of the parameter estimation (PE) runs performed in this study. Columns list the injected NR simulations or GW events, the waveform model used, the computing resources, and the runtime. All runs analyzed 8 seconds of data with a minimum frequency of 10 Hz, reference frequency of 20 Hz, and maximum frequency of 2048 Hz. Data from the Hanford (H), Livingston (L), and Virgo (V) detectors were used, except for GW150914 and GW151226, which only included HL data. For the NR injections the \phXE runs are tested with a number of mean anomaly harmonics N$_{\rm harm}=5$ and N$_{\rm harm}=13$, the rest of the runs listed for \phXE are performed with  N$_{\rm harm}=13$.}
\label{tab:pesummary}
\end{table}

\subsection{NR injections}\label{sec:PE_nr}

\begin{table*}
\renewcommand{\arraystretch}{1.3}
\begin{tabular}{c c c c c c c  c}
		\hline
        \hline
        \multirow{2}{*}{\textbf{Event}} & \multirow{2}{*}{\textbf{Parameter}} & \textbf{Injected} & \phXE &  \phXE &   \phTE  & \texttt{SEOBNRv5E}    \\
        &  & \textbf{value} & (N$_{\rm harm}=5)$ &  (N$_{\rm harm}=13)$ & \cite{Planas:2025feq}  & \cite{Gamboa:2024hli}   \\
        \hline
        \hline
        \multirow{11}{*}{\texttt{SXS:BBH:1355}}
        & $M / M_\odot$ & 70.0 &  $70.92^{+2.56}_{-2.35}$ & $70.9^{+2.59}_{-2.26}$ & $70.70^{+3.10}_{-2.67}$  &  ${71.05}^{+2.62}_{-2.35} $  \\
        & $\mathcal{M} / M_\odot$ & 30.47 &  $30.35^{+1.01}_{-0.97}$ & $30.36^{+0.98}_{-0.96}$ &  $30.34^{+1.19}_{-1.16}$   & ${30.52}^{+1.03}_{-0.97} $  \\
        & $1 / q$ & 1.0 & $0.77^{+0.18}_{-0.19}$ & $0.77^{+0.18}_{-0.2}$ & ${0.8}^{+0.16}_{-0.19} $  \\
        & $\chi_{\text{eff}}$ & 0.0 &  $0.0^{+0.09}_{-0.09}$ & $0.0^{+0.08}_{-0.09}$  & $0.00^{+0.10}_{-0.10}$   & ${0.02}^{+0.09}_{-0.09} $  \\
        & $e_{20\mathrm{Hz}}\ \left(e^{\mathrm{GW}}_{20\mathrm{Hz}}\right)$ & - (0.07)  &  $0.04^{+0.04}_{-0.03} (-)$ & $0.04^{+0.04}_{-0.03}(-)$ & $0.05^{+0.05}_{-0.04} \left(0.05^{+0.05}_{-0.04}\right)$  &    ${0.05}^{+0.04}_{-0.04}  \left( {0.06}^{+0.04}_{-0.04}  \right) $   \\
        & $l_{20\mathrm{Hz}}\ \left(l^{\mathrm{GW}}_{20\mathrm{Hz}}\right)$ & - (1.96)  & $1.93^{+1.78}_{-1.14} (-)$ & $1.92^{+1.7}_{-1.13}(-)$ & $2.04^{+2.01}_{-1.40} \left(1.94^{+2.36}_{-1.37}\right)$   &  $^*{2.27}^{+1.14}_{-1.10}  \left( {2.10}^{+1.20}_{-1.06}  \right) $ \\
        & $\iota [\text{rad}]$ & 0.0  &  $0.63^{+0.52}_{-0.38}$ & $0.62^{+0.52}_{-0.37}$ & $0.61^{+0.62}_{-0.43}$  &  ${0.62}^{+0.48}_{-0.37} $ \\
        & $d_L[\text{dMpc}]$ & 2307 &   $1828^{+381}_{-572}$ & $1829^{+379}_{-571}$ & $1827^{+442}_{-682}$  &  ${1835}^{+376}_{-569} $  \\
        & $\varphi[\text{rad}]$ & 0.0 & $3.13^{+2.48}_{-2.5}$ & $3.21^{+2.45}_{-2.53}$ & $3.11^{+2.87}_{-2.79}$   & ${3.14}^{+2.52}_{-2.51} $  \\
        & $\text{SNR}_N$ & 20.0 & $17.77^{+0.11}_{-0.21}$ & $17.77^{+0.12}_{-0.21}$  &  $19.07^{+0.10}_{-0.19}$   &  ${19.07}^{+0.09}_{-0.14} $ \\
        \bottomrule
        \multirow{11}{*}{\texttt{SXS:BBH:1359}}
        & $M / M_\odot$ & 70.0 &  $70.75^{+2.41}_{-2.23}$ & $70.8^{+2.46}_{-2.22}$ & $70.26^{+2.89}_{-2.64}$ & ${70.77}^{+2.59}_{-2.44} $  \\
        & $\mathcal{M} / M_\odot$ &  30.47 &  $30.37^{+0.98}_{-0.95}$ & $30.4^{+0.98}_{-0.96}$ &   $30.21^{+1.12}_{-1.18}$   & ${30.43}^{+1.09}_{-1.09} $  \\
        & $1 / q$ & 1.0 & $0.79^{+0.16}_{-0.19}$ & $0.8^{+0.16}_{-0.2}$  &  $0.80^{+0.18}_{-0.23}$ & ${0.8}^{+0.16}_{-0.19} $ \\
        & $\chi_{\text{eff}}$ & 0.0 &  $0.01^{+0.08}_{-0.08}$ & $0.01^{+0.08}_{-0.09}$ &  $0.00^{+0.10}_{-0.10}$  & ${0.02}^{+0.1}_{-0.11} $   \\
        & $e_{20\mathrm{Hz}}\ \left(e^{\mathrm{GW}}_{20\mathrm{Hz}}\right)$ & - (0.07)  &   $0.1^{+0.04}_{-0.04} (-)$ & $0.1^{+0.04}_{-0.04}(-)$   & $0.12^{+0.05}_{-0.05} \left(0.12^{+0.05}_{-0.05}\right)$  &  ${0.13}^{+0.03}_{-0.04} \left({0.13}^{+0.03}_{-0.04} \right) $  \\
        & $l_{20\mathrm{Hz}}\ \left(l^{\mathrm{GW}}_{20\mathrm{Hz}}\right)$ & - (1.96)  & $1.05^{+5.03}_{-0.86} (-)$ & $1.1^{+5.0}_{-0.92}(-)$ & $0.86^{+5.29}_{-0.75} \left(0.89^{+5.31}_{-0.79}\right)$   &  $^* {1.27}^{+1.83}_{-0.9} \left( {1.13}^{+4.59}_{-0.84} \right)$  \\
        & $\iota [\text{rad}]$ & 0.0  &   $0.62^{+0.51}_{-0.38}$ & $0.62^{+0.52}_{-0.38}$  & $0.61^{+0.59}_{-0.44}$  & ${0.62}^{+0.48}_{-0.38} $ \\
        & $d_L[\text{dMpc}]$ & 2307 &  $1849^{+382}_{-575}$ & $1848^{+396}_{-580}$  & $1824^{+452}_{-667}$   &  ${1827}^{+381}_{-564} $   \\
        & $\varphi[\text{rad}]$ & 0.0 & $3.12^{+2.51}_{-2.5}$ & $3.15^{+2.48}_{-2.53}$  & $3.11^{+2.84}_{-2.80}$   & ${3.14}^{+2.52}_{-2.48} $    \\
        & $\text{SNR}_N$ & 20.0 &  $17.66^{+0.12}_{-0.22}$ & $17.65^{+0.13}_{-0.22}$   &   $19.00^{+0.11}_{-0.20}$   &  ${19.05}^{+0.08}_{-0.14} $  \\
        \bottomrule
        \multirow{11}{*}{\texttt{SXS:BBH:1363}}
        & $M / M_\odot$ & 70.0 & $72.21^{+4.25}_{-3.73}$ & $74.96^{+3.92}_{-3.65}$ & $71.83^{+4.75}_{-3.46}$ &  ${71.13}^{+3.53}_{-3.25} $  \\
        & $\mathcal{M} / M_\odot$ & 30.47 & $30.77^{+1.81}_{-1.81}$ & $32.05^{+1.55}_{-1.61}$ &  $30.68^{+1.97}_{-1.59}$   & ${30.61}^{+1.53}_{-1.44} $  \\
        & $1 / q$ & 1.0 &   $0.74^{+0.2}_{-0.21}$ & $0.76^{+0.18}_{-0.2}$  &  $0.75^{+0.22}_{-0.24}$ & ${0.81}^{+0.15}_{-0.18} $ \\
        & $\chi_{\text{eff}}$ & 0.0 & $0.09^{+0.12}_{-0.12}$ & $0.15^{+0.1}_{-0.11}$  &  $0.10^{+0.12}_{-0.11}$   & ${0.03}^{+0.12}_{-0.12} $  \\
        & $e_{20\mathrm{Hz}}\ \left(e^{\mathrm{GW}}_{20\mathrm{Hz}}\right)$ & - (0.07)  &  $0.26^{+0.03}_{-0.04} (-)$ & $0.22^{+0.03}_{-0.04}(-)$  &  $0.26^{+0.03}_{-0.05}\ \left(0.26^{+0.03}_{-0.05}\right)$ &   ${0.24}^{+0.03}_{-0.03} \left( {0.24}^{+0.03}_{-0.03} \right) $  \\
        & $l_{20\mathrm{Hz}}\ \left(l^{\mathrm{GW}}_{20\mathrm{Hz}}\right)$ & - (1.96)  & $4.35^{+0.66}_{-0.68} (-)$ & $4.1^{+0.66}_{-0.66}(-)$ &  $4.58^{+0.67}_{-0.77}\ \left(4.45^{+0.66}_{-0.77}\right)$  &  $ ^*{4.01}^{+0.83}_{-0.71} \left( {4.1}^{+0.88}_{-0.91} \right)$ \\
        & $\iota [\text{rad}]$ & 0.0  &  $0.62^{+0.52}_{-0.38}$ & $0.63^{+0.51}_{-0.39}$ &  $0.61^{+0.59}_{-0.44}$  & ${0.62}^{+0.49}_{-0.38} $ \\
        & $d_L[\text{dMpc}]$ & 2307 & $1985^{+444}_{-631}$ & $2113^{+453}_{-662}$ &   $1970^{+507}_{-732}$  &  ${1910}^{+411}_{-593} $  \\
        & $\varphi[\text{rad}]$ & 0.0 & $3.15^{+2.52}_{-2.52}$ & $3.16^{+2.49}_{-2.52}$ &  $3.15^{+2.83}_{-2.81}$  &  ${3.13}^{+2.52}_{-2.52} $   \\
        & $\text{SNR}_N$ & 20.0 &  $17.38^{+0.12}_{-0.22}$ & $17.25^{+0.13}_{-0.24}$  &   $18.84^{+0.11}_{-0.20}$  &  ${18.99}^{+0.09}_{-0.15} $   \\
        \bottomrule
        \hline
        \hline
   \end{tabular}
    \caption{Injected, median values, and 90\% credible intervals for the posterior distributions shown in Fig.~\ref{fig:NRinjections} for the three NR injections (one per row), recovered with \phXE using N$_{\rm harm}=\{5,13\}$ (or equivalently $N_e=\{2,6\}$) mean anomaly harmonics. 
    The table lists the total mass $M$ and chirp mass $\mathcal{M}$ (both in solar masses), the inverse mass ratio $1/q$, the effective-spin parameter $\chi_{\mathrm{eff}}$, the reference eccentricity and mean anomaly ($e_{20\mathrm{Hz}}$ and $l_{20\mathrm{Hz}}$ respectively), the inclination angle $\iota$, the luminosity distance $d_L$, the coalescence phase $\varphi$, and the network matched-filtered SNR, $\text{SNR}_{\mathrm{N}}$, for LIGO Hanford and Livingston, and Virgo detectors.  For completeness, the injected and recovered GW eccentricity $e_{\mathrm{GW}}$ and mean anomaly $l_{\mathrm{GW}}$ are reported in brackets. For \seobE the $^*$ indicates that the reported quantity is the relativistic anomaly which differs from the mean anomaly \cite{Gamboa:2024hli}. All quantities are evaluated at a reference frequency of 20 Hz.}
    \label{tab:NRresults}
\end{table*}

In this section, we perform zero-noise injections of three publicly available eccentric NR waveforms and carry out PE studies to assess the ability of the \phXE model to recover the injected source parameters. For consistency with Refs.~\cite{Ramos-Buades:2023yhy,Gamboa:2024hli,Planas:2025feq}, we select the same three simulations from the SXS catalog: \texttt{SXS:BBH:1355}, \texttt{SXS:BBH:1359}, and \texttt{SXS:BBH:1363}, which correspond to GW eccentricities at the initial orbit-averaged frequency of $e_0^{\mathrm{GW}} = 0.077$, $0.145$, and $0.317$, respectively.

For these injections, we include all available NR modes up to $l = 8$, fix the total mass to $M = 70 M_{\odot}$, 
set the inclination angle to $\iota = 0$, choose a coalescence phase of $\varphi = 0$, and place the source at 
a luminosity distance of $d_L = 2307$ Mpc. This setup yields a network matched-filtered signal-to-noise ratio 
($\text{SNR}_\mathrm{N}$) of $\mathrm{SNR}_{\mathrm{N}} \approx 20$ in a three-detector configuration using 
the Advanced LIGO (Livingston and Hanford) and Virgo design-sensitivity 
PSDs~\cite{TheLIGOScientific:2014jea,TheVirgo:2014hva,Barsotti:2018}. The injected signals 
correspond to equal-mass, non-spinning, face-on binaries. In this geometry, only the $(2,|2|)$ modes contribute
 significantly to the signal, and therefore the inclusion of higher modes in these configurations is negligible. 

The computational efficiency of \phXE enables us to conduct systematic PE studies within practical runtimes, 
as summarized in Table~\ref{tab:pesummary}. These NR injections are used to investigate the impact of the 
inclusion of different mean anomaly harmonic content in the model with N$_{\rm harm}=5$ and N$_{\rm harm}=13$, 
or equivalently N$_e=2$ and N$_e=6$, see Eq. \eqref{eq:eq_harmonics}. 
For \texttt{SXS:BBH:1363}
 we perform additional runs varying the starting frequency. 
 All parameters of the injected signals are summarized in Table~\ref{tab:NRresults}.

We adopt the same prior distributions as in Refs.~\cite{Ramos-Buades:2023yhy,Gamboa:2024hli,Planas:2025feq}, namely $1/q \in [0.05,1]$, $\mathcal{M} \in [5,100],M_{\odot}$, and $\chi_i \in [0,0.99]$. The eccentricity prior is bounded by $e_{\max} = 0.4$.  All priors are defined at a reference frequency of 20 Hz.

\begin{figure*}
    \centering 
    \includegraphics[width=0.345\linewidth]{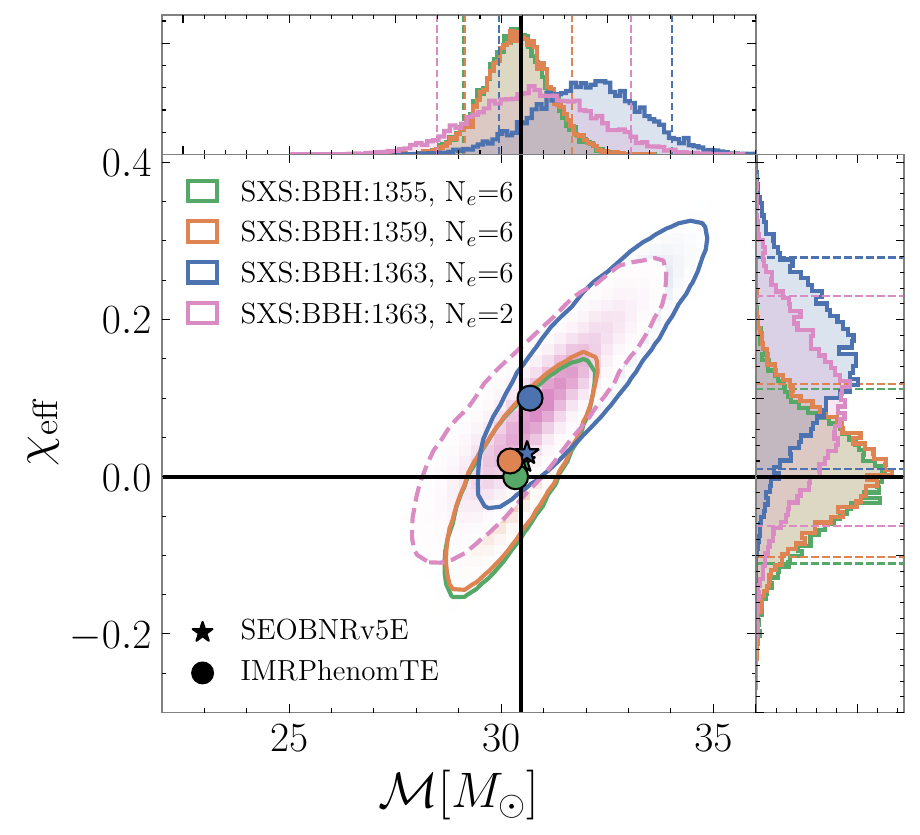}
    \includegraphics[width=0.32\linewidth]{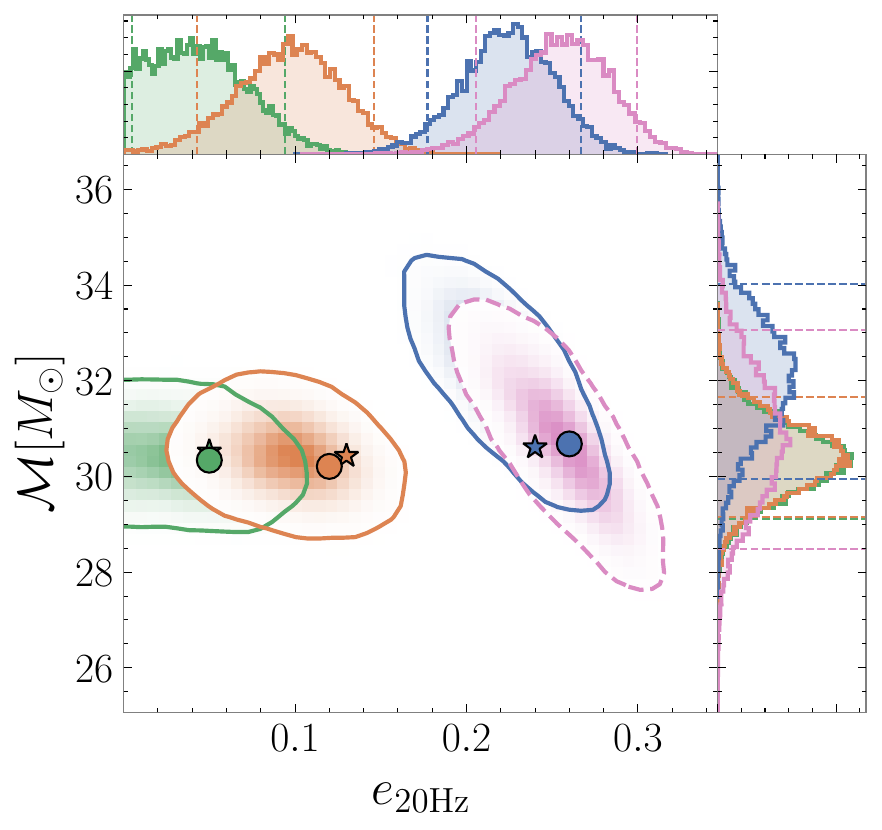}
    \includegraphics[width=0.32\linewidth]{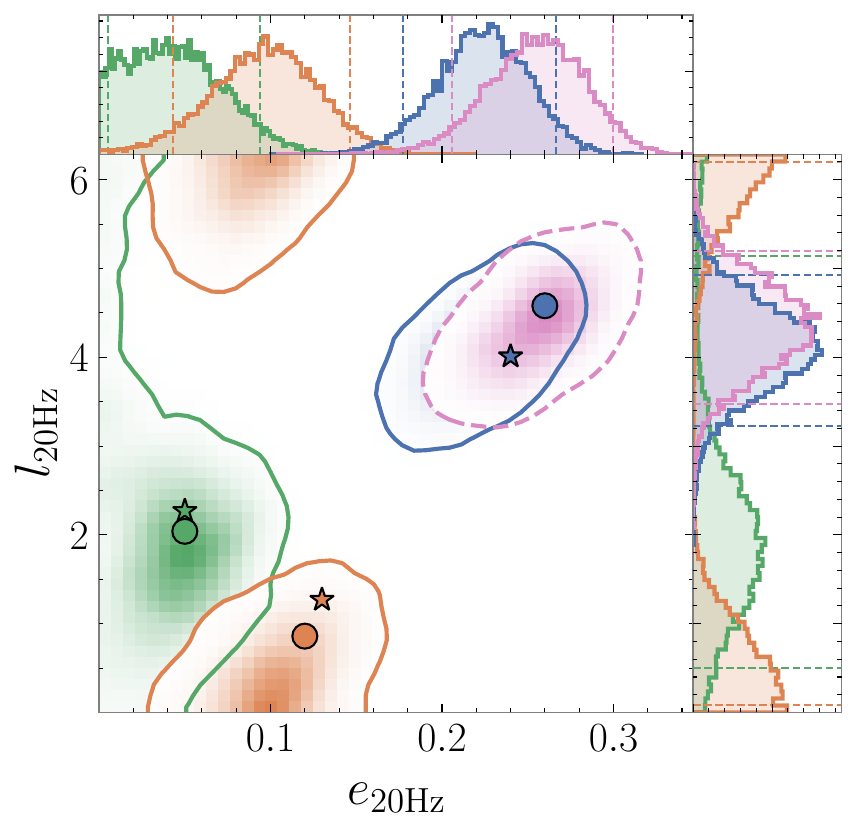}
    \caption{Posterior distributions for the three NR injections described in Table \ref{tab:NRresults}. The plots show marginalized 2D and 1D posteriors for: (i) chirp mass $\mathcal{M}$ and effective-spin $\chi_{\mathrm{eff}}$, (ii)  chirp mass $\mathcal{M}$ and reference eccentricity $e_{20\mathrm{Hz}}$ and iii) reference eccentricity $e_{20\mathrm{Hz}}$ and mean anomaly $l_{20\mathrm{Hz}}$. Injected values at $f_{\mathrm{ref}}=20$ Hz are marked by black lines. The circles and stars correspond to the median values of the \seobE and \phTE models reported in Refs. \cite{Planas:2025feq,Gamboa:2024hli}. All injections are analyzed with \phXE using $N_e=6$, i.e. N$_{\rm harm}=13$ mean anomaly harmonics, while for \texttt{SXS:BBH:1363} we additionally report the results using  $N_e=2$, i.e., N$_{\rm harm}=5$.
   }
\label{fig:NRinjections}
\end{figure*}

Figure~\ref{fig:NRinjections} shows the posterior distributions obtained for 
each NR injection, including marginalized one- and two-dimensional posteriors 
for the chirp mass $\mathcal{M}$, effective spin $\chi_{\mathrm{eff}}$, reference eccentricity 
$e_{20\mathrm{Hz}}$ and mean anomaly $l_{20\mathrm{Hz}}$. Apart from the injected values for
 the quasicircular parameters, we also include the median values of the \seobE and \phTE models
  reported in Refs. \cite{Gamboa:2024hli,Planas:2025feq}. For these models we report the GW eccentricity
   and mean anomaly parameters, which are computed using the \texttt{gw\_eccentricity} package \cite{Ramos-Buades:2022lgf,Shaikh:2023ypz,Shaikh:2025tae},
    evaluated at the same reference frequency. A summary of the injected intrinsic parameters, together with the 
    recovered median values and corresponding 90\% credible intervals, is provided in Table~\ref{tab:NRresults}.

The results presented in Fig.~\ref{fig:NRinjections} and Table~\ref{tab:NRresults} demonstrate 
that the \phXE model is able to recover the injected binary parameters with good accuracy across 
all NR injections. In all cases, the posterior distributions are Gaussian and unimodal, including 
the more eccentric configuration \texttt{SXS:BBH:1363}, indicating that the model robustly 
captures the salient features of the eccentric binary dynamics. Consistent with previous injection 
studies using the eccentric \texttt{SEOBNR} models~\cite{Ramos-Buades:2023yhy,Gamboa:2024hli}, 
our analysis restricted to the dominant harmonic exhibits small biases in the recovery of the 
luminosity distance $d_L$ and inclination $\iota$. As expected and shown in Ref. \cite{Planas:2025feq} 
the inclusion of higher-order modes (HMs) leads to a marked improvement in the recovery 
of $d_L$, as shown in Table~\ref{tab:NRresults}, while the
 inclination angle remains comparatively weakly constrained.

A more detailed interpretation is required for the injection with the highest eccentricity. 
For \texttt{SXS:BBH:1363}, a noticeable shift is observed in the effective-spin parameter 
posterior between the run including N$_{\rm harm}=5$ ($N_e=2$) and N$_{\rm harm}=13$ ($N_e=6$) mean anomaly harmonics. 

\begin{figure}
    \centering 
    \includegraphics[width=\columnwidth]{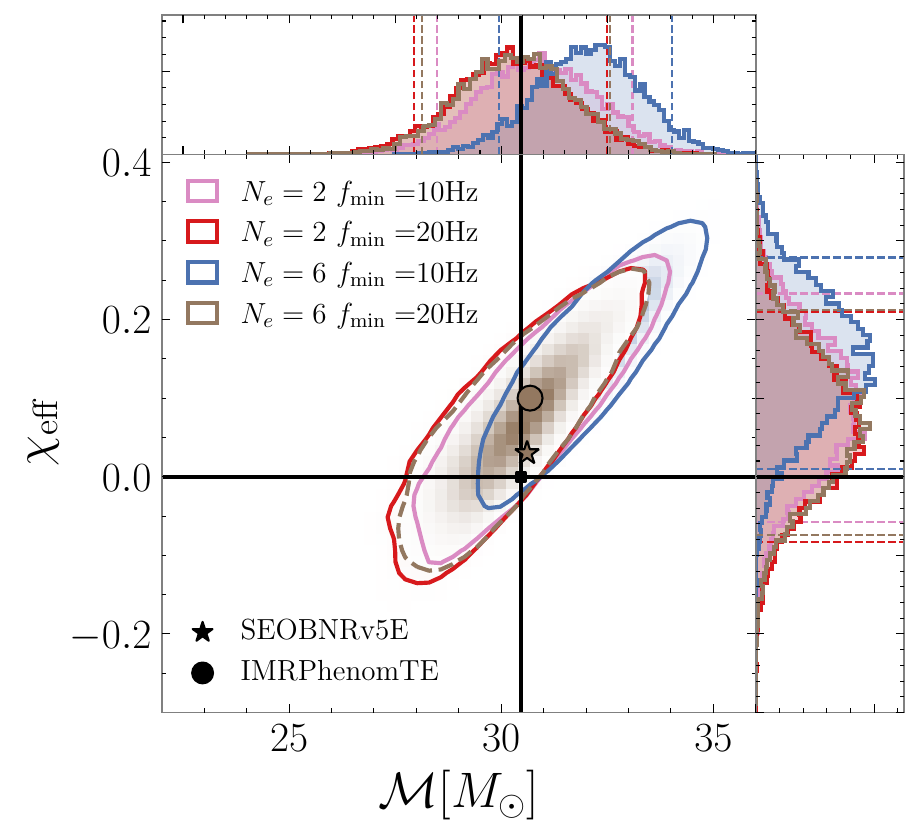}
    \includegraphics[width=\columnwidth]{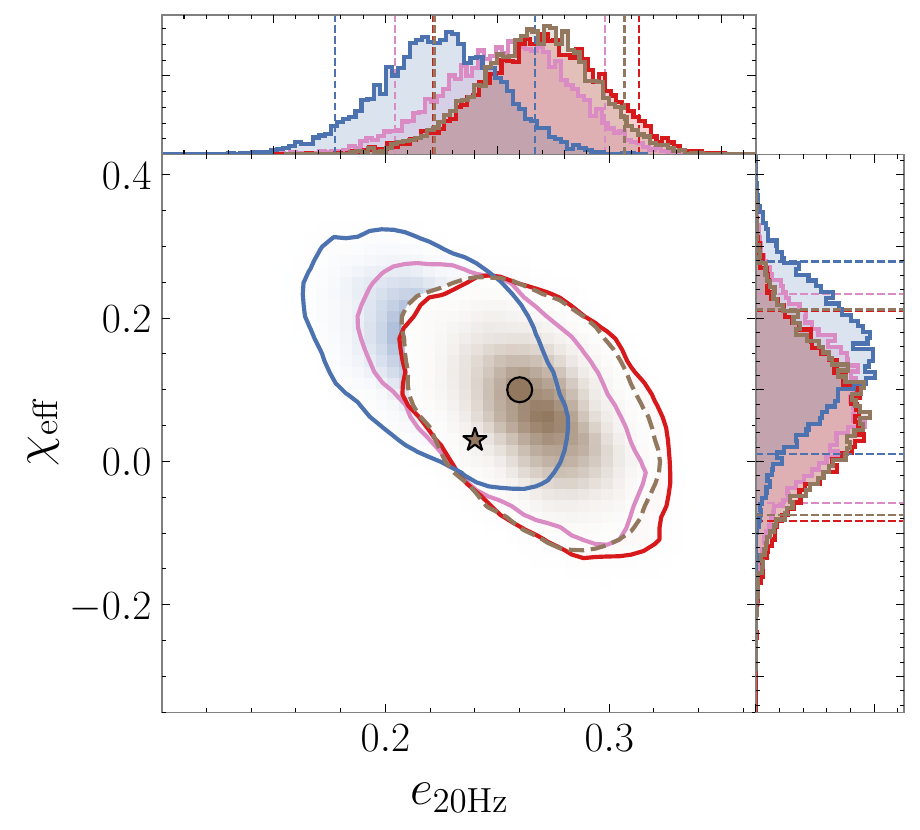} 
    \caption{Marginalized 2D and 1D posterior distributions for the \texttt{SXS:BBH:1363} NR injection described in Table \ref{tab:NRresults}. The plots show the  chirp mass $\mathcal{M}$ and effective spin $\chi_{\mathrm{eff}}$,  and the reference eccentricity $e_{20\mathrm{Hz}}$ and effective spin $\chi_{\mathrm{eff}}$. Injected values at $f_{\mathrm{ref}}=20$ Hz are marked by black lines. The circles and stars correspond to the values of the \seobE and \phTE models reported in Refs. \cite{Planas:2025feq,Gamboa:2024hli}. Injections are analyzed using \phXE with different number of mean anomaly harmonics (N$_{\rm harm}= 2N_e+1$) and starting frequency of waveform generation $f_{\rm min}$.
   }
\label{fig:NRinjections_v2}
\end{figure}

Comparing the 
posteriors for the N$_{\rm harm}=5$  run against the \seobE median values we observe that the N$_{\rm harm}=13$  
run shifts the $\chi_{\rm eff}$ posterior away from the injected value, 
but the eccentricity parameter is closer to the \seobE $e^{\rm GW}$ and $l^{\rm GW}$ posteriors. 
Note that although the definition of $e^{\rm GW}$  and $e_t$ used in \phXE are different, 
we are adopting the same eccentricity evolution equations in EOB coordinates, which are 
found to have qualitatively similar values as $e^{\rm GW}$  \cite{Gamboa:2024hli,Planas:2025feq}. 
The run with N$_{\rm harm}=5$ harmonics measures  $\chi_{\rm eff} = 0.09^{+0.11}_{-0.11}$, while 
for N$_{\rm harm}=13$ we observe a shift $\chi_{\rm eff} = 0.14^{+0.10}_{-0.11}$. In order to better 
understand these differences and the impact of the length of the waveform we perform 
additional runs with a minimum frequency of waveform generation of 20Hz. 
The results, displayed in Fig. \ref{fig:NRinjections_v2}, show that for the runs with N$_{\rm harm}=5$ harmonics a modification of the starting frequency from 10Hz to 20Hz does not change much the recovered value
of the effective-spin parameter,  $\chi_{\rm eff}^{10\rm{Hz}}=0.09^{+0.11}_{-0.11}$ and 
$\chi_{\rm eff}^{20\rm{Hz}}=0.06^{+0.11}_{-0.11}$, although the run starting at
20Hz is closer to the injected value. While for the N$_{\rm harm}$=13 harmonic runs we observe a 
more significant change,  $\chi_{\rm eff}^{20\rm{Hz}}=0.14^{+0.1}_{-0.11}$ and 
 $\chi_{\rm eff}^{20\rm{Hz}}=0.07^{+0.11}_{-0.11}$. Similar shifts are observed
  in the recovered chirp mass and reference eccentricity for the N$_{\rm harm}=13$ 
  run when moving from a starting frequency of 10Hz to 20Hz.

This fact points to the importance of finite length effects due 
to the limited duration of the injected NR waveform. A similar issue is found in 
Ref. \cite{Planas:2025feq} when performing higher order mode injections with this signal, 
and they are attributed to the limited duration of the NR waveform used to construct the injection. 
The NR waveform starts at 20~Hz and cannot be extended to lower frequencies, causing higher mean 
anomaly harmonics in \phXE to enter the analysis band and bias the recovery when templates are 
generated from 10~Hz. Additionally, we note that this particular NR simulation has been deprecated
 in the latest update of the SXS catalog \cite{Scheel:2025jct} indicating that the quality 
 of this particular NR signal is not representative of the accuracy standard of the SXS catalog. 

Overall, these NR injection studies show that \phXE achieves parameter-recovery performance comparable to that of the \seobE and \phTE models, despite exhibiting larger mismatches in some regions of parameter space. This highlights that mismatch estimates alone do not fully capture waveform performance in PE applications, and we leave for future work a detailed NR injection recovery study to fully characterize the accuracy of the \phXE model across parameter space. Moreover, the substantially improved computational efficiency of \phXE enables systematic exploration of modeling choices—such as the inclusion of eccentric mean anomaly harmonics and the choice of starting frequencies, which would be prohibitively expensive with more computationally intensive eccentric waveform models.

\subsection{GW events}\label{sec:PE_gw}

In this section, we analyze three GW events observed by the LIGO and Virgo detectors during the first and third observing runs~\cite{LIGOScientific:2018mvr,LIGOScientific:2021usb,LIGOScientific:2021djp}: GW150914, GW151226, and GW190521. We use strain data from the Gravitational Wave Open Science Center (GWOSC)~\cite{LIGOScientific:2019lzm}, together with the publicly released power spectral densities (PSDs), calibration uncertainties, and parameter-estimation products provided in the GWTC-2.1 catalog~\cite{LIGOScientific:2021usb}.

\begin{table*}
\centering
\renewcommand{\arraystretch}{1.3}
\begin{tabular}{@{}llccccccccccc@{}}
\toprule
\textbf{Event} & \textbf{Model} & $M/M_\odot$ & $\mathcal{M}/M_\odot$ & $1/q$ & $\chi_\text{eff}$ & $e_0$ & $l_0$ & $d_L$ & $\text{SNR}_{\text{N}}$ & $\log_{10}B_{\mathrm{E}/\mathrm{QC}}$ \\ 
\midrule
\multirow{3}{*}{GW150914} 
& \phXE  &  $71.52^{+2.59}_{-2.60}$  &  $30.93^{+1.15}_{-1.18}$  &  $0.86^{+0.11}_{-0.15}$  &  $-0.02^{+0.08}_{-0.09}$  &  $0.07^{+0.08}_{-0.06}$  &  $3.17^{+2.50}_{-2.54}$  &  $410^{+142}_{-144}$  &  $24.3^{+0.08}_{-0.14}$  &  $-0.41^{+0.12}_{+0.12}$  \\ 
& \phTEHM & $71.18^{+2.97}_{-2.90}$ & $30.89^{+1.28}_{-1.29}$ & $0.91^{+0.08}_{-0.16}$ & $-0.03^{+0.10}_{-0.10}$ & $^*0.06^{+0.09}_{-0.06}$ & $^*3.18^{+2.81}_{-2.88}$ & $465^{+132}_{-168}$ & $24.33^{+0.10}_{-0.15}$ & $-0.08^{+0.15}_{-0.15}$ \\ 
& \seobE  & $70.9^{+2.62}_{-2.8}$ & $30.72^{+1.15}_{-1.24}$ & $0.88^{+0.09}_{-0.14}$ & $-0.05^{+0.09}_{-0.05}$ & $^*0.06^{+0.07}_{-0.05}$ & $^*3.17^{+2.49}_{-2.54}$ & $480^{+116}_{-125}$ & -&$-0.57^{+0.13}_{-0.13}$ \\ 
\midrule
\multirow{2}{*}{GW151226} 
& \phXE &  $22.61^{+2.65}_{-0.47}$  &  $9.66^{+0.07}_{-0.1}$  &  $0.7^{+0.23}_{-0.31}$  &  $0.16^{+0.12}_{-0.06}$  &  $0.09^{+0.11}_{-0.07}$  &  $3.16^{+2.49}_{-2.56}$  &  $473^{+165}_{-179}$  &  $11.99^{+0.22}_{-0.31}$ &  $-0.45^{+0.14}_{+0.14}$ \\ 
& \texttt{SEOBNRv4E\_opt} &   ${22.82}^{+3.46}_{-0.59} $ & ${9.68}^{+0.07}_{-0.07} $ & ${0.66}^{+0.26}_{-0.32} $ & ${0.18}^{+0.13}_{-0.06} $ & ${0.04}^{+0.05}_{-0.04} $ & ${2.95}^{+2.67}_{-2.3} $ &   ${468}^{+170}_{-183} $ & - & - \\  
\midrule
\multirow{4}{*}{GW190521} 

& \phXE  &  $261.64^{+22.1}_{-19.17}$  &  $111.53^{+9.94}_{-12.09}$  &  $0.72^{+0.22}_{-0.26}$  &  $0.06^{+0.23}_{-0.22}$  &  $0.2^{+0.15}_{-0.16}$  &  $2.81^{+2.24}_{-1.98}$  &  $3993^{+1434}_{-1513}$  &  $13.45^{+0.13}_{-0.22}$   &  $0.03^{+0.1}_{+0.1}$ \\ 
& \phTEHM  & $259.1^{+26.4}_{-28.3}$ & $111.3^{+12.0}_{-15.5}$ & $0.78^{+0.20}_{-0.27}$ & $0.02^{+0.30}_{-0.34}$ & $^*0.31^{+0.13}_{-0.28}$ & $^*3.18^{+2.82}_{-2.88}$ & $4275^{+1490}_{-1732}$ & $14.44^{+0.21}_{-0.30}$ & $0.12^{+0.13}_{-0.13}$ \\
& \seobE  & $260.7^{+20.3}_{-19.5}$ & $111.4^{+9.7}_{-11.7}$ & $0.73^{+0.20}_{-0.21}$ & $0.05^{+0.20}_{-0.20}$ & $^*0.29^{+0.16}_{-0.23}$ & $^*3.14^{+2.54}_{-2.52}$ & $4786^{+1261}_{-1230}$ & - & $-0.36^{+0.11}_{-0.11}$ \\ 
\bottomrule
\end{tabular}
\caption{Median values and 90\% credible intervals for the posterior distributions shown in Fig.~\ref{fig:gwEvents} for the 3 analyzed GW events (indicated in each row), recovered with \phXE. 
For comparison, we also include results obtained using \seobE from Ref.~\cite{Gamboa:2024hli}, and \phTEHM from Ref. \cite{Planas:2025feq}.
The table reports the same parameters as Table~\ref{tab:NRresults}, as well as the log-10 Bayes factor between the eccentric (E) and the quasicircular (QC) hypothesis $\log_{10}B_{\mathrm{E}/\mathrm{QC}}$. For the QC hypothesis we have produced runs with the \phX model. All values are given at the reference frequency of 10 Hz for GW150914 and 5.5 Hz for GW190521.}
\label{tab:gwEvents}
\end{table*}

\begin{figure*}
    \centering
    GW150914
    \begin{minipage}{\textwidth}
    \includegraphics[width=0.32\linewidth]{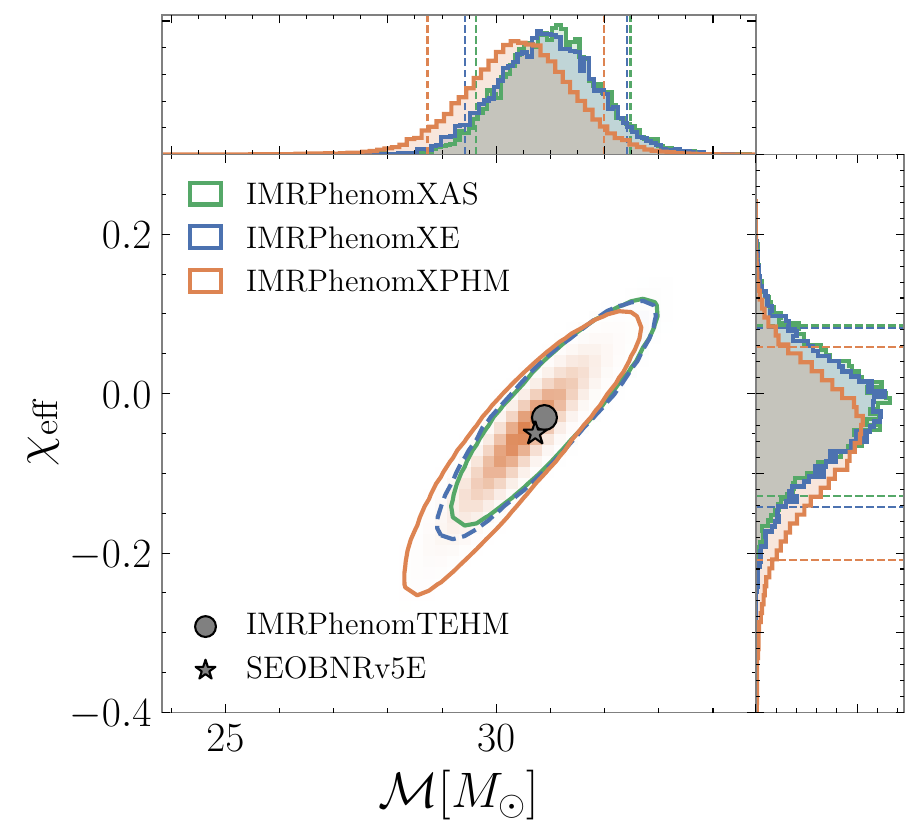}
    \includegraphics[width=0.3\linewidth]{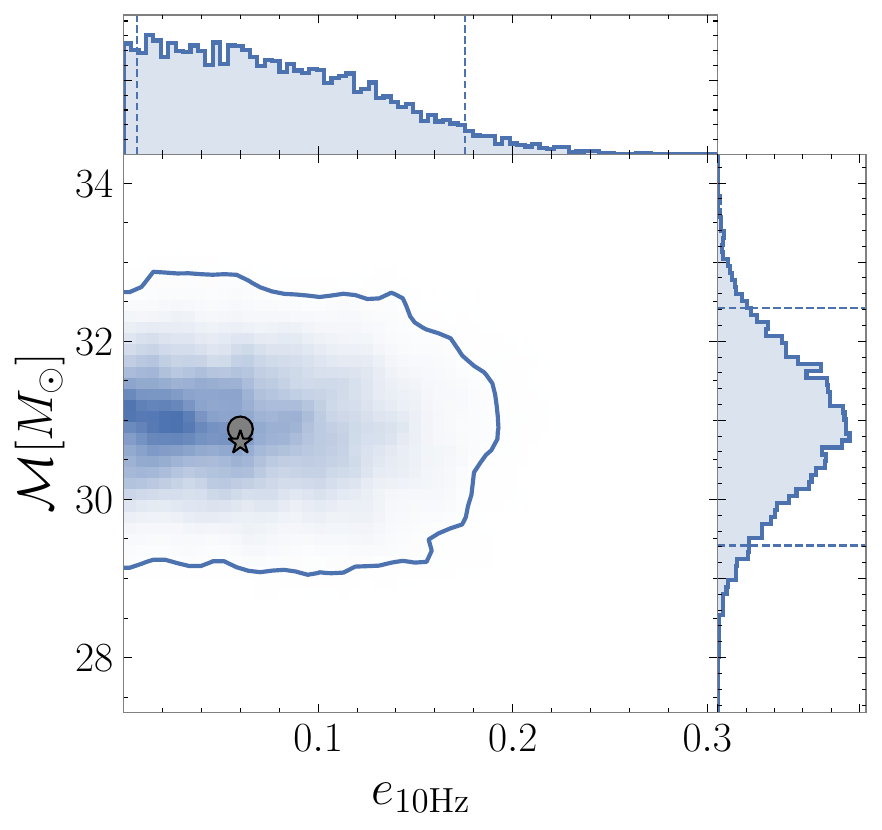}
    \includegraphics[width=0.3\linewidth]{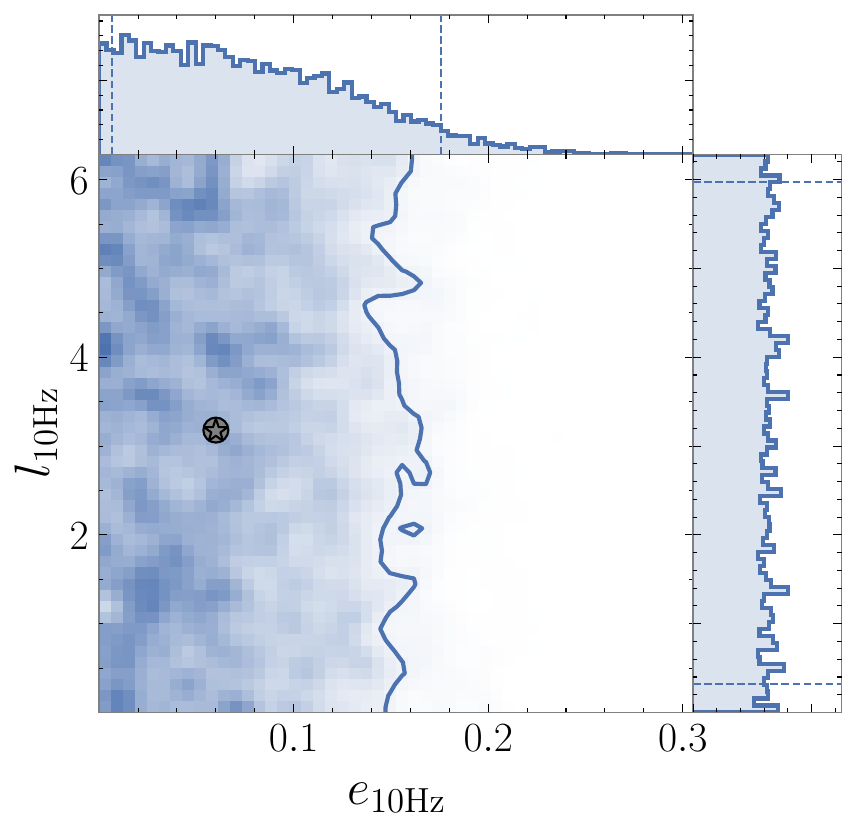}
    \vspace*{0.7em}
    \end{minipage}
    GW151226
    \begin{minipage}{\textwidth}
    \includegraphics[width=0.32\linewidth]{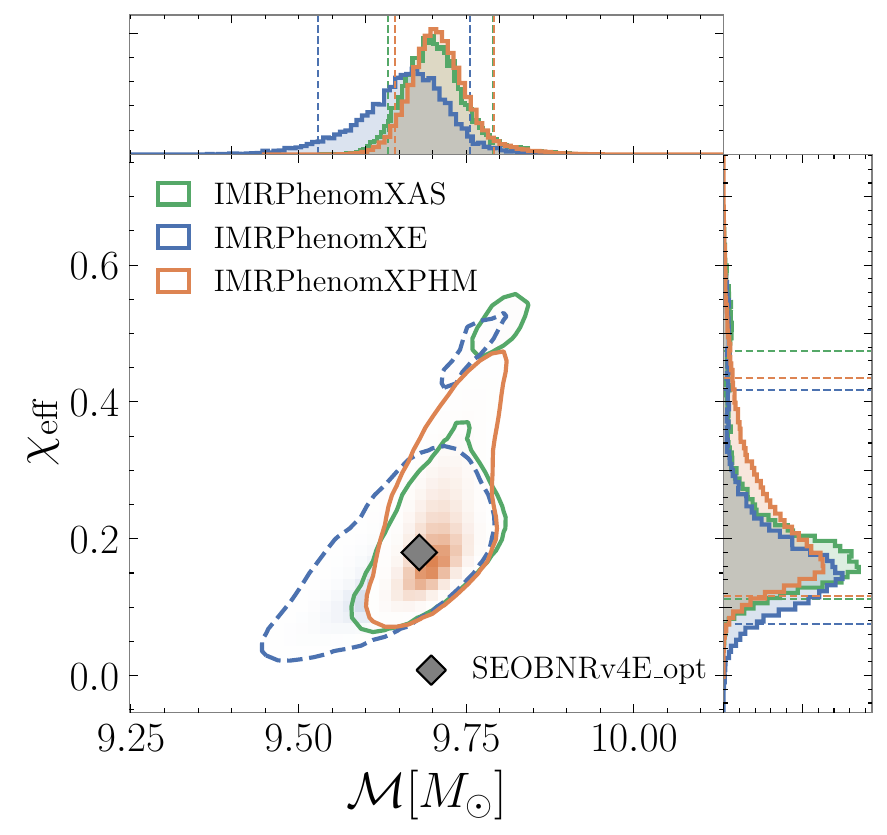}
    \includegraphics[width=0.3\linewidth]{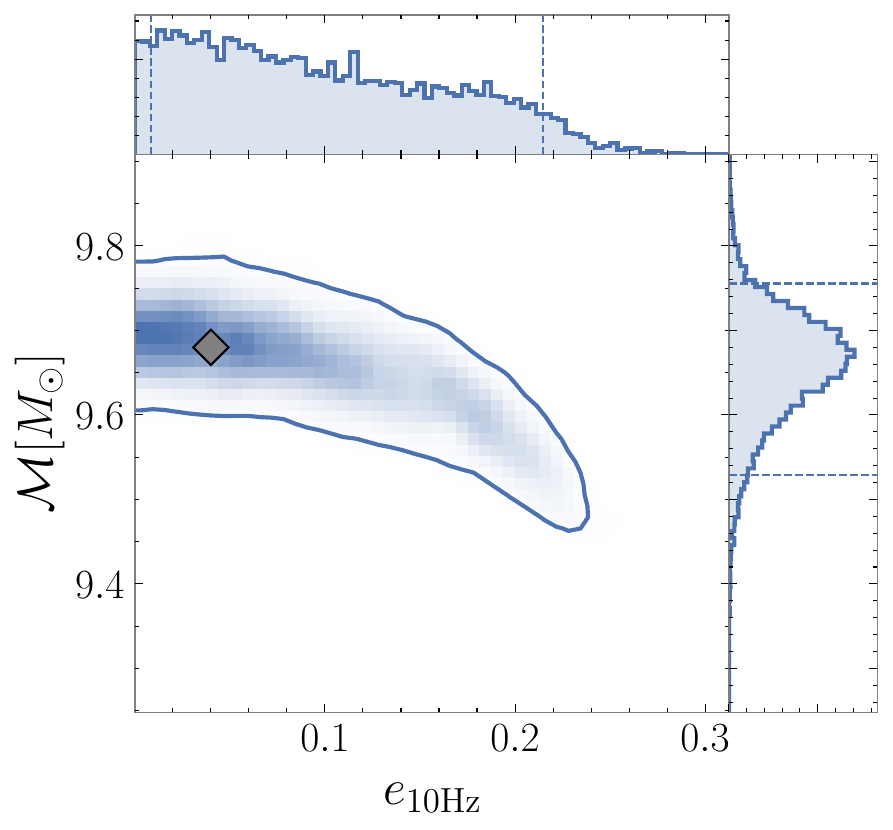}
    \includegraphics[width=0.3\linewidth]{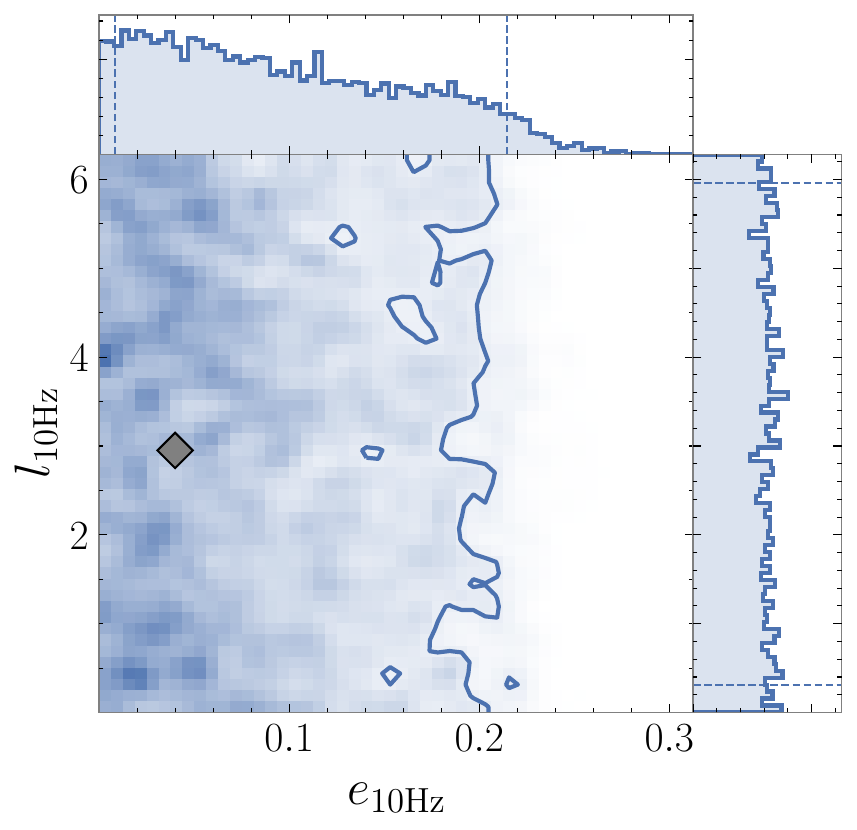}
    \vspace*{0.7em}
    \end{minipage}
    GW190521
    \begin{minipage}{\textwidth}
    \includegraphics[width=0.32\linewidth]{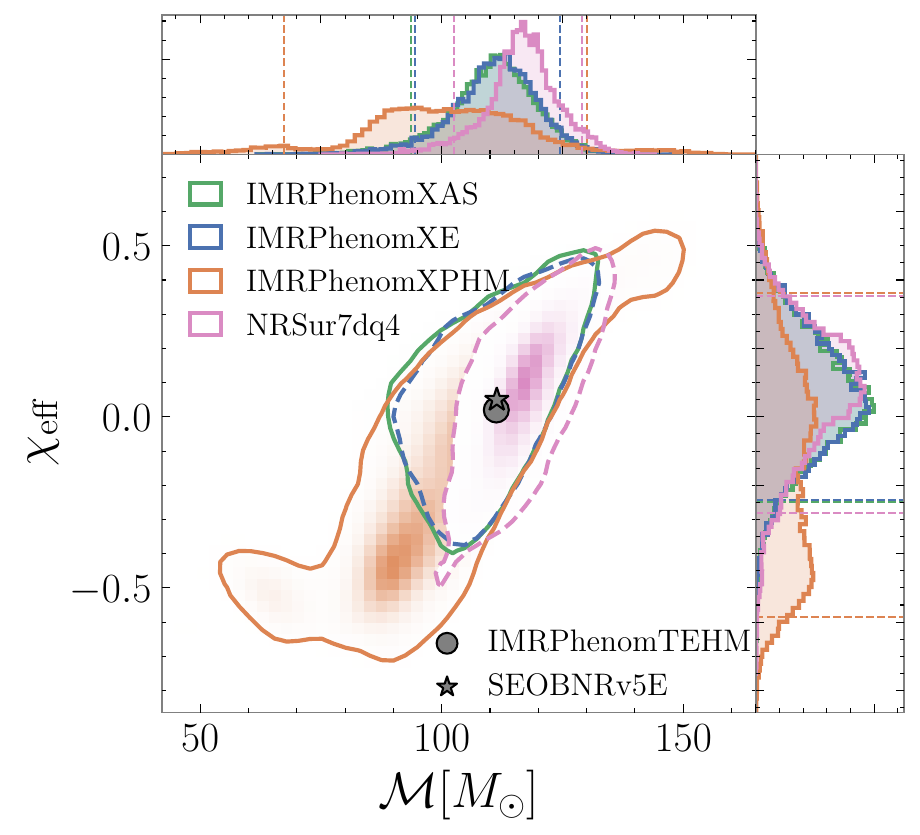}
    \includegraphics[width=0.3\linewidth]{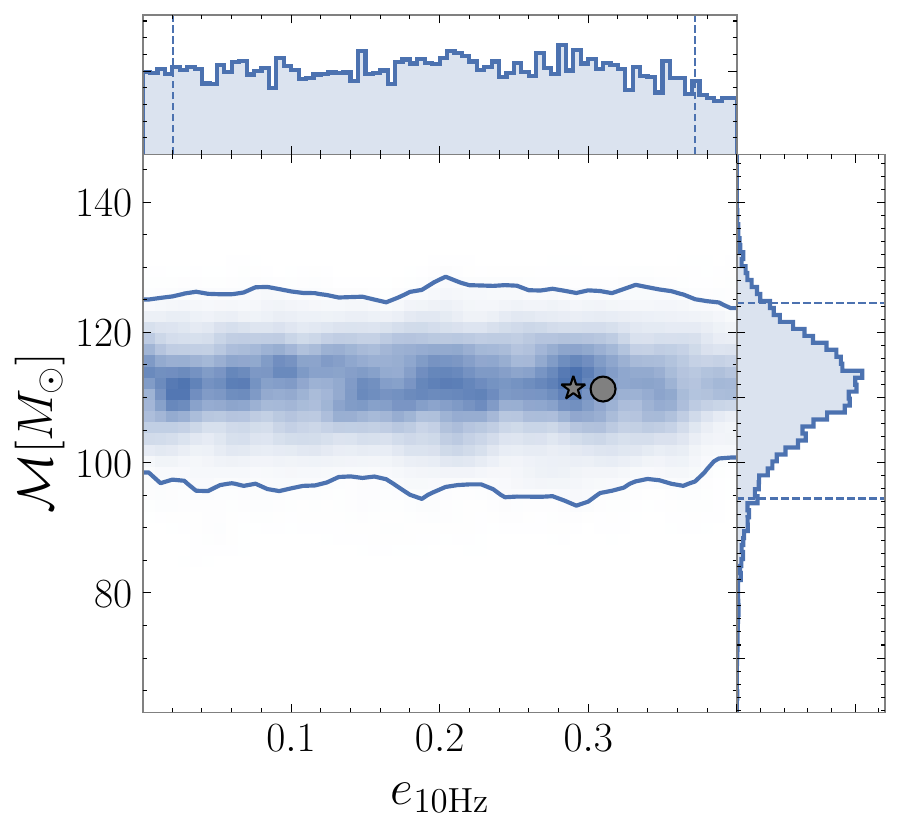}
    \includegraphics[width=0.3\linewidth]{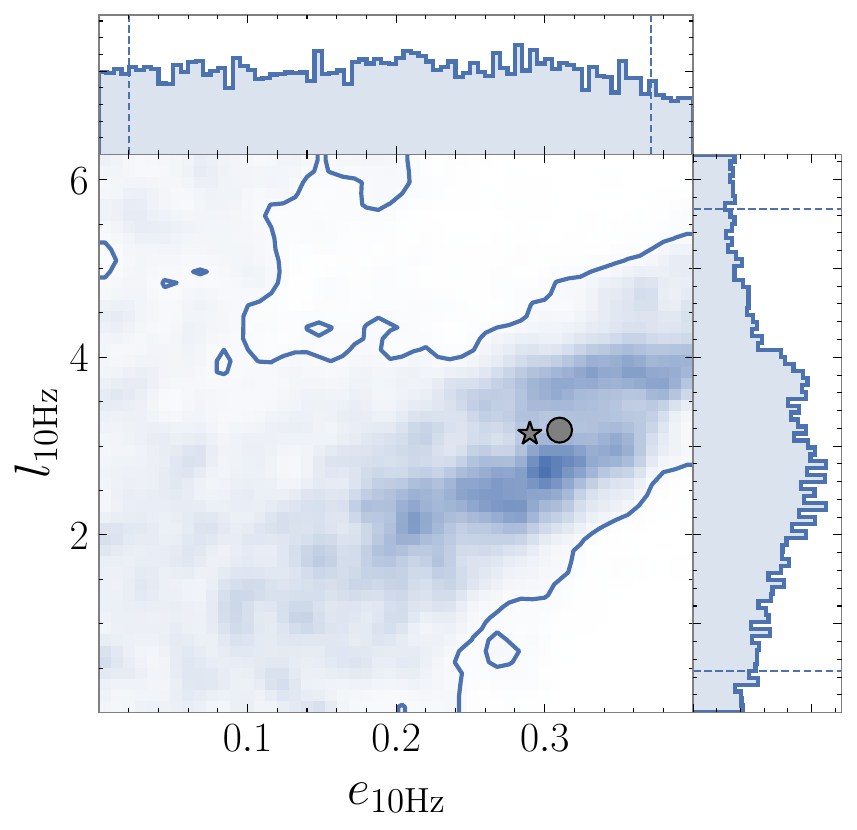}
    \end{minipage}
    \caption{Posterior distributions for 3 real GW events, GW150914 (\textit{top row}), GW151226 (\textit{mid row})  and GW190521 (\textit{bottom row}). The figure presents the posterior distributions of chirp mass and effective spin (\textit{first column}), chirp mass and reference eccentricity (\textit{second column}), and reference mean anomaly and eccentricity (\textit{third column}). All parameters are measured at a reference frequency of $f_{\mathrm{ref}}=10$ Hz. For each event we perform runs with the \phXE and \phX models, and include the \phXPHM results from the GWTC-2.1 catalog \cite{LIGOScientific:2021usb}. In the case of GW190521 we also include the NRSur7dq4 results from Ref. \cite{LIGOScientific:2020iuh}. For GW150914 and GW190521 we report the median values obtained by \phTEHM and \seobE in Refs. \cite{Gamboa:2024hli,Planas:2025feq}, and the median values of \texttt{SEOBNRv4E\_opt} for GW151226 from Ref. \cite{Ramos-Buades:2023yhy}. For the models \texttt{SEOBNRv4E\_opt}, \phTE and \seobE we indicate the GW eccentricity and mean anomaly values.}
    \label{fig:gwEvents}
\end{figure*}

\subsubsection*{GW150914}
GW150914, the first detected binary black hole (BBH) coalescence, remains one of the highest-SNR events ($\mathrm{SNR}\simeq23.7$) observed during the first three LVK observing runs~\cite{LIGOScientific:2016aoc,LIGOScientific:2021usb}. Its inferred source properties are consistent with a comparable-mass, weakly spinning binary~\cite{LIGOScientific:2016ebw}.

We analyze GW150914 using \phXE with priors uniform in inverse mass ratio, $1/q\in[0.05,1]$, and chirp mass, $\mathcal{M}\in[20,50],M_\odot$, resulting in uniform priors on the component masses. Uniform priors are also adopted for the initial eccentricity, $e_0\in[0,0.4]$, and mean anomaly, $\l_0\in[0,2\pi]$. All remaining priors follow those described in Sec.~\ref{sec:PE_nr}. Waveforms are generated starting at 10~Hz, where $e_0$ and $l_0$ are defined, ensuring that higher mean anomaly harmonics up to $j\geq +2$ included in \phXE are fully within the detector band when the likelihood evaluation begins at 20~Hz. 

Posterior distributions for the chirp mass, effective spin $\chi_{\rm eff}$, eccentricity, and  mean anomaly are shown in the top row of Fig.~\ref{fig:gwEvents}. Median values and 90\% credible intervals for additional parameters are summarized in Table~\ref{tab:gwEvents}. For comparison, we include posterior samples obtained with the quasi-circular precessing-spin model \phXPHM from the GWTC-2.1 catalog~\cite{LIGOScientific:2021usb}. The intrinsic parameters recovered with \phXE are consistent with those inferred using \phXPHM, as expected given the weakly spinning nature of GW150914. Additionally, we also report in Fig. \ref{fig:gwEvents} the median values obtained with the \seobE and \phTEHM models in Refs. \cite{Gamboa:2024hli,Planas:2025feq}. For the eccentricity and mean anomaly values of \seobE and \phTEHM we quote the GW eccentricity and mean anomaly values.

Although the median eccentricity inferred with \phXE is nonzero, $e_{\rm 10Hz}=0.07^{+0.09}_{-0.06}$, the posterior distribution shows strong support at zero eccentricity, in agreement with previous analyses using eccentric waveform models~\cite{LIGOScientific:2016ebw,Romero-Shaw:2019itr,Bonino:2022hkj,Iglesias:2022xfc,Gamboa:2024hli,Planas:2025feq}.  A comparison of Bayesian evidences through the log10 Bayes factor $\log_{10} \mathcal{B}_{\rm E/QC} = -0.42^{+0.12}_{-0.12}$, further disfavors the non-precessing eccentric hypothesis relative to the precessing quasi-circular one, indicating that GW150914 is consistent with a quasi-circular BBH merger.

\subsubsection*{GW151226}

GW151226 is among the lowest-mass BBH mergers detected during the first observing run and exhibits statistically significant support for a nonzero effective spin~\cite{LIGOScientific:2018mvr}. Previous studies constrained its eccentricity to be small at frequencies near 10~Hz~\cite{Romero-Shaw:2019itr,OShea:2021faf}.

Our analysis employs \phXE with uniform priors on the initial eccentricity, $e_0\in[0,0.4]$, and mean anomaly, $l_0\in[0,2\pi]$, as well as priors uniform in component masses via $1/q\in[0.125,1]$ and $\mathcal{M}\in[5,100],M_\odot$. Waveforms are generated starting at 10~Hz, reflecting the lower total mass of the system in order to include in band the $j\geq +2$  mean anomaly harmonics. For comparison, we include posterior samples obtained with the quasi-circular precessing-spin model \phXPHM from GWTC-2.1~\cite{LIGOScientific:2021usb}, and the median values obtained with the \texttt{SEOBNRv4E\_opt} model obtained in Ref. \cite{Ramos-Buades:2023yhy}. For the eccentricity and mean anomaly values of \texttt{SEOBNRv4E\_opt} we report the GW eccentricity and mean anomaly values.

The inferred intrinsic parameters in the mid row of Fig. \ref{fig:gwEvents} show broad agreement between \phXE and \texttt{IMRPhenomXPHM}, with differences attributable to the distinct physical effects included in each model. In particular, \texttt{IMRPhenomXPHM} incorporates spin precession and higher order modes, while \phXE describes the dominant-mode of non-precessing eccentric binaries. The eccentricity posterior inferred with \phXE peaks near zero, with $e_{\rm 10Hz}=0.09^{+0.11}_{-0.07}$, indicating that GW151226 is consistent with a quasi-circular binary. This obtained value of eccentricity is consistent with the GW eccentricity value obtained with \texttt{SEOBNRv4E\_opt} in Ref. \cite{Ramos-Buades:2023yhy}. 
We note that the \texttt{SEOBNRv4E\_opt} values are quoted at a reference frequency of 20Hz,
while the \phXE values at a reference frequency of 10Hz. This explains the difference in the 
obtained median values.
  Bayesian model comparison of the non-precessing spin quasicircular and non-precessing eccentric hypothesis yields a moderate preference for the quasicircular hypothesis with a $\log_{10} \mathcal{B}_{\rm E/QC} = -0.45^{+0.14}_{+0.14}$, consistent with previous results in the literature \cite{Ramos-Buades:2023yhy}.

\subsubsection*{GW190521}

GW190521 is an exceptional event characterized by only a few GW cycles in band, making it largely merger–ringdown dominated~\cite{LIGOScientific:2020iuh}. Its interpretation remains debated, with proposed scenarios ranging from eccentric mergers to head-on collisions~\cite{CalderonBustillo:2020fyi,Gamba:2021gap,Gayathri:2020coq,Romero-Shaw:2020thy}, although recent studies find limited evidence for eccentricity~\cite{Iglesias:2022xfc,Ramos-Buades:2023yhy,Gamboa:2024hli,Planas:2025feq}.

We analyze GW190521 using \phXE with uniform priors on $e_0\in[0,0.4]$ and $l_0\in[0,2\pi]$, and component-mass–uniform priors induced by $1/q\in[0.05,1]$ and $\mathcal{M}\in[60,200],M_\odot$. Waveforms are generated from 5.5~Hz to ensure that higher mean anomaly harmonics up to $j\geq +2$  are in band when the likelihood evaluation begins at 11~Hz. 

The resulting posteriors, shown in the bottom row of Fig.~\ref{fig:gwEvents}, exhibit large uncertainties in the eccentricity and mean anomaly. While the median eccentricity is $e^{\rm 10Hz}=0.2^{+0.15}_{-0.16}$, the posterior remains largely uninformative, reflecting the short duration of the signal. This limitation is expected, as eccentric effects in \phXE primarily enter during the inspiral, while the merger–ringdown is modeled assuming effective circularization. Additionally, we include the \phXPHM results from GWTC-2.1~\cite{LIGOScientific:2021usb}, and we observe some discrepancies in the quasicircular parameters, thus, we display the results obtained with quasicircular precessing spin \texttt{NRSur7dq4} \cite{Varma:2019csw}, for which we observe better agreement with our \phXE and \phX results. The quasicircular and eccentric parameters are consistent with the eccentric analysis using the  \seobE and \phTEHM  models in Refs. \cite{Gamboa:2024hli,Planas:2025feq}. 

Bayesian evidence comparison of the quasicircular aligned-spin and the eccentric aligned-spin hypothesis using the \phX and \phXE models shows that the eccentric hypothesis is slightly favored with a $\log_{10} \mathcal{B}_{\rm E/QC} = 0.03^{+0.10}_{-0.10}$. This value of Bayes factor is consistent with zero and shows that the non-precessing eccentric hypothesis is not strongly favored by the data, consistent with previous analyses~\cite{LIGOScientific:2020iuh,Ramos-Buades:2023yhy,Gamboa:2024hli,Planas:2025feq}. These results highlight the difficulty of measuring eccentricity in high-mass, merger-dominated signals and emphasize the need for waveform models that consistently incorporate eccentricity and spin precession through merger and ringdown.

Finally, the computational efficiency of \phXE enables full Bayesian inference for GW190521 on timescales of minutes using \texttt{Bilby} \cite{Ashton:2018jfp}. This efficiency makes \phXE a practical tool for systematic analyses of LVK catalogs and motivates future extensions incorporating spin precession and improved merger-ringdown modeling.


\section{Conclusions}\label{sec:conclusions}

In this work we developed \phXE, a frequency-domain phenomenological inspiral–merger–ringdown waveform model for non-precessing binary black holes
 on eccentric orbits, describing the dominant $(\ell,|m|)=(2,2)$ modes. The model extends the quasi-circular \phX framework 
 by incorporating eccentric inspiral dynamics through orbit-averaged quasi-Keplerian equations of motion evolved up to 
 third post-Newtonian order including spin effects, combined with a stationary phase approximation
  applied to eccentricity-expanded waveform expressions up to $\mathcal{O}(e^{12})$. 
  The merger–ringdown part assumes circularization and is constructed using the underlying \phX prescription,
   ensuring a consistent and well-defined quasi-circular limit.

The validation of the quasicircular limit of \phXE is performed by comparing to the quasicircular frequency-domain \phX  \cite{Pratten:2020fqn} 
and time-domain \phT  \cite{Estelles:2020twz} models. We find that for regions of parameter space where NR simulations 
are available the unfaithfulness against \phX and \phT is small and comparable, however, 
for high mass ratios and high spins where NR information is scarce, the \phXE inherits the waveform systematics 
between the \texttt{IMRPhenomX} and \phT families and is more accurate to the \phT model due 
to the construction of the inspiral of \phXE based on the time evolution of the \phT frequency. 

The accuracy of \phXE in the eccentric case is assessed through comparisons with 186 public
 eccentric NR simulations from the SXS catalog \cite{Scheel:2025jct},  see Fig.~\ref{fig:nr_mismatches}. For systems 
 with eccentricities below $e \lesssim 0.4$, we find unfaithfulness values below $3\%$ for $72\%$ of cases, 
 demonstrating that the model captures the dominant eccentric effects relevant for current 
 ground-based observations. At larger eccentricities, the performance degrades progressively, 
 reaching unfaithfulness values of order $\sim 10-20\%$, consistent with the expected limitations 
 of the small-eccentricity expansions and the stationary phase approximation used in the inspiral construction.

A defining feature of \phXE is its computational efficiency (see Fig. \ref{fig:benchmark}). Implemented within the \texttt{phenomxpy} infrastructure \cite{Garcia-Quiros:2025usi}, 
\phXE enables waveform generation and likelihood evaluations at speeds exceeding those of existing inspiral–merger–ringdown eccentric models. 
This efficiency facilitates systematic Bayesian inference studies, including injections into zero noise and analyses of 
observed GW events. We perform 3 equal mass NR injections with increasing eccentricity up to 0.3 and find that \phXE
is able to recover both the quasicircular and eccentric parameters accurately. For future work we leave the performance of a 
NR injection study with a large number of NR simulations  
to assess the biases in the recovered parameters of \phXE 
across parameter space, as well as the adaption of the code to be able to extract accurately
GW eccentricity and mean anomaly using the \texttt{gw\_eccentricity} \cite{Shaikh:2023ypz}. 

Besides NR injections, we investigate three GW events (GW150914, GW151226 and GW190521) from the first and third observing runs 
of the LVK detectors. Despite the restriction to the dominant harmonic, we demonstrate that \phXE 
can recover accurately source parameters for the three events, for which consistenly with previous results 
in the literature \cite{Ramos-Buades:2023yhy,Gamboa:2024hli,Planas:2025feq} we find no evidence of eccentricity. 
The analysis of these three events 
shows that \phXE can provide results in the timescale of minutes for high mass events such as GW150914 and GW190521, and 
of around 2 hours for lower mass events such as GW151226, using serial Bilby \cite{Ashton:2018jfp}. 
This salient feature of computationally efficiency enables the analysis of large number of GW events with 
a moderate computational cost, and we are currently conducting a study to analyze every single GW event
observed by the LVK up to the latest O4a catalog release \cite{LIGOScientific:2025slb}. 

The current implementation of \phXE is limited to aligned-spin systems 
and does not include higher-order modes or eccentric merger–ringdown effects. 
While these approximations are sufficient for a broad class of observed signals, 
they may become limiting for high-mass, high-inclination, or high–signal-to-noise-ratio binaries. 
Extensions to include additional physical effects, such as higher harmonics and spin precession, can be naturally 
incorporated within the phenomenological framework adopted here.

In conclusion, \phXE provides an accurate and efficient frequency-domain description 
of eccentric BBH coalescences and is immediately applicable to GW data analysis. 
The model offers a practical tool for current searches and parameter-estimation studies, 
and establishes a foundation for the development of more general and accurate
frequency-domain eccentric waveform models in the future.

\section*{Acknowledgements}
We would like to thank Aldo Gambo for helpful comments on the manuscript.
We would also like to thank Cecilio Garcia-Quiros and Hector Estelles Estrella for useful discussions about the model development, and Maria Rossello for helpful comments on the manuscript. 
A. Ramos-Buades is supported by the Veni research programme which is (partly) financed by the Dutch Research Council (NWO) under the grant VI.Veni.222.396; acknowledges support from the Spanish Agencia Estatal de Investigación grant PID2022-138626NB-I00 funded by MICIU/AEI/10.13039/501100011033 and the ERDF/EU,  PID2024-157460NA-I00; and the Spanish Ministerio de Ciencia, Innovación y Universidades (Beatriz Galindo, BG23/00056), co-financed by UIB.
This work was supported by the Universitat de les Illes Balears (UIB); the Spanish Agencia Estatal de Investigación grants PID2022-138626NB-I00, RED2024-153978-E, RED2024-153735-E, funded by MICIU/AEI/10.13039/501100011033 and the ERDF/EU; and the Comunitat Autònoma de les Illes Balears through the Conselleria d'Educació i Universitats with funds from the European Union - NextGenerationEU/PRTR-C17.I1 (SINCO2022/6719) and from the European Union - European Regional Development Fund (ERDF) (SINCO2022/18146). 
Authors also acknowledge the computational resources at the cluster CIT provided by LIGO Laboratory and supported by National Science Foundation Grants PHY-0757058 and PHY-0823459, as well as the cluster HAWK provided by Cardiff University and supported by STFC grant ST/I006285/1.

We thankfully acknowledge the computer resources from the Dutch national e-infrastructure with the support of the SURF Cooperative using grant no. EINF-7366 and NWO-2024.002, from the Red Española de Supercomputación (RES) and the computer resources (Picasso Supercomputer), technical expertise and assistance provided by the SCBI (Supercomputing and Bioinformatics) center of the University of Málaga (AECT-2025-1-0017, AECT-2025-2-0004, AECT-2025-3-0015).

This research has made use of data obtained from the Gravitational Wave Open Science Center~\cite{gwosc12,gwosc3}, a service of LIGO Laboratory, the LIGO Scientific Collaboration and the Virgo Collaboration. LIGO is funded by the U.S. National Science Foundation. Virgo is funded by the French Centre National de Recherche Scientifique (CNRS), the Italian Istituto Nazionale della Fisica Nucleare (INFN) and the Dutch Nikhef, with contributions by Polish and Hungarian institutes.

\appendix



\section{NR simulations}
\label{sec:nr_list}
In Table \ref{tab:nr} we list the NR simulations from the public SXS catalog \cite{Scheel:2025jct}, which are used in Sec. \ref{sec:ecc_NRcomparison} to assess the accuracy of the \phXE model.

\newcounter{tabletemp}
\setcounter{tabletemp}{\value{table}}

{
\renewcommand{\arraystretch}{1.1} 
\begin{table}[!]
\centering
\caption{NR simulations used in this work.
  Columns 2--7 give the mass ratio,$q$, z-component of the dimensionless spin vectors $\chi_{1,2;z}$, 
  the initial eccentricity, $e_0$, and mean anomaly $l_0$, the initial orbital frequency $M_0 \Omega_0$,
   and the number of orbits $N_\text{orbits}$. All these quantities are obtained from the metadata files 
   of the simulations \cite{Scheel:2025jct}. Last column reports the maximum mismatch across the 
   mass range $[20,200]M_\odot$ of \phXE against the NR simulation.
  }
\vspace{8pt}
\label{tab:nr}
\begin{tabular}{ c | c c c c c c | c}
\hline 
\hline 
$\text{SXS ID}$ & q & $\chi_{1,z}$ & $\chi_{2,z}$ & $e_0$ & $l_0$ & $M_0 \Omega_0$ & $\mathcal{M}_{\rm max} [\%]$ \\
\hline 
\hline 
SXS:BBH:0069 & 1.0 & 0.0 & 0.0 & 0.023 & 0.74 & 0.012 & \cellcolor{DarkSeaGreen2}{\textbf{0.1}} \\
SXS:BBH:0087 & 1.0 & 0.0 & 0.0 & 0.027 & 0.66 & 0.012 & \cellcolor{DarkSeaGreen2}{\textbf{0.1}} \\
SXS:BBH:0089 & 1.0 & -0.5 & 0.0 & 0.078 & 1.71 & 0.011 & \cellcolor{DarkSeaGreen2}{\textbf{0.21}} \\
SXS:BBH:0091 & 1.0 & 0.0 & 0.0 & 0.027 & 6.07 & 0.011 & \cellcolor{DarkSeaGreen2}{\textbf{0.18}} \\
SXS:BBH:0106 & 5.0 & 0.0 & 0.0 & 0.046 & 3.59 & 0.017 & \cellcolor{DarkSeaGreen2}{\textbf{0.21}} \\
SXS:BBH:0109 & 5.0 & -0.5 & 0.0 & 0.001 & 1.32 & 0.02 & \cellcolor{DarkSeaGreen2}{\textbf{0.29}} \\
SXS:BBH:0111 & 5.0 & -0.5 & 0.0 & 0.005 & 0.88 & 0.02 & \cellcolor{DarkSeaGreen2}{\textbf{0.3}} \\
SXS:BBH:0175 & 1.0 & 0.75 & 0.75 & 0.003 & 3.72 & 0.015 & \cellcolor{DarkSeaGreen2}{\textbf{0.06}} \\
SXS:BBH:0177 & 1.0 & 0.99 & 0.99 & 0.001 & 4.07 & 0.014 & \cellcolor{DarkSeaGreen2}{\textbf{0.16}} \\
SXS:BBH:0306 & 1.31 & 0.96 & -0.9 & 0.001 & 3.44 & 0.018 & \cellcolor{DarkSeaGreen2}{\textbf{0.38}} \\
SXS:BBH:0309 & 1.22 & 0.33 & -0.44 & 0.036 & 3.41 & 0.017 & \cellcolor{DarkSeaGreen2}{\textbf{0.11}} \\
SXS:BBH:0319 & 1.22 & 0.33 & -0.44 & 0.016 & 1.06 & 0.018 & \cellcolor{DarkSeaGreen2}{\textbf{0.09}} \\
SXS:BBH:0320 & 1.22 & 0.33 & -0.44 & 0.03 & 1.63 & 0.018 & \cellcolor{DarkSeaGreen2}{\textbf{0.09}} \\
SXS:BBH:0321 & 1.22 & 0.33 & -0.44 & 0.086 & 3.3 & 0.018 & \cellcolor{DarkSeaGreen2}{\textbf{0.25}} \\
SXS:BBH:0322 & 1.22 & 0.33 & -0.44 & 0.099 & 2.23 & 0.016 & \cellcolor{DarkSeaGreen2}{\textbf{0.2}} \\
SXS:BBH:0323 & 1.22 & 0.33 & -0.44 & 0.152 & 3.31 & 0.014 & \cellcolor{DarkSeaGreen2}{\textbf{0.58}} \\
SXS:BBH:0324 & 1.22 & 0.33 & -0.44 & 0.31 & 2.02 & 0.011 & \cellcolor{LightCyan1}{\textbf{2.23}} \\
SXS:BBH:0616 & 2.0 & 0.75 & 0.5 & 0.001 & 1.78 & 0.022 & \cellcolor{DarkSeaGreen2}{\textbf{0.1}} \\
SXS:BBH:0620 & 5.0 & -0.8 & 0.0 & 0.004 & 0.18 & 0.023 & \cellcolor{DarkSeaGreen2}{\textbf{0.4}} \\
SXS:BBH:0621 & 7.0 & -0.8 & 0.0 & 0.003 & 0.38 & 0.025 & \cellcolor{DarkSeaGreen2}{\textbf{0.69}} \\
SXS:BBH:1107 & 10.0 & 0.0 & 0.0 & 0.001 & 3.83 & 0.019 & \cellcolor{DarkSeaGreen2}{\textbf{0.45}} \\
SXS:BBH:1136 & 1.0 & -0.75 & -0.75 & 0.127 & 4.47 & 0.015 & \cellcolor{DarkSeaGreen2}{\textbf{0.77}} \\
SXS:BBH:1144 & 1.0 & -0.44 & -0.44 & 0.009 & 1.93 & 0.015 & \cellcolor{DarkSeaGreen2}{\textbf{0.13}} \\
SXS:BBH:1149 & 3.0 & 0.7 & 0.6 & 0.068 & 3.88 & 0.019 & \cellcolor{DarkSeaGreen2}{\textbf{0.19}} \\
SXS:BBH:1164 & 2.0 & 0.0 & 0.0 & 0.001 & 5.28 & 0.01 & \cellcolor{DarkSeaGreen2}{\textbf{0.17}} \\
SXS:BBH:1165 & 2.0 & 0.0 & 0.0 & 0.001 & 4.59 & 0.01 & \cellcolor{DarkSeaGreen2}{\textbf{0.08}} \\
SXS:BBH:1168 & 1.0 & 0.0 & 0.0 & 0.008 & 0.07 & 0.01 & \cellcolor{DarkSeaGreen2}{\textbf{0.15}} \\
SXS:BBH:1169 & 3.0 & -0.7 & -0.6 & 0.062 & 0.7 & 0.015 & \cellcolor{DarkSeaGreen2}{\textbf{0.69}} \\
SXS:BBH:1170 & 3.0 & -0.7 & -0.6 & 0.014 & 1.95 & 0.015 & \cellcolor{DarkSeaGreen2}{\textbf{0.27}} \\
SXS:BBH:1171 & 3.0 & -0.7 & -0.6 & 0.002 & 4.42 & 0.015 & \cellcolor{DarkSeaGreen2}{\textbf{0.25}} \\
SXS:BBH:1176 & 3.0 & 0.0 & 0.0 & 0.025 & 3.62 & 0.019 & \cellcolor{DarkSeaGreen2}{\textbf{0.04}} \\
SXS:BBH:1177 & 3.0 & 0.0 & 0.0 & 0.003 & 1.7 & 0.019 & \cellcolor{DarkSeaGreen2}{\textbf{0.03}} \\
SXS:BBH:1180 & 3.0 & 0.0 & 0.0 & 0.03 & 1.68 & 0.019 & \cellcolor{DarkSeaGreen2}{\textbf{0.02}} \\
SXS:BBH:1181 & 3.0 & 0.0 & 0.0 & 0.015 & 0.18 & 0.019 & \cellcolor{DarkSeaGreen2}{\textbf{0.03}} \\
SXS:BBH:1182 & 3.0 & 0.0 & 0.0 & 0.009 & 6.25 & 0.019 & \cellcolor{DarkSeaGreen2}{\textbf{0.02}} \\
SXS:BBH:1183 & 3.0 & 0.0 & 0.0 & 0.009 & 6.25 & 0.019 & \cellcolor{DarkSeaGreen2}{\textbf{0.02}} \\
SXS:BBH:1355 & 1.0 & 0.0 & 0.0 & 0.095 & 0.61 & 0.02 & \cellcolor{DarkSeaGreen2}{\textbf{0.43}} \\
SXS:BBH:1356 & 1.0 & 0.0 & 0.0 & 0.164 & 0.79 & 0.011 & \cellcolor{DarkSeaGreen2}{\textbf{0.57}} \\
SXS:BBH:1357 & 1.0 & 0.0 & 0.0 & 0.181 & 2.24 & 0.013 & \cellcolor{DarkSeaGreen2}{\textbf{0.61}} \\
SXS:BBH:1358 & 1.0 & 0.0 & 0.0 & 0.165 & 2.6 & 0.014 & \cellcolor{DarkSeaGreen2}{\textbf{0.7}} \\
SXS:BBH:1359 & 1.0 & 0.0 & 0.0 & 0.173 & 4.17 & 0.014 & \cellcolor{DarkSeaGreen2}{\textbf{0.48}} \\
SXS:BBH:1360 & 1.0 & 0.0 & 0.0 & 0.268 & 2.08 & 0.013 & \cellcolor{LightCyan1}{\textbf{1.22}} \\
SXS:BBH:1362 & 1.0 & 0.0 & 0.0 & 0.353 & 2.02 & 0.011 & \cellcolor{LightCyan1}{\textbf{2.54}} \\
SXS:BBH:1363 & 1.0 & 0.0 & 0.0 & 0.351 & 2.31 & 0.011 & \cellcolor{LightCyan1}{\textbf{2.18}} \\
SXS:BBH:1364 & 2.0 & 0.0 & 0.0 & 0.08 & 2.27 & 0.016 & \cellcolor{DarkSeaGreen2}{\textbf{0.42}} \\
SXS:BBH:1365 & 2.0 & 0.0 & 0.0 & 0.105 & 2.79 & 0.015 & \cellcolor{DarkSeaGreen2}{\textbf{0.41}} \\
SXS:BBH:1366 & 2.0 & 0.0 & 0.0 & 0.159 & 4.62 & 0.014 & \cellcolor{DarkSeaGreen2}{\textbf{0.86}} \\
SXS:BBH:1367 & 2.0 & 0.0 & 0.0 & 0.172 & 3.37 & 0.014 & \cellcolor{DarkSeaGreen2}{\textbf{0.46}} \\
SXS:BBH:1368 & 2.0 & 0.0 & 0.0 & 0.176 & 2.68 & 0.014 & \cellcolor{DarkSeaGreen2}{\textbf{0.8}} \\
SXS:BBH:1369 & 2.0 & 0.0 & 0.0 & 0.314 & 0.57 & 0.011 & \cellcolor{LightCyan1}{\textbf{1.96}} \\
SXS:BBH:1370 & 2.0 & 0.0 & 0.0 & 0.292 & 2.53 & 0.011 & \cellcolor{LightCyan1}{\textbf{2.59}} \\
SXS:BBH:1371 & 3.0 & 0.0 & 0.0 & 0.106 & 4.66 & 0.015 & \cellcolor{DarkSeaGreen2}{\textbf{0.24}} \\
SXS:BBH:1372 & 3.0 & 0.0 & 0.0 & 0.173 & 2.67 & 0.014 & \cellcolor{DarkSeaGreen2}{\textbf{0.64}} \\
SXS:BBH:1373 & 3.0 & 0.0 & 0.0 & 0.171 & 2.58 & 0.014 & \cellcolor{LightCyan1}{\textbf{1.13}} \\
SXS:BBH:1374 & 3.0 & 0.0 & 0.0 & 0.302 & 3.78 & 0.011 & \cellcolor{LightCyan1}{\textbf{2.25}} \\
SXS:BBH:1382 & 3.0 & 0.7 & 0.6 & 0.01 & 0.34 & 0.018 & \cellcolor{DarkSeaGreen2}{\textbf{0.32}} \\
\hline 
\hline 
\end{tabular}
\end{table}
}

\setcounter{table}{\value{tabletemp}}
{
\renewcommand{\arraystretch}{1.1} 
\begin{table}
\centering
\caption{ \emph{Continued}.}
\vspace{8pt}
\begin{tabular}{ c | c c c c c c | c}
\hline 
\hline 
$\text{SXS ID}$ & q & $\chi_{1,z}$ & $\chi_{2,z}$ & $e_0$ & $l_0$ & $M_0 \Omega_0$ & $\mathcal{M}_{\rm max} [\%]$ \\
\hline 
\hline 
SXS:BBH:1382 & 3.0 & 0.7 & 0.6 & 0.01 & 0.34 & 0.018 & \cellcolor{DarkSeaGreen2}{\textbf{0.32}} \\
SXS:BBH:1503 & 1.0 & 0.73 & 0.14 & 0.001 & 2.1 & 0.016 & \cellcolor{DarkSeaGreen2}{\textbf{0.12}} \\
SXS:BBH:2517 & 1.0 & 0.0 & 0.0 & 0.04 & 5.97 & 0.014 & \cellcolor{DarkSeaGreen2}{\textbf{0.16}} \\
SXS:BBH:2518 & 1.0 & 0.0 & 0.0 & 0.068 & 2.35 & 0.009 & \cellcolor{DarkSeaGreen2}{\textbf{0.18}} \\
SXS:BBH:2519 & 1.0 & 0.0 & 0.0 & 0.064 & 1.36 & 0.01 & \cellcolor{DarkSeaGreen2}{\textbf{0.12}} \\
SXS:BBH:2520 & 1.0 & 0.0 & 0.0 & 0.176 & 5.5 & 0.011 & \cellcolor{DarkSeaGreen2}{\textbf{0.61}} \\
SXS:BBH:2521 & 1.0 & 0.0 & 0.0 & 0.309 & 6.12 & 0.006 & \cellcolor{LightCyan1}{\textbf{1.31}} \\
SXS:BBH:2522 & 1.0 & 0.0 & 0.0 & 0.406 & 3.26 & 0.005 & \cellcolor{LightCyan1}{\textbf{1.36}} \\
SXS:BBH:2523 & 1.0 & 0.0 & 0.0 & 0.403 & 5.36 & 0.004 & \cellcolor{LightCyan1}{\textbf{2.39}} \\
SXS:BBH:2524 & 1.0 & 0.0 & 0.0 & 0.702 & 4.96 & 0.001 & \cellcolor{Wheat1}{\textbf{8.2}} \\
SXS:BBH:2525 & 1.0 & 0.0 & 0.0 & 0.611 & 6.21 & 0.002 & \cellcolor{Wheat1}{\textbf{3.95}} \\
SXS:BBH:2526 & 1.0 & 0.0 & 0.0 & 0.712 & 0.8 & 0.001 & \cellcolor{Wheat1}{\textbf{9.05}} \\
SXS:BBH:2527 & 1.0 & 0.0 & 0.0 & 0.799 & 3.78 & 0.0 & \cellcolor{DarkSalmon}{\textbf{22.16}} \\
SXS:BBH:2528 & 1.0 & 0.0 & 0.0 & 0.616 & 1.02 & 0.001 & \cellcolor{DarkSalmon}{\textbf{13.64}} \\
SXS:BBH:2529 & 2.0 & 0.0 & 0.0 & 0.046 & 5.95 & 0.012 & \cellcolor{DarkSeaGreen2}{\textbf{0.06}} \\
SXS:BBH:2530 & 2.0 & 0.0 & 0.0 & 0.186 & 5.34 & 0.01 & \cellcolor{DarkSeaGreen2}{\textbf{0.49}} \\
SXS:BBH:2531 & 2.0 & 0.0 & 0.0 & 0.307 & 0.36 & 0.006 & \cellcolor{LightCyan1}{\textbf{1.72}} \\
SXS:BBH:2532 & 2.0 & 0.0 & 0.0 & 0.406 & 4.51 & 0.005 & \cellcolor{LightCyan1}{\textbf{1.28}} \\
SXS:BBH:2533 & 2.0 & 0.0 & 0.0 & 0.504 & 4.66 & 0.004 & \cellcolor{LightCyan1}{\textbf{2.81}} \\
SXS:BBH:2534 & 2.0 & 0.0 & 0.0 & 0.702 & 5.15 & 0.001 & \cellcolor{DarkSalmon}{\textbf{10.29}} \\
SXS:BBH:2535 & 2.0 & 0.0 & 0.0 & 0.58 & 4.23 & 0.001 & \cellcolor{DarkSalmon}{\textbf{22.68}} \\
SXS:BBH:2536 & 3.0 & 0.0 & 0.0 & 0.058 & 5.21 & 0.009 & \cellcolor{DarkSeaGreen2}{\textbf{0.07}} \\
SXS:BBH:2537 & 3.0 & 0.0 & 0.0 & 0.191 & 5.46 & 0.008 & \cellcolor{DarkSeaGreen2}{\textbf{0.4}} \\
SXS:BBH:2538 & 3.0 & 0.0 & 0.0 & 0.306 & 5.33 & 0.006 & \cellcolor{DarkSeaGreen2}{\textbf{0.85}} \\
SXS:BBH:2539 & 3.0 & 0.0 & 0.0 & 0.254 & 5.59 & 0.008 & \cellcolor{LightCyan1}{\textbf{1.12}} \\
SXS:BBH:2540 & 3.0 & 0.0 & 0.0 & 0.203 & 5.76 & 0.012 & \cellcolor{DarkSeaGreen2}{\textbf{0.97}} \\
SXS:BBH:2541 & 3.0 & 0.0 & 0.0 & 0.305 & 5.29 & 0.006 & \cellcolor{LightCyan1}{\textbf{1.28}} \\
SXS:BBH:2542 & 3.0 & 0.0 & 0.0 & 0.306 & 4.53 & 0.006 & \cellcolor{LightCyan1}{\textbf{2.16}} \\
SXS:BBH:2543 & 3.0 & 0.0 & 0.0 & 0.405 & 5.48 & 0.005 & \cellcolor{Wheat1}{\textbf{4.27}} \\
SXS:BBH:2544 & 3.0 & 0.0 & 0.0 & 0.702 & 5.41 & 0.002 & \cellcolor{DarkSalmon}{\textbf{11.08}} \\
SXS:BBH:2545 & 4.0 & 0.0 & 0.0 & 0.043 & 5.62 & 0.012 & \cellcolor{DarkSeaGreen2}{\textbf{0.23}} \\
SXS:BBH:2546 & 4.0 & 0.0 & 0.0 & 0.184 & 5.01 & 0.01 & \cellcolor{DarkSeaGreen2}{\textbf{0.81}} \\
SXS:BBH:2547 & 4.0 & 0.0 & 0.0 & 0.306 & 4.8 & 0.006 & \cellcolor{Wheat1}{\textbf{3.63}} \\
SXS:BBH:2548 & 4.0 & 0.0 & 0.0 & 0.406 & 4.52 & 0.005 & \cellcolor{Wheat1}{\textbf{4.9}} \\
SXS:BBH:2549 & 4.0 & 0.0 & 0.0 & 0.505 & 4.9 & 0.004 & \cellcolor{Wheat1}{\textbf{5.86}} \\
SXS:BBH:2550 & 4.0 & 0.0 & 0.0 & 0.702 & 5.27 & 0.002 & \cellcolor{DarkSalmon}{\textbf{14.04}} \\
SXS:BBH:2551 & 4.0 & 0.0 & 0.0 & 0.587 & 4.96 & 0.001 & \cellcolor{DarkSalmon}{\textbf{29.53}} \\
SXS:BBH:2552 & 6.0 & 0.0 & 0.0 & 0.043 & 1.15 & 0.012 & \cellcolor{DarkSeaGreen2}{\textbf{0.37}} \\
SXS:BBH:2553 & 6.0 & 0.0 & 0.0 & 0.189 & 3.64 & 0.01 & \cellcolor{LightCyan1}{\textbf{1.67}} \\
SXS:BBH:2554 & 6.0 & 0.0 & 0.0 & 0.307 & 5.63 & 0.006 & \cellcolor{Wheat1}{\textbf{4.74}} \\
SXS:BBH:2555 & 6.0 & 0.0 & 0.0 & 0.405 & 5.69 & 0.005 & \cellcolor{Wheat1}{\textbf{5.31}} \\
SXS:BBH:2556 & 6.0 & 0.0 & 0.0 & 0.505 & 3.43 & 0.004 & \cellcolor{DarkSalmon}{\textbf{14.76}} \\
SXS:BBH:2557 & 6.0 & 0.0 & 0.0 & 0.603 & 5.29 & 0.002 & \cellcolor{DarkSalmon}{\textbf{26.39}} \\
SXS:BBH:2558 & 6.0 & 0.0 & 0.0 & 0.426 & 5.23 & 0.001 & \cellcolor{DarkSalmon}{\textbf{41.23}} \\
SXS:BBH:2559 & 8.0 & 0.0 & 0.0 & 0.016 & 5.69 & 0.016 & \cellcolor{DarkSeaGreen2}{\textbf{0.32}} \\
SXS:BBH:2560 & 8.0 & 0.0 & 0.0 & 0.187 & 5.8 & 0.01 & \cellcolor{LightCyan1}{\textbf{2.62}} \\
SXS:BBH:2561 & 8.0 & 0.0 & 0.0 & 0.305 & 5.34 & 0.006 & \cellcolor{Wheat1}{\textbf{7.74}} \\
SXS:BBH:2562 & 8.0 & 0.0 & 0.0 & 0.405 & 5.47 & 0.005 & \cellcolor{DarkSalmon}{\textbf{14.23}} \\
SXS:BBH:2563 & 8.0 & 0.0 & 0.0 & 0.405 & 5.14 & 0.005 & \cellcolor{DarkSalmon}{\textbf{22.93}} \\
SXS:BBH:2564 & 10.0 & 0.0 & 0.0 & 0.02 & 5.79 & 0.016 & \cellcolor{DarkSeaGreen2}{\textbf{0.46}} \\
SXS:BBH:2565 & 10.0 & 0.0 & 0.0 & 0.025 & 2.94 & 0.016 & \cellcolor{DarkSeaGreen2}{\textbf{0.42}} \\
SXS:BBH:2566 & 10.0 & 0.0 & 0.0 & 0.504 & 5.75 & 0.004 & \cellcolor{DarkSalmon}{\textbf{32.97}} \\
SXS:BBH:2567 & 10.0 & 0.0 & 0.0 & 0.505 & 2.21 & 0.005 & \cellcolor{DarkSalmon}{\textbf{27.7}} \\
SXS:BBH:2568 & 10.0 & 0.0 & 0.0 & 0.603 & 5.29 & 0.002 & \cellcolor{DarkSalmon}{\textbf{54.43}} \\
SXS:BBH:2570 & 1.0 & 0.0 & 0.0 & 0.02 & 3.73 & 0.013 & \cellcolor{DarkSeaGreen2}{\textbf{0.13}} \\
SXS:BBH:2571 & 1.0 & 0.0 & 0.0 & 0.017 & 4.76 & 0.013 & \cellcolor{DarkSeaGreen2}{\textbf{0.16}} \\
SXS:BBH:2572 & 1.0 & 0.0 & 0.0 & 0.017 & 0.64 & 0.012 & \cellcolor{DarkSeaGreen2}{\textbf{0.19}} \\
SXS:BBH:2573 & 1.0 & 0.0 & 0.0 & 0.016 & 2.7 & 0.013 & \cellcolor{DarkSeaGreen2}{\textbf{0.09}} \\
SXS:BBH:2574 & 1.0 & 0.0 & 0.0 & 0.041 & 3.56 & 0.014 & \cellcolor{DarkSeaGreen2}{\textbf{0.09}} \\
SXS:BBH:2575 & 1.0 & 0.0 & 0.0 & 0.036 & 5.25 & 0.013 & \cellcolor{DarkSeaGreen2}{\textbf{0.23}} \\
SXS:BBH:2576 & 1.0 & 0.0 & 0.0 & 0.038 & 1.08 & 0.012 & \cellcolor{DarkSeaGreen2}{\textbf{0.12}} \\
\hline 
\hline 
\end{tabular}
\end{table}
}

\setcounter{table}{\value{tabletemp}}
{
\renewcommand{\arraystretch}{1.1} 
\begin{table}
\centering
\caption{ \emph{Continued}.}
\vspace{8pt}
\begin{tabular}{ c | c c c c c c | c}
\hline 
\hline 
$\text{SXS ID}$ & q & $\chi_{1,z}$ & $\chi_{2,z}$ & $e_0$ & $l_0$ & $M_0 \Omega_0$ & $\mathcal{M}_{\rm max} [\%]$ \\
\hline 
\hline 
SXS:BBH:2578 & 1.0 & 0.0 & 0.0 & 0.093 & 4.27 & 0.014 & \cellcolor{DarkSeaGreen2}{\textbf{0.13}} \\
SXS:BBH:2579 & 1.0 & 0.0 & 0.0 & 0.095 & 5.32 & 0.013 & \cellcolor{DarkSeaGreen2}{\textbf{0.13}} \\
SXS:BBH:2580 & 1.0 & 0.0 & 0.0 & 0.091 & 0.14 & 0.011 & \cellcolor{DarkSeaGreen2}{\textbf{0.22}} \\
SXS:BBH:2581 & 1.0 & 0.0 & 0.0 & 0.095 & 2.51 & 0.012 & \cellcolor{DarkSeaGreen2}{\textbf{0.16}} \\
SXS:BBH:2582 & 1.0 & 0.0 & 0.0 & 0.126 & 4.28 & 0.016 & \cellcolor{DarkSeaGreen2}{\textbf{0.43}} \\
SXS:BBH:2583 & 1.0 & 0.0 & 0.0 & 0.133 & 5.64 & 0.012 & \cellcolor{DarkSeaGreen2}{\textbf{0.37}} \\
SXS:BBH:2584 & 1.0 & 0.0 & 0.0 & 0.13 & 0.02 & 0.01 & \cellcolor{DarkSeaGreen2}{\textbf{0.51}} \\
SXS:BBH:2585 & 1.0 & 0.0 & 0.0 & 0.13 & 2.47 & 0.012 & \cellcolor{DarkSeaGreen2}{\textbf{0.23}} \\
SXS:BBH:2586 & 1.0 & 0.0 & 0.0 & 0.155 & 4.26 & 0.017 & \cellcolor{DarkSeaGreen2}{\textbf{0.45}} \\
SXS:BBH:2587 & 1.0 & 0.0 & 0.0 & 0.16 & 6.05 & 0.012 & \cellcolor{DarkSeaGreen2}{\textbf{0.36}} \\
SXS:BBH:2588 & 1.0 & 0.0 & 0.0 & 0.161 & 0.1 & 0.009 & \cellcolor{DarkSeaGreen2}{\textbf{0.42}} \\
SXS:BBH:2589 & 1.0 & 0.0 & 0.0 & 0.161 & 2.54 & 0.012 & \cellcolor{DarkSeaGreen2}{\textbf{0.29}} \\
SXS:BBH:2590 & 1.0 & 0.0 & 0.0 & 0.162 & 4.23 & 0.017 & \cellcolor{DarkSeaGreen2}{\textbf{0.54}} \\
SXS:BBH:2591 & 1.0 & 0.0 & 0.0 & 0.169 & 5.93 & 0.012 & \cellcolor{DarkSeaGreen2}{\textbf{0.33}} \\
SXS:BBH:2592 & 1.0 & 0.0 & 0.0 & 0.17 & 0.08 & 0.009 & \cellcolor{DarkSeaGreen2}{\textbf{0.61}} \\
SXS:BBH:2593 & 1.0 & 0.0 & 0.0 & 0.168 & 2.52 & 0.011 & \cellcolor{DarkSeaGreen2}{\textbf{0.37}} \\
SXS:BBH:2594 & 1.0 & 0.0 & 0.0 & 0.21 & 3.66 & 0.018 & \cellcolor{DarkSeaGreen2}{\textbf{0.5}} \\
SXS:BBH:2595 & 1.0 & 0.0 & 0.0 & 0.211 & 6.02 & 0.012 & \cellcolor{DarkSeaGreen2}{\textbf{0.47}} \\
SXS:BBH:2596 & 1.0 & 0.0 & 0.0 & 0.209 & 1.44 & 0.009 & \cellcolor{DarkSeaGreen2}{\textbf{0.43}} \\
SXS:BBH:2597 & 1.0 & 0.0 & 0.0 & 0.21 & 2.15 & 0.011 & \cellcolor{DarkSeaGreen2}{\textbf{0.57}} \\
SXS:BBH:2598 & 1.0 & 0.0 & 0.0 & 0.004 & 3.49 & 0.013 & \cellcolor{DarkSeaGreen2}{\textbf{0.12}} \\
SXS:BBH:2599 & 1.0 & 0.0 & 0.0 & 0.205 & 5.1 & 0.011 & \cellcolor{DarkSeaGreen2}{\textbf{0.59}} \\
SXS:BBH:2600 & 1.0 & 0.0 & 0.0 & 0.21 & 5.68 & 0.012 & \cellcolor{DarkSeaGreen2}{\textbf{0.63}} \\
SXS:BBH:2601 & 1.0 & 0.0 & 0.0 & 0.25 & 1.77 & 0.008 & \cellcolor{DarkSeaGreen2}{\textbf{0.55}} \\
SXS:BBH:2602 & 1.0 & 0.0 & 0.0 & 0.273 & 1.06 & 0.009 & \cellcolor{DarkSeaGreen2}{\textbf{0.64}} \\
SXS:BBH:2603 & 1.0 & 0.0 & 0.0 & 0.283 & 5.51 & 0.01 & \cellcolor{DarkSeaGreen2}{\textbf{0.85}} \\
SXS:BBH:2604 & 1.0 & 0.0 & 0.0 & 0.257 & 0.28 & 0.012 & \cellcolor{DarkSeaGreen2}{\textbf{0.75}} \\
SXS:BBH:2605 & 1.0 & 0.0 & 0.0 & 0.282 & 2.27 & 0.008 & \cellcolor{DarkSeaGreen2}{\textbf{0.7}} \\
SXS:BBH:2606 & 1.0 & 0.0 & 0.0 & 0.297 & 2.39 & 0.01 & \cellcolor{DarkSeaGreen2}{\textbf{0.82}} \\
SXS:BBH:2607 & 1.0 & 0.0 & 0.0 & 0.31 & 5.56 & 0.01 & \cellcolor{DarkSeaGreen2}{\textbf{0.93}} \\
SXS:BBH:2608 & 1.0 & 0.0 & 0.0 & 0.306 & 5.06 & 0.01 & \cellcolor{LightCyan1}{\textbf{1.19}} \\
SXS:BBH:2609 & 1.0 & 0.0 & 0.0 & 0.308 & 0.46 & 0.007 & \cellcolor{LightCyan1}{\textbf{1.12}} \\
SXS:BBH:2610 & 1.0 & 0.0 & 0.0 & 0.307 & 1.91 & 0.01 & \cellcolor{LightCyan1}{\textbf{1.08}} \\
SXS:BBH:2611 & 1.0 & 0.0 & 0.0 & 0.307 & 4.96 & 0.011 & \cellcolor{LightCyan1}{\textbf{1.17}} \\
SXS:BBH:2612 & 1.0 & 0.0 & 0.0 & 0.306 & 5.09 & 0.01 & \cellcolor{DarkSeaGreen2}{\textbf{0.8}} \\
SXS:BBH:2613 & 1.0 & 0.0 & 0.0 & 0.21 & 6.07 & 0.012 & \cellcolor{DarkSeaGreen2}{\textbf{0.52}} \\
SXS:BBH:2614 & 1.0 & 0.0 & 0.0 & 0.21 & 5.91 & 0.011 & \cellcolor{DarkSeaGreen2}{\textbf{0.6}} \\
SXS:BBH:2615 & 1.0 & 0.0 & 0.0 & 0.288 & 3.06 & 0.007 & \cellcolor{DarkSeaGreen2}{\textbf{0.84}} \\
SXS:BBH:2616 & 1.0 & 0.0 & 0.0 & 0.286 & 2.22 & 0.007 & \cellcolor{DarkSeaGreen2}{\textbf{0.77}} \\
SXS:BBH:3703 & 1.35 & 0.04 & 0.01 & 0.001 & 2.17 & 0.025 & \cellcolor{DarkSeaGreen2}{\textbf{0.09}} \\
SXS:BBH:3933 & 2.0 & 0.0 & 0.0 & 0.709 & 4.88 & 0.001 & \cellcolor{Wheat1}{\textbf{9.0}} \\
SXS:BBH:3934 & 2.0 & 0.0 & 0.0 & 0.709 & 5.17 & 0.001 & \cellcolor{Wheat1}{\textbf{7.07}} \\
SXS:BBH:3935 & 2.0 & 0.0 & 0.0 & 0.709 & 5.08 & 0.001 & \cellcolor{Wheat1}{\textbf{8.62}} \\
SXS:BBH:3936 & 2.0 & 0.0 & 0.0 & 0.702 & 5.07 & 0.001 & \cellcolor{Wheat1}{\textbf{8.99}} \\
SXS:BBH:3937 & 2.0 & 0.0 & 0.0 & 0.702 & 4.86 & 0.001 & \cellcolor{DarkSalmon}{\textbf{11.92}} \\
SXS:BBH:3938 & 2.0 & 0.0 & 0.0 & 0.702 & 4.86 & 0.001 & \cellcolor{DarkSalmon}{\textbf{11.57}} \\
SXS:BBH:3939 & 6.0 & 0.0 & 0.0 & 0.602 & 5.78 & 0.002 & \cellcolor{DarkSalmon}{\textbf{33.1}} \\
SXS:BBH:3940 & 6.0 & 0.0 & 0.0 & 0.602 & 5.78 & 0.002 & \cellcolor{DarkSalmon}{\textbf{31.51}} \\
SXS:BBH:3941 & 6.0 & 0.0 & 0.0 & 0.602 & 5.67 & 0.002 & \cellcolor{DarkSalmon}{\textbf{34.33}} \\
SXS:BBH:3942 & 6.0 & 0.0 & 0.0 & 0.602 & 5.73 & 0.002 & \cellcolor{DarkSalmon}{\textbf{31.62}} \\
SXS:BBH:3943 & 6.0 & 0.0 & 0.0 & 0.602 & 5.83 & 0.002 & \cellcolor{DarkSalmon}{\textbf{31.0}} \\
SXS:BBH:3944 & 6.0 & 0.0 & 0.0 & 0.602 & 5.82 & 0.002 & \cellcolor{DarkSalmon}{\textbf{43.7}} \\
SXS:BBH:3945 & 2.0 & 0.0 & 0.0 & 0.702 & 5.11 & 0.001 & \cellcolor{DarkSalmon}{\textbf{11.93}} \\
\hline 
\hline 
\end{tabular}
\end{table}
}

\setcounter{table}{\value{tabletemp}}
{
\renewcommand{\arraystretch}{1.1} 
\begin{table}
\centering
\caption{ \emph{Continued}.}
\vspace{8pt}
\begin{tabular}{ c | c c c c c c | c}
\hline 
\hline 
$\text{SXS ID}$ & q & $\chi_{1,z}$ & $\chi_{2,z}$ & $e_0$ & $l_0$ & $M_0 \Omega_0$ & $\mathcal{M}_{\rm max} [\%]$ \\
\hline 
\hline 
SXS:BBH:3947 & 2.0 & 0.0 & 0.0 & 0.702 & 5.09 & 0.001 & \cellcolor{DarkSalmon}{\textbf{15.18}} \\
SXS:BBH:3948 & 2.0 & 0.0 & 0.0 & 0.702 & 5.09 & 0.001 & \cellcolor{DarkSalmon}{\textbf{12.68}} \\
SXS:BBH:3949 & 2.0 & 0.0 & 0.0 & 0.702 & 5.22 & 0.001 & \cellcolor{DarkSalmon}{\textbf{13.95}} \\
SXS:BBH:3950 & 2.0 & 0.0 & 0.0 & 0.61 & 4.82 & 0.001 & \cellcolor{DarkSalmon}{\textbf{13.27}} \\
SXS:BBH:3951 & 2.0 & 0.0 & 0.0 & 0.612 & 5.08 & 0.001 & \cellcolor{DarkSalmon}{\textbf{13.44}} \\
SXS:BBH:3952 & 2.0 & 0.0 & 0.0 & 0.702 & 5.12 & 0.001 & \cellcolor{DarkSalmon}{\textbf{10.1}} \\
SXS:BBH:3953 & 2.0 & 0.0 & 0.0 & 0.702 & 5.14 & 0.001 & \cellcolor{Wheat1}{\textbf{8.81}} \\
SXS:BBH:3954 & 2.0 & 0.0 & 0.0 & 0.702 & 5.13 & 0.001 & \cellcolor{DarkSalmon}{\textbf{11.74}} \\
SXS:BBH:3955 & 2.0 & 0.0 & 0.0 & 0.702 & 3.66 & 0.001 & \cellcolor{Wheat1}{\textbf{6.43}} \\
SXS:BBH:3956 & 2.0 & 0.0 & 0.0 & 0.702 & 3.63 & 0.001 & \cellcolor{Wheat1}{\textbf{8.76}} \\
SXS:BBH:3957 & 2.0 & 0.0 & 0.0 & 0.702 & 3.75 & 0.001 & \cellcolor{DarkSalmon}{\textbf{11.85}} \\
SXS:BBH:3958 & 2.0 & 0.0 & 0.0 & 0.702 & 5.02 & 0.001 & \cellcolor{Wheat1}{\textbf{6.06}} \\
SXS:BBH:3959 & 2.0 & 0.0 & 0.0 & 0.702 & 3.62 & 0.001 & \cellcolor{DarkSalmon}{\textbf{12.0}} \\
SXS:BBH:3971 & 1.0 & 0.0 & 0.0 & 0.041 & 2.88 & 0.014 & \cellcolor{DarkSeaGreen2}{\textbf{0.06}} \\
SXS:BBH:3972 & 1.0 & 0.0 & 0.0 & 0.403 & 5.36 & 0.004 & \cellcolor{LightCyan1}{\textbf{2.39}} \\
\hline 
\hline 
\end{tabular}
\end{table}
}
 
\clearpage
 
\bibliography{references} 

\end{document}